\documentclass[a4paper,11pt]{article}
\makeatletter
\g@addto@macro\bfseries{\boldmath}
\makeatother
\usepackage[english]{babel}
\usepackage{jheppub}
\pdfoutput=1
\usepackage[T1]{fontenc} 
\usepackage{bbold}
\usepackage{amsmath}
\usepackage{empheq}
\usepackage{booktabs}
\usepackage[dvipsnames]{xcolor}
\usepackage[utf8]{inputenc}
\usepackage{xspace}
\usepackage{scalerel}
\usepackage[most]{tcolorbox}
\usepackage{multirow}
\usepackage{scalefnt}
\usepackage{bold-extra}
\usepackage{afterpage}

\DeclareRobustCommand{\ensuremathrm}[1]{\ensuremath{\mathrm{#1}}\xspace}

\DeclareRobustCommand{\GeV}{\ensuremathrm{GeV}\xspace}

\def\sss{\scriptscriptstyle}

\newcommand\F{${\rm F}$}
\newcommand\FJ{${\rm FJ}$}
\newcommand\FJJ{${\rm FJJ}$}
\newcommand\PhiB{\Phi_{\scriptscriptstyle \rm F}}

\newcommand\PhiBJ{\Phi_{\scriptscriptstyle \rm FJ}}

\newcommand\PhiBJJ{\Phi_{\scriptscriptstyle \rm FJJ}}
\newcommand{\Fcorr}{F^{\tmop{corr}}}

\newcommand{\as}{\alpha_s}

\newcommand{\cO}[1]{{\cal O}\left(#1\right)}

\newcommand{\pt}{p_{\text{\scalefont{0.77}T}}}
\newcommand{\ptvec}{{\vec{p}_{\text{\scalefont{0.77}T}}}}
\newcommand{\ptrad}{{p_{\text{\scalefont{0.77}T,rad}}}}

\newcommand{\muF}{{\mu_{\text{\scalefont{0.77}F}}}}
\newcommand{\muR}{{\mu_{\text{\scalefont{0.77}R}}}}
\newcommand{\muRzero}{{\mu_{\text{\scalefont{0.77}R}}^{\text{\scalefont{0.77}(0)}}}}
\newcommand{\KF}{K_{\text{\scalefont{0.77}F}}}
\newcommand{\KR}{K_{\text{\scalefont{0.77}R}}}
\newcommand{\KQ}{K_{\text{\scalefont{0.77}Q}}}

\newcommand{\mQQ}{{m_{Q\bar{Q}}}}

\newcommand{\noun}[1]{{\scshape #1}}

\newcommand{\POWHEG}{\noun{Powheg}\xspace}
\newcommand{\MadSpin}{\noun{MadSpin}\xspace}

\newcommand{\POWHEGBOX}{\noun{Powheg-Box}\xspace}
\newcommand{\POWHEGBOXVTWO}{\noun{Powheg-Box-V2}\xspace}

\newcommand{\minlobare}{{\noun{MiNLO}}\xspace}
\newcommand{\minlo}{{\noun{MiNLO$^{\prime}$}}\xspace}
\newcommand{\minnlo}{{\noun{MiNNLO$_{\rm PS}$}}\xspace}

\newcommand{\PYTHIA}[1]{\noun{Pythia{#1}}\xspace}

\newcommand{\geneva}{\noun{Geneva}\xspace}
\newcommand{\unnlops}{\noun{Unnlops}\xspace}

\newcommand{\setupinclusive}{\texttt{setup-inclusive}\xspace}
\newcommand{\setupleptonic}{\texttt{setup-leptonic}\xspace}
\newcommand{\setupsemi}{\texttt{setup-semi-leptonic}\xspace}
\newcommand{\setuphadronic}{\texttt{setup-hadronic}\xspace}

\newcommand{\abar}{\frac{\as}{2\pi}}
\newcommand{\abarmu}[1]{\frac{\as(#1)}{2\pi}}

\newcommand{\citere}[1]{ref.\,\cite{#1}}
\newcommand{\citeres}[1]{refs.\,\cite{#1}}
\newcommand{\eqn}[1]{eq.\,(\ref{#1})}
\newcommand{\neqn}[1]{eqs.\,(\ref{#1})}
\newcommand{\fig}[1]{figure\,\ref{#1}}

\newcommand{\sct}[1]{section\,\ref{#1}}

\newcommand{\app}[1]{appendix\,\ref{#1}}

\newcommand{\LambdaPWG}{\Lambda_{\rm pwg}}

\newcommand{\ccbar}{{c\bar{c}}}

\usepackage{xcolor}
\newcommand{\mathd}{\mathrm{d}}
\newcommand{\tmop}[1]{\ensuremath{\operatorname{#1}}}

\newenvironment{itemizedot}{\begin{itemize} }{\end{itemize}}

\DeclareMathOperator{\Tr}{Tr}

\newtcolorbox{empheqboxed}{colback=white!35, 
 colframe=black,
 width=\textwidth,
 sharpish corners,
 top=-2mm, 
 bottom=0pt
}

\title{Top-pair production at the LHC with \minnlo{}}

\author[a]{Javier Mazzitelli,}
\author[b]{Pier Francesco Monni,}
\author[c]{Paolo Nason,}
\author[c,d]{Emanuele Re,}
\author[a]{\qquad\qquad Marius Wiesemann}
\author[a,e]{and Giulia Zanderighi}

\emailAdd{jmazzi@mpp.mpg.de}
\emailAdd{pier.monni@cern.ch}
\emailAdd{paolo.nason@mib.infn.it}
\emailAdd{emanuele.re@mib.infn.it}
\emailAdd{marius.wiesemann@mpp.mpg.de}
\emailAdd{zanderi@mpp.mpg.de}

\affiliation[a]{Max-Planck-Institut f\"ur Physik, F\"ohringer Ring 6,
  80805 M\"unchen, Germany}

\affiliation[b]{CERN, Theoretical Physics Department, CH-1211 Geneva 23, Switzerland}

\affiliation[c] {Universit\`a di Milano\,-\,Bicocca and INFN, Sezione di
  Milano\,-\,Bicocca, Piazza della Scienza 3, 20126 Milano, Italy}

\affiliation[d]{LAPTh, Universit\'e Grenoble Alpes, Universit\'e Savoie Mont Blanc, CNRS, 74940 Annecy, France}

\affiliation[e]{Physik-Department, Technische Universit\"at M\"unchen, James-Franck-Strasse 1, 85748 Garching, Germany}

\abstract{
  We consider the production of a pair of heavy quarks and illustrate
  the derivation of the \minnlo{} method to match
  next-to-next-to-leading order calculations with parton showers
  (NNLO+PS) for this class of processes.  As a first application, we
  construct an event generator for the fully differential simulation
  of hadronic top-quark pair production at NNLO+PS and discuss all
  details of its implementation in a parton shower Monte Carlo
  framework.
  We present new phenomenological results for the Large Hadron
  Collider obtained by including the tree-level decays of the top
  quarks, while accounting for spin-correlation effects. A
  comprehensive comparison to LHC measurements shows an excellent
  description of experimental data across multiple hadronic and
  leptonic particle-level observables.
  The computer code is available for download within the \POWHEGBOX{}.
}

\keywords{Perturbative QCD}

\preprint{
\vspace{-24pt}
  \begin{flushright}
  CERN-TH-2021-220\\
  MPP-2021-211\\
  LAPTH-046/21
  \end{flushright}
}

\begin{document}

\maketitle

\section{Introduction}
\label{sec:intro}

The experimental precision reached by present and future LHC
experiments demands theoretical simulations with an accuracy at the
edge of (or beyond) what can be achieved with current technology.
The data--theory comparisons used in the precise studies of the SM and
the extraction of its SM parameters directly profit from smaller
experimental and theoretical uncertainties. Furthermore,
high-precision predictions and measurements play a fundamental role in
the quest for new-physics phenomena at the LHC when searching for
small deviations from the SM picture. Theoretical predictions for the
LHC are commonly computed as a perturbative expansion in the coupling
constants, with corrections in the strong coupling of QCD being the
most important ones at a hadron collider.
For typical LHC production processes the computation of
next-to-next-to-leading order (NNLO) QCD corrections (or even beyond
in some cases) on the theoretical side are crucial to keep up with the
experimental precision. Moreover, the matching of such fixed-order
calculations at NNLO QCD with increasingly accurate parton-shower
generators, which can provide a description of full-fledged hadronic
LHC events, is indispensable to fully exploit the vast potential of
data taken at the LHC.

The matching to a QCD parton-shower simulation becomes crucial not
only for a direct comparison at hadron level between theoretical
predictions and experimental measurements, but also to guarantee
physical predictions in corners of phase space where fixed-order QCD
calculations become unreliable. The latter generally fail to provide
physical results for observables sensitive to soft/collinear QCD
radiation due to a hierarchy between two or more scales that lead to
large logarithms of their ratios. The inclusion of soft/collinear QCD
emissions to all orders in perturbation theory, as provided for
instance by a parton shower, effectively resums the large logarithmic
corrections, leading to a physical description.
The recent progress in this field has been remarkable, both on the
matching technology to combine higher-order QCD corrections with
parton shower Monte Carlo
generators~\cite{Frixione:2002ik,Nason:2004rx,Jadach:2015mza,Nason:2021xke,Hamilton:2012rf,Hamilton:2013fea,Alioli:2013hqa,Alioli:2015toa,Alioli:2021qbf,Hoeche:2014aia,Monni:2019whf,Monni:2020nks,Campbell:2021svd}, 
as well as on the development of
parton-shower algorithms whose logarithmic accuracy can be formally
established and systematically improved for a wide class of
observables (see e.g.\
\citeres{Hoche:2017hno,Dulat:2018vuy,Dasgupta:2018nvj,Bewick:2019rbu,Dasgupta:2020fwr,Forshaw:2020wrq,Hamilton:2020rcu,Gellersen:2021eci}).

The problem of matching NNLO QCD corrections to (leading logarithmic)
parton showers (NNLO+PS) has been addressed with different approaches:
reweighting of \minlo
generators~\cite{Hamilton:2012rf,Hamilton:2013fea},
\geneva~\cite{Alioli:2013hqa,Alioli:2015toa,Alioli:2021qbf},
\unnlops~\cite{Hoeche:2014aia},
\minnlo~\cite{Monni:2019whf,Monni:2020nks}, and Sector Showers
matching~\cite{Campbell:2021svd}.
All processes involving only two massless coloured legs (either in the
initial or final state) at the Born level can be described with
NNLO+PS accuracy by now, and many results have been already
obtained~\cite{Hamilton:2013fea,Karlberg:2014qua,Frederix:2015fyz,Astill:2016hpa,Astill:2018ivh,Re:2018vac,Bizon:2019tfo,Alioli:2015toa,Alioli:2019qzz,Alioli:2020fzf,Alioli:2020qrd,Alioli:2021qbf,Alioli:2021egp,Cridge:2021hfr,Hoeche:2014aia,Hoche:2014dla,Hoche:2018gti,Monni:2019whf,Monni:2020nks,Lombardi:2020wju,Lombardi:2021rvg,Buonocore:2021fnj,Lombardi:2021wug,Zanoli:2021iyp,Hu:2021rkt}.
On the other hand, reactions with final-state coloured particles cannot be
straightforwardly simulated with the same theoretical accuracy.

Recently, the \minnlo method was extended to more complicated collider
reactions containing colour charges in both the initial and the
final state already at the Born level, and specifically to the NNLO+PS
simulation of top-quark pair production at hadron colliders~\cite{Mazzitelli:2020jio}.
%
%
Besides being an important background in Higgs-physics measurements
and new-physics searches (see e.g.\
\citeres{deFlorian:2016spz,Brivio:2019ius,Ellis:2020unq}), this
process is used for high-precision studies of the properties of the
top quark, which have by now been performed at the level of both
inclusive and multi-differential observables~\cite{Aad:2019ntk,
  ATLAS:2019hau,Aad:2020tmz,ATLAS:2020ccu,
  Khachatryan:2016kzg,Sirunyan:2017uhy,
  Sirunyan:2018ucr,
  Sirunyan:2018ptc,CMS:2018htd, 
  Sirunyan:2017mzl}. 
These investigations have led to the precise extraction of fundamental
parameters of the Standard Model, such as $\alpha_s$,
parton density functions (PDFs), and the top-quark mass (see
e.g.\ \citeres{Chatrchyan:2013haa,
  Klijnsma:2017eqp,
  Sirunyan:2019zvx,
  Aad:2019mkw,
  Cooper-Sarkar:2020twv
}).

The crucial role of top-quark pair production for the precision
physics programme of the LHC has motivated a significant progress in
the associated theoretical calculations.
Specifically, fixed-order computations of the perturbative expansion
in the strong coupling constant $\alpha_s$ are known up to
NNLO~\cite{Baernreuther:2012ws,Czakon:2012zr,Czakon:2012pz,Czakon:2013goa,Czakon:2015owf,Czakon:2016ckf,Catani:2019iny,Catani:2019hip,Catani:2020tko}
(also including the top-quark decays in the narrow-width
approximation~\cite{Behring:2019iiv,Czakon:2020qbd}), full off-shell
effects have been studied extensively at the NLO
order~\cite{Bevilacqua:2010qb,Denner:2012yc,Bevilacqua:2015qha,Bevilacqua:2016jfk},
and electroweak (EW) corrections have been computed up to
next-to-leading order
(NLO)~\cite{Beenakker:1993yr,Bernreuther:2005is,Kuhn:2006vh,Denner:2016jyo,Czakon:2017wor}.
In certain kinematic regimes, a reliable perturbative description
requires the all-order resummation of large logarithmic
corrections~\cite{Zhu:2012ts,Li:2013mia,Catani:2014qha,Catani:2018mei,%
  Beneke:2011mq,Beneke:2012wb,Ju:2020otc,Alioli:2021ggd}.  Some of the above
calculations have been consistently combined to obtain  
state-of-the-art perturbative predictions for top-quark pair production at the LHC~\cite{Czakon:2019txp}.  The
striking accuracy of experimental measurements of the top-quark mass
requires pushing theoretical calculations to the edge of what can be
achieved with perturbative methods (for recent reviews
see\ \citeres{Nason:2017cxd,Hoang:2020iah}), and motivates new studies of
non-perturbative aspects of top-quark physics (see
e.g.\ \citeres{Beneke:2016cbu,Hoang:2017btd,Hoang:2018zrp,FerrarioRavasio:2018ubr}).

In this article, we present a detailed derivation of the method 
presented in \citere{Mazzitelli:2020jio} and describe all necessary steps to
build an event generator with NNLO+PS accuracy for top-quark pair production.
In particular, besides the construction of the event generator, we
discuss the implementation of dynamical perturbative (i.e.\
renormalisation, factorisation and resummation) scales in the
simulation, which we use to provide a robust estimate of the
associated theoretical errors.
Moreover, we describe how the top-quark decays and spin correlations are included.
We exploit the newly developed event generator to carry out an
extensive phenomenological study of several differential observables,
for which we present a comparison to recent LHC experimental
measurements, both for inclusive observables extrapolated to the
$t\bar{t}$ phase space and in the fiducial phase space of all relevant
top-decay modes (i.e.\ leptonic, semi-leptonic and hadronic).
Overall, we find a remarkable agreement with data within the 
substantially reduced uncertainties compared to lower-order Monte-Carlo 
predictions.
The computer code is publicly released with the article in the \POWHEGBOXVTWO{} framework.\footnote{Instructions 
to download the code can be found on the webpage \url{http://powhegbox.mib.infn.it/}.}

The paper is organised as follows:
Section~\ref{sec:minnlo_colour_singlet} presents an introduction to the
\minnlo method for colour-singlet production at hadron colliders, and
in \sct{sec:ttbar} we derive its extension to the production of
two heavy quarks (specifically a top-quark pair).
In \sct{sec:comput} several computational aspects are discussed,
including the scale dependence of the \minnlo{} formulae and the
inclusion of the top-quark decays and spin correlations at tree-level.
Phenomenological results are presented in \sct{sec:results} and a
detailed comparison of the \minnlo{} predictions to experimental data
from the LHC experiments is performed, for on-shell top quarks as well
as leptonic, semi-leptonic, and hadronic decays of the two top quarks.
Section~\ref{sec:summary} contains our summary and conclusions. 
In \app{app:resum} we provide technical details on the additional
resummation ingredients necessary to
implement the \minnlo method for top-quark pair production. 
In \app{app:additionalplots} we show further observables 
comparing \minnlo{} predictions with data that were not discussed 
in the main text for fully leptonic top-decays including
(\ref{app:leptonictau}) and excluding (\ref{app:leptonicnotau}) $\tau$
decays to electrons and muons, and for the semi-leptonic (\ref{app:semi}) and fully-hadronic
(\ref{app:hadronic}) decay mode.

\section{From \POWHEG{} to \minnlo{}: the colour singlet case}
\label{sec:minnlo_colour_singlet}
In this section we summarise in a schematic way the main features of
the \minnlo{} method.
We consider the production of a colour-singlet final state \F{} with
transverse momentum $\pt{}$. For sake of illustration we assume \F{}
to be a Higgs boson produced in gluon fusion. To define the goals of
the method developed in this article, we start by stating what the
requirements of a NNLO+PS prediction are:
\begin{itemize}
\item to reach NNLO accuracy for observables that are inclusive over
  QCD radiation (possibly without introducing any extra
  resolution/slicing scales);
\item to reach NLO(LO) accuracy for the production of \F{} in association
  with one (two) final-state \text{hard} jets, possibly with a scale
  setting that is appropriate to the treatment of each kinematic
  regime;
\item to preserve the (leading-logarithmic) accuracy of common
  parton-shower generators in the relevant kinematic
  regions.\footnote{These commonly include also simple next-to-leading logarithmic (NLL) $\alpha_s^2 L^2$
    corrections to the Sudakov form factor~\cite{Catani:1990rr}, which should be preserved as well.}
\end{itemize}
Our starting point for the construction of an event generator with NNLO+PS accuracy is
a \POWHEG{}~\cite{Nason:2004rx,Frixione:2007vw,Alioli:2010xd} calculation for the production of a
general colour-singlet system in association with a light
jet (\FJ{}) at NLO QCD matched to parton
showers (NLO+PS).
The \POWHEG{} formula for \FJ{} production can be schematically written as
\begin{equation}
  \mathd\sigma= \mathd \PhiBJ\ {\bar B}(\PhiBJ)\
  \bigg\{\Delta_{\rm pwg} (\LambdaPWG) + \mathd \Phi_{\sss\rm rad}
  \Delta_{\rm pwg} (\ptrad)  \frac{R (\PhiBJ{}, \Phi_{\sss\rm rad})}{B
    (\PhiBJ{})}\bigg\}\,,
\label{eq:BasicPowheg}
\end{equation}
which describes the production of \F{} in association with either one
or two real partons. In \eqn{eq:BasicPowheg}, $\PhiBJ{}$ refers 
to the phase space of the \FJ{} final state, while $\Delta_{\rm pwg} (\mu)$ denotes the \POWHEG{} Sudakov form
factor~\cite{Nason:2004rx}, representing the no-emission probability
for the real radiation down to a transverse-momentum scale $\mu$, and
the scale $\LambdaPWG$ represents an infrared cutoff chosen to avoid
the Landau singularity in the QCD coupling. We have also defined
\begin{equation}
  \bar  B(\PhiBJ) = \int \mathd \Phi_{\rm rad}
  \tilde{B}(\PhiBJ,\Phi_{\rm rad}),\,\qquad \tilde{B}(\PhiBJ,\Phi_{\rm rad})\equiv B(\PhiBJ) + V(\PhiBJ) + R(\PhiBJ,\Phi_{\rm rad}) \,,  
\label{eq:Btilde}
\end{equation}
where $B(\PhiBJ)$, $V(\PhiBJ)$ and $R(\PhiBJ,\Phi_{\rm rad})$ denote
the Born, virtual and real corrections, respectively.  The real and
virtual corrections include local counter-terms~\cite{Frixione:2007vw}
such that the $\bar B(\PhiBJ) $ function is finite and can be
evaluated numerically.
The radiation phase space $\mathd\Phi_{\rm rad}$ is normalised such that
$\int \mathd\Phi_{\rm rad} = 1$, and the corresponding
Jacobian factor is included in $R(\PhiBJ,\Phi_{\rm rad})$. 
Eq.\,\eqref{eq:BasicPowheg} is used to generate events with either one
or two real partons, which are then fed into a parton shower that
completes them with the addition of soft and collinear radiation at
all perturbative orders, with the requirement that the radiation from
the parton shower is softer than that generated by
\eqn{eq:BasicPowheg}.
The original \POWHEG{} formulation is restricted to the generation of
the colour-singlet system \F{} with a relatively hard associated jet
and has NLO accuracy for the production of the \FJ{} final state. In
the following we denote this accuracy by NLO$^{(1)}$, where the
superscript $(1)$ refers to the jet multiplicity.\footnote{We use the
  notation ((N)N)LO$^{(j)}$ to denote ((N)N)LO accuracy for
  observables defined for the production of the system \F{} in
  association with $j$ hard jets.}
The \POWHEG{} generator based upon \eqn{eq:BasicPowheg} contains
parts of the NNLO corrections to the inclusive production of the system \F{},
i.e.\ the double-real and real-virtual corrections of relative order
$\as^2$ above a resolution scale set by the hardness of the resolved
jet.
In particular, if we introduce any transverse-momentum cut on the jet,
the \POWHEG{} generator based upon \eqn{eq:BasicPowheg} has the
same perturbative accuracy as a NNLO+PS generator for \F{} above
the cut, while it is divergent below the cut.
Therefore, the missing ingredients to achieve NNLO accuracy for the
production of the system \F{} are given by the contributions with \F{} kinematics 
$\PhiB$ (i.e\ Born, one- and two-loop virtual corrections for \F{} production)
as well as a calculation of the \FJ{} and \FJJ{} contributions below the cut.

The \minlobare{} procedure~\cite{Hamilton:2012np} is a modification of
Eq~\eqref{eq:Btilde} such that the corresponding \POWHEG{}
generator gives meaningful results also below the transverse-momentum
cut.
This is achieved by multiplying the $\tilde B$-function by an appropriate 
Sudakov form factor that acts as a regulator of the infrared singularity in the limit of a 
vanishing transverse momentum cut and applying
appropriate (transverse-momentum dependent) scale choices.
For common (leading-logarithmic) parton showers, such
Sudakov form factor can be extracted from for the transverse-momentum
resummation for the system \F{}. The inclusion of this correction in
\eqn{eq:Btilde} must be accompanied by a suitable 
compensating term in order to avoid double counting. With this
prescription, one also achieves LO accuracy, i.e.\ LO$^{(0)}$, for 
observables defined on the phase space of \F{} (inclusive over radiation).

In \citere{Hamilton:2012rf} it was shown that the Sudakov form
factor can be corrected in such a way that observables 
inclusive over radiation acquire $\cO{\as}$ accuracy relative to 
the Born process of \F{} production, i.e.\ NLO$^{(0)}$ accuracy, 
and the corresponding method was dubbed \minlo{}.
This work also pointed out that NNLO$^{(0)}$ accuracy for the
inclusive production of \F{} can be achieved using an
\textit{a-posteriori} reweighing of the \minlo{} events differential
in the \F{} kinematics $\PhiB$. This procedure was applied to several
processes~\cite{Hamilton:2013fea,Karlberg:2014qua,Astill:2016hpa,Astill:2018ivh,Re:2018vac,Bizon:2019tfo}. 
However, for complex processes this approach becomes prohibitively
intensive from a computational viewpoint.

In \citeres{Monni:2019whf,Monni:2020nks} an alternative approach
was introduced to achieve NNLO$^{(0)}$ accuracy by including the
appropriate higher-order terms directly during the generation of events,
circumventing the issues of the previous approach based on reweighting. 
The precise expression of these terms can be extracted from the following formula for the
differential cross section
\begin{equation}
  \label{eq:master1}
     \frac{\mathd\sigma}{\mathd\pt \mathd\PhiB} = \frac{\mathd}{\mathd\pt} \Big{\{}  \mathcal{L}(\Phi_{\rm F},\pt) \exp[-\tilde{S}(\PhiB, \pt)] \Big{\}} +  R_{\rm finite}(\pt)\,.
\end{equation}
The right-hand side of \eqn{eq:master1} can be derived from the
resummation of the transverse momentum of \F{} at next-to-next-to-NLL
(N$^3$LL) matched to fixed-order.
Here $\mathcal{L}(\PhiB,\pt)$ denotes a \textit{luminosity} factor
that includes the PDFs, the hard-virtual corrections and the collinear
coefficient functions for the production of the system \F{} expanded
in powers of $\as(\pt)$ up to two loops. The factor
$\exp[-\tilde{S}(\pt)]$ denotes the Sudakov form factor for the
resummation of logarithms $\ln(Q/\pt)$ (with $Q$ being a scale of the
order of the invariant mass of \F{}), and $R_{\rm finite}(\pt)$
denotes the regular part of the differential cross section, also
expanded in powers of $\as(\pt)$.
To obtain \eqn{eq:master1}, we have simplified the original N$^3$LL
transverse-momentum resummation formula to retain only terms that are
relevant up to ${\cal O}(\as^2)$ relative to the \F{} Born while
preserving the LL accuracy (including also the
$\alpha_s^2 \ln^2 (Q/\pt)$ corrections to the Sudakov radiator
$\tilde{S}(\pt)$, which are accounted for in common parton showers we
will match onto).
In its present form, the integral of \eqn{eq:master1} over $\pt$ up to
any perturbative value is NNLO$^{(0)}$ accurate.

In order to include NNLO$^{(0)}$ accuracy in \eqn{eq:BasicPowheg} we
need to modify the expression for $\tilde B(\PhiBJ,\Phi_{\rm rad})$ in \eqn{eq:Btilde}
in the following way~\cite{Monni:2019whf,Monni:2020nks}:
\begin{align}
  \tilde B(\PhiBJ,\Phi_{\rm rad}) &=& \exp[-\tilde{S}(\PhiB,\pt)] \bigg[ B(\PhiBJ) \left(1+\abarmu{\pt} [\tilde{S}(\PhiB,\pt)]^{(1)}\right)  + V(\PhiBJ) \nonumber \\
    &&  + R(\PhiBJ,\Phi_{\rm rad}) + D^{{\scriptscriptstyle (\geq 3)}}(\PhiB,\pt){\Fcorr(\PhiBJ)} \bigg]\,,
  \label{eq:BtildeMinnlo}
\end{align}  
where we have defined
\begin{align}
  \label{eq:Ddef}
  D^{{\scriptscriptstyle (\geq 3)}}(\PhiB,\pt) =& D(\PhiB,\pt) -\abarmu{\pt} [D(\PhiB,\pt)]^{(1)}- \left(\abarmu{\pt}\right)^2[D(\PhiB,\pt)]^{(2)}\,,\notag\\
       D(\PhiB,\pt) =& -\frac{\mathd \tilde{S}(\PhiB,\pt)}{\mathd \pt} {\cal L}(\PhiB,\pt)+\frac{\mathd {\cal L}(\PhiB,\pt)}{\mathd \pt}\,.
\end{align}
In the above equations we use the notation $[X]^{(k)}$ to denote the
coefficient of the expansion $X=\sum_k (\abar(\pt))^k [X]^{(k)}$.  We
note that the term $[\tilde{S}(\PhiB,\pt)]^{(1)}$ in
\eqn{eq:BtildeMinnlo} cancels the
corresponding $\cO{\as}$ term arising from the expansion of the
Sudakov form factor, which is required to avoid double counting and to preserve NLO$^{(1)}$
accuracy.
Although
for simplicity in
  \eqn{eq:BtildeMinnlo} we have factorised the Sudakov form
  factor, $\pt$ is actually evaluated in the $\PhiBJ$ configuration
  for all terms except $R(\PhiBJ,\Phi_{\rm rad})$, for which it is
  taken equal to the transverse momentum of the system \F{} in the
  full $\PhiBJJ$ phase space.
Without the last term, \eqn{eq:BtildeMinnlo} corresponds exactly to the \minlo{}
formula, yielding both NLO$^{(0)}$ and NLO$^{(1)}$ accuracy.

The quantity $D^{{\scriptscriptstyle (\geq 3)}}(\PhiB,\pt)$ guarantees
that the integral of $\tilde B$ at fixed $\PhiB$ matches the expansion
of \eqn{eq:master1}, up to the accuracy that we require, namely
NNLO$^{(0)}$.
This quantity depends on $\PhiB$ and $\pt$.
On the other hand, $\tilde B(\PhiBJ,\Phi_{\rm rad})$ depends on the full
$\PhiBJ$. Thus, we must spread the
$D^{{\scriptscriptstyle (\geq 3)}}(\PhiB,\pt)$ term in the $\PhiBJ$
phase space. This is done by the function $\Fcorr(\PhiBJ)$, which is defined
such that its integral at fixed $\PhiB$ and $\pt$ equals one (see
\citere{Monni:2019whf} for a precise definition).

We note that the correction term
$D^{{\scriptscriptstyle (\geq 3)}}(\PhiB,\pt)$, starting formally at
order $\as^3(\pt)$, after integration over $\pt$ at fixed $\PhiB$
contributes starting at relative order $\as^2(Q)$.
This is because $D^{{\scriptscriptstyle (\geq 3)}}(\PhiB,\pt)$
contains logarithmic singularities that are regulated by the Sudakov
form factor. When integrated over $\pt$, their product gives rise to a
Gaussian integral of the form 
\begin{equation}
  \int_0^{\infty} \mathd L \exp[-\as L^2] \as^m L^n \propto \as^{m-\frac{n+1}{2}}\,,
\end{equation}
where we schematically defined $L=\ln(Q/\pt)$ and ignored the running
of $\as$ (cf. appendix\,C of \citere{Hamilton:2012rf} for a detailed
discussion on this point).
Therefore, to reach NNLO$^{(0)}$ accuracy one must ensure that the
perturbative truncation error is of relative order $\alpha_s^3(Q)$,
which in turn requires including all singular terms at least up to
$m=3$ and $n=0$.\footnote{We stress that the function $D(\PhiB,\pt)$
  contains at most single logarithmic terms (i.e.\ only $n=1$ and
  $n=0$) as a consequence of the fact that the coupling and the PDFs
  are evaluated at scales of order $\pt$.}
However, we also stress that the definition of the function
$D^{{\scriptscriptstyle (\geq 3)}}(\PhiB,\pt)$ does not require any
actual computation at order ${\cal O}(\alpha_s^3)$. The higher-order
terms beyond $\as^2(\pt)$ in
$\tilde B(\PhiBJ,\Phi_{\rm rad})$ are simply generated by taking the
total derivative in \eqn{eq:master1}, leading to \eqn{eq:Ddef}.

In \citeres{Monni:2019whf,Monni:2020nks,Lombardi:2020wju,Lombardi:2021rvg,Buonocore:2021fnj,Lombardi:2021wug,Zanoli:2021iyp}
the \minnlo{} formalism was applied to several processes where the
system \F{} is a colour singlet.  Strictly speaking,
\eqn{eq:master1} holds in this form only in this case.
The production of a colour-charged heavy system involves
additional conceptual complications related to the fact that the final
state will exchange soft gluons within itself and with the initial
state.
This leads to a more complicated structure of the transverse-momentum
resummation formula, which now involves colour matrices (see
e.g.~\cite{Zhu:2012ts,Li:2013mia,Catani:2014qha,Catani:2018mei}).
Thus, in order to deal with heavy-quark pair production one
must first generalize the procedure adopted in the colour singlet case
by finding the appropriate extension of \eqn{eq:master1}. This
will be discussed in the next section.

\section{The \minnlo{} method for heavy-quark pair production}
\label{sec:ttbar}
In this section we present an extension of the \minnlo method to the
production of a pair of heavy quarks in hadronic collisions, having in
mind the concrete application to top-quark pair production.
Our goal is to achieve NNLO$^{(0)}$ accuracy for all observables
defined in the phase space of the pair of heavy quarks, and inclusive
in the additional QCD radiation. At the same time, we will retain
NLO$^{(1)}$ (LO$^{(2)}$) accuracy for observables which require one
(two) resolved hard jet(s).
We also demand that our matching procedure preserves the leading logarithmic 
(LL) accuracy
(plus the simple NLL corrections to the Sudakov form factor captured
by the CMW scheme~\cite{Catani:1990rr}) of the parton shower the
calculation is matched onto, which in this case is assumed to be
ordered in a transverse-momentum variable. Specifically, we will
consider the dipole shower \PYTHIA{8}~\cite{Sjostrand:2014zea} in all
phenomenological applications presented in this article.\footnote{For
  different ordering variables, preserving the accuracy of the
  shower is more subtle. A common procedure is to resort to truncated
  showers~\cite{Nason:2004rx,Bahr:2008pv} to compensate for missing
  collinear and soft radiation. Failing to do so spoils the shower
  accuracy at leading-logarithmic level (in fact, at the
  double-logarithmic level).}
The above accuracy goals will be taken into account when performing
some of the approximations necessary in the extension of the \minnlo
method. 
\subsection{Structure of heavy-quark pair production at small $\pt$}
We consider the production of a pair of heavy quarks,
$\F{}\equiv Q\bar{Q}$ from now on, differential in the phase space of
the pair
$\mathd\Phi_{Q\bar{Q}}\equiv \mathd {\bar x}_1 \mathd {\bar x}_2
[\mathd\Phi_2]$.
Here ${\bar x}_{1,2}=m_{Q\bar{Q}}/\sqrt{s} \,e^{\pm y_{Q\bar{Q}}}$,
with $y_{Q\bar{Q}}$ being the rapidity of the $Q\bar{Q}$ system,
$m_{Q\bar{Q}}$ its invariant mass, $[\mathd\Phi_2]$ denotes the
Lorentz-invariant two-body phase space and $\sqrt{s}$ is the collider
centre-of-mass energy.
Our starting point is the well-known formula for transverse-momentum
resummation for a $Q\bar{Q}$ pair, which
reads~\cite{Zhu:2012ts,Li:2013mia,Catani:2014qha,Catani:2018mei} (we
denote $b_0=2\,e^{-\gamma_E}$, $b=|\vec{b}|$)
\begin{align}
\label{eq:bspace}
\frac{\mathd\sigma}{ \mathd^2\ptvec\, \mathd \Phi_{Q\bar{Q}}}&=\sum_{c=q,\bar{q},g}
  \frac{|M^{(0)}_{c\bar{c}}|^2}{2 m_{Q\bar{Q}}^2}\int\frac{ \mathd^2\vec{b}}{(2\pi)^2} e^{i \vec{b}\cdot
  \ptvec } e^{-S_{c\bar{c}} \left(\frac{b_0}{b}\right)}\sum_{i,j}\Tr({\mathbf H}_{c\bar{c}}{\mathbf \Delta})\,
 \,({C}_{ci}\otimes f_i) \,({C}_{\bar{c} j}\otimes f_j) \,.
\end{align}
The first sum in \eqn{eq:bspace} runs over all possible flavour
configurations of the incoming partons $p_1$ of flavour $c$ and $p_2$
of flavour $\bar c$.
In the above equation, the quantity $S_{c\bar{c}}$ represents the Sudakov
radiator,
\begin{equation}
\label{eq:sudakov}
  S_\ccbar\left(\frac{b_0}{b}\right) \equiv \int_{\frac{b_0^2}{b^2}}^{m_{Q\bar{Q}}^2}\frac{\mathd{}q^2}{q^2} \left[A_\ccbar(\alpha_s(q))\ln\frac{m_{Q\bar{Q}}^2}{q^2}+B_\ccbar(\alpha_s(q))\right]\,,
\end{equation}
which encodes the resummation of logarithms of the impact parameter
$b$ (and thus $\pt$ of the heavy-quark pair) of collinear and
soft-collinear origin (i.e.\ double and single logarithms), that
originate entirely from initial-state radiation.

The coefficient functions
$C_{ij}\equiv C_{ij}(z,p_1,p_2,\vec{b};\alpha_s(b_0/b))$ encode
constant contributions to the factorisation theorem that originate
from initial-state collinear radiation. The symbol
$\otimes$ denotes the usual convolution,
\begin{equation}
  (f \otimes g)(z) = \int_z^1 \frac{\mathd{}x}{x} f(x) g\left(\tfrac{z}{x}\right)\,.
\end{equation}
Both $S_\ccbar$ and $C_{ij}$ are universal, and are identical to the ones
featuring in the resummation formula describing the low transverse
momentum behaviour of a heavy colorless final state.

In \eqn{eq:bspace} and throughout the paper, operators in colour
space are denoted in bold face. The colour-space operator
$\mathbf \Delta$ encodes the resummation of single-logarithmic
corrections that arise from soft radiation exchanged with large angles
between the final state heavy-quark legs and between the initial and
final state (we refer the reader to \citere{Catani:2014qha} for a
more thorough discussion).

The trace $\Tr({\mathbf H}_{c\bar{c}}{\mathbf \Delta})$ runs over the
colour indices and encodes the difference between the factorisation
formula~\eqref{eq:bspace} and the corresponding one for colour-singlet
production. It can be given an explicit representation by using the
colour space formalism of \citere{Catani:1996vz}. In order to do so,
we will denote by $|M_{c\bar c}\rangle$ the all-orders
\textit{infrared-subtracted} amplitude for the production of the
heavy-quark pair from the incoming partons $p_1$ of flavour $c$ and
$p_2$ of flavour $\bar c$. All amplitudes here are considered to be
renormalised, specifically in the $\overline{\rm MS}$ scheme.
The object $|M_{c\bar c}\rangle$ represents a vector in colour space,
and can be obtained from the un-subtracted (IR divergent) scattering
amplitude $|M^{\text{IR-div.}}_{c\bar c}\rangle$ by applying an
appropriate subtraction operator~\cite{Catani:2014qha}
\begin{equation}
|M_{c\bar c}\rangle = \left(
\mathbb{1}-{\mathbf{I}}_{c\bar{c}}
\right) |M^{\text{IR-div.}}_{c\bar c}\rangle \,.
\end{equation}
In terms of $|M_{c\bar c}\rangle$, the operator ${\mathbf H}_{c\bar{c}}$ can be
written as
${\mathbf H}_{c\bar{c}} = |M_{c\bar c}\rangle \langle M_{c\bar
  c}|/|M^{(0)}_{c\bar{c}}|^2$, and therefore we
have~\cite{Catani:2014qha}
\begin{equation}
\label{eq:trace}
\Tr({\mathbf H}_{c\bar{c}}{\mathbf \Delta} ) ({C}_{ci}\otimes f_i)
\,({C}_{\bar{c} j}\otimes f_j) \equiv \frac{\langle M_{c\bar c}|{\mathbf \Delta}| M_{c\bar c}
  \rangle}{|M^{(0)}_{c\bar{c}}|^2}({C}_{ci}\otimes f_i)
\,({C}_{\bar{c} j}\otimes f_j)\,,
\end{equation}
where a sum over the colours (and spins) of the external legs is
understood.
The symbolic object
$\Tr({\mathbf H}_{c\bar{c}}{\mathbf \Delta})\, \,({C}_{ci}\otimes f_i)
\,({C}_{\bar{c} j}\otimes f_j)$ takes a different form in the
$q\bar{q}$ and $gg$ channels. In particular, while
\eqn{eq:trace} is strictly valid for $q\bar{q}$ annihilation, in
the gluon-initiated $gg$ channel this factor has a rich Lorentz
structure.
Firstly, the object
$({C}_{ci}\otimes f_i) \,({C}_{\bar{c} j}\otimes f_j)$ now contains
the contribution from additional coefficient functions that encode
azimuthal correlations of collinear origin (commonly denoted with
$G_{ij}$~\cite{Catani:2010pd}). Furthermore, \eqn{eq:trace}
should be now understood as a tensor contraction between
$\Tr({\mathbf H}_{c\bar{c}}{\mathbf \Delta})$ and
$({C}_{ci}\otimes f_i) \,({C}_{\bar{c} j}\otimes f_j)$.
This leads to additional azimuthal correlations of soft nature that,
for simplicity, are kept implicit in \eqn{eq:bspace}. For more
details, we refer the reader to eq.\.(13) of
\citere{Catani:2014qha}, and subsequent discussions.
We note that in the case in which the produced final state is a
colour singlet, we have ${\mathbf \Delta} = {\mathbb 1}$ and
\eqn{eq:trace} simply reduces to $\Tr({\mathbf H}_{c\bar{c}}) = H _{c\bar{c}}$,
where $H _{c\bar{c}}$ is the usual process-dependent hard-virtual coefficient.

It is worth pointing out that while the pole structure of the
subtraction operator ${\mathbf{I}}_{c\bar{c}}$ is uniquely defined,
its finite ${\cal O}(\epsilon^0)$ piece is, in principle,
arbitrary. Its exact definition is tied to the choice of the
resummation scheme. In the current work we adopt the resummation
scheme of \citere{Catani:2014qha} (which in the colour-singlet case
reduces to the so-called hard scheme~\cite{Catani:2013tia}). The
corresponding finite terms can only be obtained by means of an
explicit calculation of the expansion coefficients
${\mathbf{I}}^{(i)}_{c\bar{c}}$, defined by the perturbative series
\begin{equation}
{\mathbf{I}}_{c\bar{c}} = \sum_{i} \left(\frac{\alpha_s(\mu)}{2\pi}\right)^i{\mathbf{I}}^{(i)}_{c\bar{c}}\,.
\end{equation}
The expression for ${\mathbf{I}}^{(1)}_{c\bar{c}}$ can be found in
\citere{Catani:2014qha}, while the NNLO operator
${\mathbf{I}}^{(2)}_{c\bar{c}}$ has been taken
from~\citere{SoftFunction}, which has been used in the context of
$q_T$-subtraction~\cite{Catani:2007vq} for heavy-quark
production~\cite{Catani:2019hip,Catani:2019iny,Catani:2020kkl}.

As already mentioned, the operator ${\mathbf \Delta}$ encodes the main
source of difference between the colour-singlet and the $Q\bar{Q}$
final states. It can be written as
${\mathbf \Delta} = {\mathbf V}^\dagger {\mathbf D} {\mathbf V}$,
where the operator ${\mathbf V}$ is obtained by exponentiating the
soft anomalous dimension matrix for
heavy-quark pair production ${\mathbf \Gamma}_t(\Phi_{Q\bar{Q}};\alpha_s(q))$, i.e.\ \cite{Catani:2014qha}
\begin{equation}
\label{eq:V}
{\mathbf V} = {\cal
  P}\exp\left\{-\int_{b_0^2/b^2}^{m_{Q\bar{Q}}^2}\frac{\mathd{}q^2}{q^2}{\mathbf
  \Gamma}_t(\Phi_{Q\bar{Q}};\alpha_s(q))\right\}\,.
\end{equation}
In the above equation, the symbol ${\cal P}$ denotes the path ordering
(with increasing scales from left to right) of the exponential matrix
with respect to the integration variable $q^2$.
The operator
${\mathbf D}\equiv{\mathbf
  D}(\Phi_{Q\bar{Q}},\vec{b};\alpha_s(b_0/b))$,
on the other hand, encodes the azimuthal correlations of the produced
$Q\bar{Q}$ system in the low-$\pt$ limit due to soft radiation at
large angles. It is normalised in such a way that
$[{\mathbf D}]_\phi={\mathbb 1}$, where $[\cdots]_\phi$ denotes the
average over the azimuthal angle $\phi$ of $\ptvec$. As mentioned
above, additional sources of azimuthal correlations, this time of
collinear origin, are present in the coefficient functions $C_{ij}$
corresponding to the gluon-initiated channels (the two are combined in
the contraction of \eqn{eq:trace} to produce non-trivial correlations
that are typical of the process under consideration).
We note that ${\mathbf V}$ and ${\mathbf D}$ are also
resummation-scheme dependent quantities. In the adopted resummation
scheme all the dependence on the azimuthal angle is absorbed in
${\mathbf D}$ and absent from ${\mathbf H}_{c\bar{c}}$ and ${\mathbf V}$, but
other schemes could be defined in which this is not the
case~\cite{Catani:2014qha}.

Having introduced the main features of the resummation formula
\eqref{eq:bspace}, we focus now on its connection to the derivation of
a NNLO+PS generator for $Q\bar{Q}$ production. In the
colour-singlet case~\cite{Monni:2019whf,Monni:2020nks}, the
resummation formula (in particular its formulation in direct
space~\cite{Monni:2016ktx,Bizon:2017rah}) is the basis to constructing
\eqn{eq:master1}. The low-$q_T$ structure of the $Q\bar{Q}$
cross section is, however, clearly more involved. In particular,
writing the equivalent to \eqn{eq:master1} starting from
\eqn{eq:bspace} is not straightforward due to the colour
interference effects present in ${\mathbf \Delta}$ and due to the
non-trivial azimuthal correlations in \eqn{eq:trace}.
However, as in the colour singlet case, we are not seeking a formula
that is N$^3$LL accurate for the transverse-momentum distribution of
the $Q\bar{Q}$ pair (which through matching to the shower reduces 
to its accuracy regardless), 
but instead we only need to ensure that its integral at fixed $\Phi_{Q\bar{Q}}$ is accurate at the NNLO level,
while preserving the logarithmic accuracy of the shower algorithms we
match to.
This allows us to perform several approximations in
\eqn{eq:bspace}, which are described in what follows, with the
goal of transforming it into a NNLO-accurate expression which
resembles the (considerably simpler) structure of the colour-singlet
case.

We start by defining the perturbative expansion of the different
ingredients entering \eqn{eq:bspace},
\begin{eqnarray}
\label{eq:series}
Z(\{x\} ;\as(\mu))&=&
\sum_{i}\left(\frac{\as(\mu)}{2\pi}\right)^i Z^{(i)}\left(\{x\}\right)\,,\notag\\
|M_{c{\bar c}}\rangle &=& \sum_{i}\left(\frac{\as(\mQQ)}{2\pi}\right)^i
|M_{c{\bar c}}^{(i)}\rangle\,,
\end{eqnarray}
where $Z$ represents any of the quantities
$Z\equiv\{A_\ccbar,\,B_\ccbar,\,C_{ij},\,{\mathbf H}_{c\bar{c}},\,{\mathbf D},\,{\mathbf \Gamma}_t\}$,
and $\{x\}$ stands for any additional set of arguments. The scale $\mu$ at which
the expansion is performed is the one explicitly indicated in the corresponding
function $Z$.
Explicit results for the coefficients $Z^{(i)}$ up to two loops can be
extracted from the results presented in
\citeres{Czakon:2008zk,Catani:2010pd,Catani:2011kr,Czakon:2011xx,Catani:2012qa,Gehrmann:2012ze,Baernreuther:2013caa,Li:2013mia,Catani:2014qha,Gehrmann:2014yya,Echevarria:2016scs,Angeles-Martinez:2018mqh,Luo:2019bmw,Catani:2019hip,Luo:2019hmp,SoftFunction}.
We note that ${\mathbf D}^{(2)}$ is still unknown at present, however
due to the property $[{\mathbf D}]_\phi={\mathbb 1}$ its contribution
for azimuthally-averaged observables exactly vanishes at NNLO (this is
not the case for ${\mathbf D}^{(1)}$, as discussed in the
following). In the case of $|M_{c{\bar c}}\rangle$, as stated in
\eqn{eq:series}, the expansion is performed in powers of
$\as(\mQQ)$.  The additional power of $\as$ present already at LO is
included in the definition of $|M_{c{\bar c}}^{(i)}\rangle$, and is
evaluated at a hard scale $\muRzero \sim m_{Q\bar{Q}}$.  The
evaluation of $|M_{c{\bar c}}^{(i)}\rangle$ up to NNLO requires the
knowledge of the complete two-loop virtual corrections for
$p_1 p_2\to Q\bar{Q}$, which we take from the numerical implementation
in \citere{Baernreuther:2013caa}, as well as the corresponding
subtraction operator
${\mathbf I}^{(2)}_{c\bar{c}}$~\cite{SoftFunction}.

Let us now consider the matrix exponential in ${\mathbf V}$. Up to
N$^3$LL accuracy, we can expand the ${\mathbf \Gamma}_t^{(2)}$ term
inside the path ordering sign in the following way,
\begin{equation}
\label{eq:soft_recast}
{\mathbf V} = {\cal
 P}\Bigg[\exp\left\{-\int_{b_0^2/b^2}^{m_{Q\bar{Q}}^2}\frac{\mathd{}q^2}{q^2}\frac{\alpha_s(q)}{2\pi}{\mathbf
   \Gamma}^{(1)}_t\right\}
 \times\left(1-\int_{b_0^2/b^2}^{m_{Q\bar{Q}}^2}\frac{\mathd{}q^2}{q^2}\frac{\alpha^2_s(q)}{(2\pi)^2}{\mathbf
 \Gamma}^{(2)}_t \right)\Bigg]
  + {\cal O}({\rm N}^3{\rm LL})\,.
\end{equation}
The path ordered expression must be now evaluated exactly to retain
next-to-NLL (NNLL) accuracy. However, we recall that our goal is to
achieve NNLO$^{(0)}$ and we do not necessarily need to preserve the
full logarithmic accuracy of the resummation formula, which allows us
to make the following approximation.
Without spoiling the NNLO$^{(0)}$ accuracy of the expression, we can
take the ${\mathbf \Gamma}^{(2)}_t$ contribution out of the
path-ordering symbol in \eqn{eq:soft_recast}.
This is because ${\mathbf \Gamma}^{(2)}_t$ enters at relative order
${\cal O}(\alpha_s^2)$ and therefore the action of the path-ordering
would only start at ${\cal O}(\alpha_s^3)$.
Then we have
\begin{equation}
\label{eq:soft_recast2}
{\mathbf V} = {\mathbf V}_{\rm NLL}
 \times\left(1-\int_{b_0^2/b^2}^{m_{Q\bar{Q}}^2}\frac{\mathd{}q^2}{q^2}\frac{\alpha^2_s(q)}{(2\pi)^2}{\mathbf
 \Gamma}^{(2)}_t \right)
  + {\cal O}(\as^3)\,,
\end{equation}
where we have defined the object ${\mathbf V}_{\rm NLL}$ entering $q_T$-resummation
at NLL accuracy,
\begin{equation}
\label{eq:V_NLL}
{\mathbf V}_{\rm NLL} = {\cal
 P}\Bigg[\exp\left\{-\int_{b_0^2/b^2}^{m_{Q\bar{Q}}^2}\frac{\mathd{}q^2}{q^2}\frac{\alpha_s(q)}{2\pi}{\mathbf
   \Gamma}^{(1)}_t\right\} \Bigg]\,.
\end{equation}

We focus now on the contribution to \eqn{eq:bspace} proportional
to ${\mathbf \Gamma}^{(2)}_t$. Firstly, we observe that said
contribution is no longer in an exponential form, due to the
approximations made in \eqn{eq:soft_recast2}.  Furthermore, it
is accompanied by an $\as^2$ suppression.  This implies that it
contributes to the trace $\Tr({\mathbf H}_{c\bar{c}}{\mathbf \Delta})$ with the
following NNLO term
\begin{equation}
\Tr({\mathbf H}_{c\bar{c}}{\mathbf \Delta}) \supset
-\frac{\langle
    M_{c\bar{c}}^{(0)} | {\mathbf \Gamma}^{(2)\,\dagger}_t +
  {\mathbf \Gamma}^{(2)}_t|M_{c\bar{c}}^{(0)}
  \rangle}{|M^{(0)}_{c\bar{c}}|^2}
  \int_{b_0^2/b^2}^{m_{Q\bar{Q}}^2}\frac{\mathd{}q^2}{q^2}\frac{\alpha^2_s(q)}{(2\pi)^2}
  \,,
\end{equation}
where, to simplify the discussion, the notation $A \supset B$
indicates that $A$ contains the summand $B$ in its perturbative
expansion.
We now note that this contribution has the functional
(single-logarithmic) form of the ${\cal O}(\as^2)$ term stemming from
the $B^{(2)}$ coefficient in the Sudakov radiator,
\begin{equation}
\label{eq:B2terms}
e^{-S_\ccbar \left(\frac{b_0}{b}\right)}
\supset
-B_\ccbar^{(2)}
  \int_{b_0^2/b^2}^{m_{Q\bar{Q}}^2}\frac{\mathd{}q^2}{q^2}\frac{\alpha^2_s(q)}{(2\pi)^2}
  \,.
\end{equation}
This implies that the ${\mathbf \Gamma}^{(2)}_t$ contribution can be
completely absorbed into the $B_\ccbar^{(2)}$ coefficient at the desired accuracy
by performing the replacement
\begin{equation}
\label{eq:B2repl1}
B_\ccbar^{(2)} \to B_\ccbar^{(2)} + \frac{\langle
    M_{c\bar{c}}^{(0)} | {\mathbf \Gamma}^{(2)\,\dagger}_t +
  {\mathbf \Gamma}^{(2)}_t|M_{c\bar{c}}^{(0)}
  \rangle}{|M^{(0)}_{c\bar{c}}|^2} \,.
\end{equation}

We turn now to the term coming from ${\mathbf V}_{\rm NLL}$, which
appears in the combination
${\mathbf V}_{\rm NLL}^\dagger {\mathbf D}\,{\mathbf V}_{\rm NLL}$. 
As stated before, the function ${\mathbf D}$ depends upon the
azimuthal angle $\phi$ of $\ptvec$ and it is an important source of
azimuthal correlations.
However, due to the fact that ${\mathbf \Gamma}_t^{(1)}$ does not
depend on $\phi$, the ${\cal O}(\as^2)$ contribution proportional to
${\mathbf \Gamma}_t^{(1)} {\mathbf D}^{(1)}$ present in the product
${\mathbf V}_{\rm NLL}^\dagger {\mathbf D}\,{\mathbf V}_{\rm NLL}$
vanishes upon taking the azimuthal average since
$[{\mathbf D}]_\phi={\mathbb 1}$. We can therefore perform the
approximation
 \begin{equation}
\frac{\langle M_{c\bar{c}} | 
{\mathbf V}_{\rm NLL}^\dagger {\mathbf D}\, {\mathbf V}_{\rm NLL} |M_{c\bar{c}}\rangle}{|M^{(0)}_{c\bar{c}}|^2}
=
\frac{\langle M_{c\bar{c}} | 
{\mathbf V}_{\rm NLL}^\dagger {\mathbf V}_{\rm NLL} |M_{c\bar{c}}\rangle}
{|M^{(0)}_{c\bar{c}}|^2}
\Tr({\mathbf H}_{c\bar{c}}{\mathbf D})
+ E_{c\bar{c}}(\Phi_{Q\bar{Q}},\vec{b})+{\cal O}(\alpha_s^3)\,,
 \end{equation}
 where the remainder term $E_{c\bar{c}}(\Phi_{Q\bar{Q}},\vec{b})$ contributes
 at order $\alpha_s^2\ln(m_{Q\bar{Q}} \,b)$, but for the reasons just
 discussed vanishes upon azimuthal integration and therefore can be
 safely ignored.
 We now focus on the object
 $\langle M_{c\bar{c}} | {\mathbf V}_{\rm NLL}^\dagger {\mathbf
   V}_{\rm NLL} |M_{c\bar{c}}\rangle$,
 and deal with the NNLO contribution from the function ${\mathbf D}$
 afterwards.

 In order to obtain a colour-singlet-like resummation formula, we
 would like to factorize the loop corrections in the following way,
\begin{equation}\label{eq:soft_recast3}
\frac{\langle M_{c\bar{c}} | 
{\mathbf V}_{\rm NLL}^\dagger {\mathbf V}_{\rm NLL} |M_{c\bar{c}}\rangle}
{|M^{(0)}_{c\bar{c}}|^2}
\to
\frac{\langle M_{c\bar{c}}^{(0)} | 
{\mathbf V}_{\rm NLL}^\dagger {\mathbf V}_{\rm NLL} |M_{c\bar{c}}^{(0)}\rangle}
{|M^{(0)}_{c\bar{c}}|^2}
\frac{\langle M_{c\bar{c}} |M_{c\bar{c}}\rangle}
{|M^{(0)}_{c\bar{c}}|^2}\,,
\end{equation}
however, the replacement in \eqn{eq:soft_recast3} is not NNLO
accurate due to the colour structure of the operator
$ {\mathbf V}_{\rm NLL}$ and the amplitude $|M_{c\bar{c}}\rangle$.  We
can however subtract the incorrect ${\cal O}(\as^2)$ term that is
induced by such replacement, and add back the correct one in order to
match the l.h.s. of \eqn{eq:soft_recast3} up to NNLO
accuracy. This yields:
\begin{eqnarray}
\label{eq:soft_recast4}
&&\frac{\langle M_{c\bar{c}} | 
{\mathbf V}_{\rm NLL}^\dagger {\mathbf V}_{\rm NLL} |M_{c\bar{c}}\rangle}
{|M^{(0)}_{c\bar{c}}|^2}
\to 
\frac{\langle M_{c\bar{c}}^{(0)} | 
{\mathbf V}_{\rm NLL}^\dagger {\mathbf V}_{\rm NLL} |M_{c\bar{c}}^{(0)}\rangle}
{|M^{(0)}_{c\bar{c}}|^2}
\frac{\langle M_{c\bar{c}} |M_{c\bar{c}}\rangle}
{|M^{(0)}_{c\bar{c}}|^2} \\
&&\phantom{a\qquad\qquad a}+ 2 \Re\left[\frac{\langle
    M_{c\bar{c}}^{(1)} |M_{c\bar{c}}^{(0)}
  \rangle}{|M^{(0)}_{c\bar{c}}|^2}\right]
\frac{\langle
    M_{c\bar{c}}^{(0)} | {\mathbf \Gamma}^{(1)\,\dagger}_t +
  {\mathbf \Gamma}^{(1)}_t|M_{c\bar{c}}^{(0)}
  \rangle}{|M^{(0)}_{c\bar{c}}|^2}
\frac{\as(\mQQ)}{2\pi}\int_{b_0^2/b^2}^{m_{Q\bar{Q}}^2}\frac{\mathd{}q^2}{q^2}\frac{\as(q)}{2\pi}  
  \notag \\
  &&\phantom{a\qquad\qquad a}- 2\,\Re\left[\frac{\langle
    M_{c\bar{c}}^{(1)} | {\mathbf \Gamma}^{(1)\,\dagger}_t +
  {\mathbf \Gamma}^{(1)}_t|M_{c\bar{c}}^{(0)}
  \rangle}{|M^{(0)}_{c\bar{c}}|^2}\right]
\frac{\as(\mQQ)}{2\pi}\int_{b_0^2/b^2}^{m_{Q\bar{Q}}^2}\frac{\mathd{}q^2}{q^2}\frac{\as(q)}{2\pi}  
+ {\cal O}(\as^3)  \,.
  \notag
\end{eqnarray}
If in addition we replace $\as(\mQQ)=\as(q) + {\cal O}(\as^2)$ in the
above equation, which does not affect the NNLO accuracy of the
expression, we can notice that the additional terms in the second and
third line of \eqn{eq:soft_recast4} have again the same structure as
those generated by $B_\ccbar^{(2)}$ in \eqn{eq:B2terms}. Therefore, we
can perform the simplification in \eqn{eq:soft_recast3} accompanied by
the further replacement, in addition to that in \eqn{eq:B2repl1},
\begin{eqnarray}
\label{eq:B2repl2}
B_\ccbar^{(2)} \to B_\ccbar^{(2)}
&-& 2 \Re\left[\frac{\langle
    M_{c\bar{c}}^{(1)} |M_{c\bar{c}}^{(0)}
  \rangle}{|M^{(0)}_{c\bar{c}}|^2}\right]
\frac{\langle
    M_{c\bar{c}}^{(0)} | {\mathbf \Gamma}^{(1)\,\dagger}_t +
  {\mathbf \Gamma}^{(1)}_t|M_{c\bar{c}}^{(0)}
  \rangle}{|M^{(0)}_{c\bar{c}}|^2}
  \notag \\
  &+& 2\,\Re\left[\frac{\langle
    M_{c\bar{c}}^{(1)} | {\mathbf \Gamma}^{(1)\,\dagger}_t +
  {\mathbf \Gamma}^{(1)}_t|M_{c\bar{c}}^{(0)}
  \rangle}{|M^{(0)}_{c\bar{c}}|^2}\right]
  \,.
\end{eqnarray}

The exponential factor present in
$\langle M_{c\bar{c}}^{(0)} | {\mathbf V}_{\rm NLL}^\dagger {\mathbf
  V}_{\rm NLL} |M_{c\bar{c}}^{(0)}\rangle$
can now be evaluated explicitly by performing a rotation in colour
space into a basis in which ${\mathbf \Gamma}_t^{(1)}$ is
diagonal. Explicit expressions for the corresponding colour matrices
can be found in \citere{hayk} and are reported in
\app{app:resum}.

After performing the approximations described above, we arrive to the
following expression:
\begin{eqnarray}
\label{eq:starting}
\frac{\mathd\sigma}{ \mathd^2\ptvec\, \mathd \Phi_{Q\bar{Q}}}&=&\frac{1}{2 m_{Q\bar{Q}}^2}\sum_{c=q,\bar{q},g}
  \int\frac{ \mathd^2\vec{b}}{(2\pi)^2} e^{i \vec{b}\cdot
  \ptvec } e^{-\hat{S}_{\ccbar} \left(\frac{b_0}{b}\right)}
 \langle
    M_{c\bar{c}}^{(0)} | \left({\mathbf V}_{\rm NLL}\right)^\dagger{\mathbf V}_{\rm NLL} |M_{c\bar{c}}^{(0)}
  \rangle \notag\\
&\times & \sum_{i,j}\left[\Tr({\mathbf H}_{c\bar{c}}{\mathbf D})\, \,({C}_{ci}\otimes f_i)
  \,({C}_{\bar{c} j}\otimes f_j)\right]+{\cal O}(\alpha_s^5)\,,
\end{eqnarray}
where the effective Sudakov radiator $\hat{S}_\ccbar$ is obtained from the
usual colour-singlet case by replacing $B_\ccbar^{(2)} \to \hat{B}_\ccbar^{(2)}$,
with
\begin{eqnarray}\label{eq:B2hat}
\hat{B}_\ccbar^{(2)} = B_\ccbar^{(2)} &+& \frac{\langle
    M_{c\bar{c}}^{(0)} | {\mathbf \Gamma}^{(2)\,\dagger}_t +
  {\mathbf \Gamma}^{(2)}_t|M_{c\bar{c}}^{(0)}
  \rangle}{|M^{(0)}_{c\bar{c}}|^2} \notag\\
  &-& 2 \Re\left[\frac{\langle
    M_{c\bar{c}}^{(1)} |M_{c\bar{c}}^{(0)}
  \rangle}{|M^{(0)}_{c\bar{c}}|^2}\right]
\frac{\langle
    M_{c\bar{c}}^{(0)} | {\mathbf \Gamma}^{(1)\,\dagger}_t +
  {\mathbf \Gamma}^{(1)}_t|M_{c\bar{c}}^{(0)}
  \rangle}{|M^{(0)}_{c\bar{c}}|^2}
  \notag \\
  &+& 2\,\Re\left[\frac{\langle
    M_{c\bar{c}}^{(1)} | {\mathbf \Gamma}^{(1)\,\dagger}_t +
  {\mathbf \Gamma}^{(1)}_t|M_{c\bar{c}}^{(0)}
  \rangle}{|M^{(0)}_{c\bar{c}}|^2}\right]
  \,.
\end{eqnarray}

We finally need to address the contribution coming from the
azimuthally-dependent operator ${\mathbf D}$. As mentioned before, due
to the property $[{\mathbf D}]_\phi={\mathbb 1}$, the second-order
coefficient ${\mathbf D}^{(2)}$ gives a vanishing contribution to
azimuthally-averaged observables at NNLO. On the other hand, in the
case of ${\mathbf D}^{(1)}$ we still have an ${\cal O}(\as^2)$
contribution after averaging over $\phi$, due to the interference
between ${\mathbf D}^{(1)}$ and the collinear functions $C_{ij}$ in
\eqn{eq:bspace} (see \citere{Catani:2014qha} for a detailed
discussion). More specifically, this non-vanishing contribution arises
only in the gluon-initiated channel, and it originates from
the product of ${\mathbf D}^{(1)}$ with the polarization-dependent
coefficient functions $G_{ij}$~\cite{Catani:2010pd}. 
Therefore, the term in square brackets in \eqn{eq:starting} can
be written, upon azimuthal averaging, as
\begin{eqnarray}\label{eq:Dapprox}
\left[\Tr({\mathbf H}_{c\bar{c}}{\mathbf D})\, \,({C}_{ci}\otimes f_i)
  \,({C}_{\bar{c} j}\otimes f_j)\right]_\phi
  &\equiv&
  \left[
  H_{c\bar{c}} \,({C}_{ci}\otimes f_i)
  \,({C}_{\bar{c} j}\otimes f_j)\right]_\phi   \\
  && \hspace{-5.5cm}+ \left(\frac{\as(b_0/b)}{2\pi}\right)^2 \left[
  \left( 
  \frac{\langle M_{c\bar c}^{(0)}|{\mathbf D}^{(1)}| M_{c\bar c}^{(0)}
  \rangle}{|M^{(0)}_{c\bar{c}}|^2}
   \right) \left(
   \,({C}_{ci}^{(1)}\otimes f_i)
  \,(f_{\bar{c}}) 
  +
  \,(f_c)
  \,({C}_{\bar{c} j}^{(1)}\otimes f_j) 
  \right)\right]_\phi + {\cal O}(\as^3)\,.
  \notag  
\end{eqnarray}
The first term in the r.h.s.\ of \eqn{eq:Dapprox} is analogous to
what features in colour-singlet production, where the quantity
$H_{c\bar{c}}$ is defined as
\begin{equation}
\label{eq:Hcdef}
H_{c\bar{c}} \equiv
\frac{|M_{c\bar{c}}|^2}{|M^{(0)}_{c\bar{c}}|^2}=1+\frac{\alpha_s}{2\pi}
H_{c\bar{c}}^{(1)} + \frac{\alpha^2_s}{(2\pi)^2} H_{c\bar{c}}^{(2)} + {\cal O}(\alpha^3_s)\,.
\end{equation}
We recall, however, that the subtraction operator needed to compute
the IR-subtracted amplitudes defining $H_{c\bar{c}}$ is not the same as
in the colour-singlet case.
The analytic result for the integrals over the azimuthal angle that
are needed to compute the additional term in \eqn{eq:Dapprox} can be
found in \citere{hayk}, see e.g.\ eq.~(4.85) of that reference.

\subsection{The \minnlo{} master formulae for heavy-quark pair production}
The result obtained in \eqn{eq:starting} can be regarded as the
starting point for the derivation of the \minnlo formalism for
$Q\bar{Q}$ production. If we compare \eqn{eq:starting} with the
factorisation formula for colour-singlet production (obtained from
\eqn{eq:bspace} by setting ${\mathbf \Delta}={\mathbb 1}$), we
can see that the only difference, besides the extra term taken into
account by \eqn{eq:Dapprox}, is the replacement of the usual
Sudakov radiator by the more complicated expression
$\exp(-\hat{S}_\ccbar ) \langle M_{c\bar{c}}^{(0)} | \left({\mathbf V}_{\rm
    NLL}\right)^\dagger{\mathbf V}_{\rm NLL} |M_{c\bar{c}}^{(0)}
\rangle$,
where $\hat{S}_\ccbar $ is defined by \eqn{eq:sudakov} with the
replacement~\eqref{eq:B2hat}.
In the colour basis in which ${\mathbf \Gamma}_t^{(1)}$ is diagonal,
this expression reduces to a linear combination of complex exponential
terms, whose coefficients are related to the rotation needed for the
change of basis. More specifically, the structure of each of the terms
in the sum is exactly that of a Sudakov radiator with a
\textit{complex} coefficient $\hat{B}_\ccbar^{(1)}$, obtained from the
one in $\hat{S}_\ccbar$ by adding the contribution from the
eigenvalues of the ${\mathbf \Gamma}_t^{(1)}$ operator. The details of
the linear combination are given in \app{app:resum}.

We are now in a position to write an expression for the $\pt$
distribution of the $Q\bar{Q}$ system, differential in the born phase
space $\Phi_{Q\bar{Q}}$, that is NNLO-accurate upon integration
over $\pt$ and LL accurate (plus the simple NLL corrections associated with the coefficient $A^{(2)}$ in the
Sudakov captured by common parton showers) in the limit $\pt \to 0$.
Each summand of this formula will have the same structure as in the colour-singlet
case discussed in \eqn{eq:master1}. Since the new starting point
given in \eqn{eq:starting} is a linear combination of
colour-singlet-like terms, we can follow the same steps used for the
colour-singlet \minnlo formulation that lead from the
$q_T$-resummation formula to
\eqn{eq:master1}~\cite{Monni:2019whf}. 
Specifically, the Fourier integral of \eqn{eq:starting} can be
performed as described in appendix\,E of \citere{Monni:2019whf},
leading to the result
\begin{eqnarray}
\label{eq:master}
  \frac{\mathd\sigma}{\mathd{} \pt\,\mathd{} \Phi_{Q\bar{Q}}}  &=& \frac{\mathd{}}{\mathd{}
  \pt}\bigg\{\sum_{c}\,\frac{e^{-\tilde{S}_{\ccbar}(\pt)}}{2 m_{Q\bar{Q}}^2}\,\langle
  M_{c\bar{c}}^{(0)} | \left({\mathbf V}_{\rm NLL}\right)^\dagger{\mathbf V}_{\rm NLL} |M_{c\bar{c}}^{(0)}
  \rangle\, \notag
 \\  
 &\times& \sum_{i,j}\left[\Tr(\tilde{\mathbf H}_{c\bar{c}}{\mathbf D})\, \,(\tilde{C}_{ci}\otimes f_i)
  \,(\tilde{C}_{\bar{c} j}\otimes f_j)\right]_\phi \bigg\}
  + R_{\rm finite}(\pt) + {\cal O}(\alpha_s^5)\,.
\end{eqnarray}
The quantities $\tilde{S}_\ccbar$, $\tilde{\mathbf H}_{c\bar{c}}$ and
$\tilde{C}_{ij}$ in \eqn{eq:master} are now all evaluated at the
scale $\pt$, and can be obtained from the ones in
\eqn{eq:starting} by means of the following additional
replacements~\cite{Monni:2019whf}:
\begin{eqnarray}
\label{eq:B2tilde}
\hat{B}_\ccbar^{(2)} \to \tilde{B}_\ccbar^{(2)} &\equiv& \hat{B}_\ccbar^{(2)} + 2\zeta_3 (A_\ccbar^{(1)})^2
+ 2\pi\beta_0\, H_{c\bar{c}}^{(1)}\,, \\
\label{eq:H2tilde}
  H_{c\bar{c}}^{(2)}  \to \tilde{H}_{c\bar{c}}^{(2)}  &\equiv& H _{c\bar{c}}^{(2)} - 2\zeta_3 A_\ccbar^{(1)} B_\ccbar^{(1)} \,, \\
\label{eq:C2tilde}
  C_{ci}^{(2)}(z)  \to \tilde{C}_{ci}^{(2)}(z) &\equiv& C_{ci}^{(2)}(z) -2\zeta_3 A_\ccbar^{(1)} \hat{P}_{ci}^{(0)}(z)\,,
\end{eqnarray}
where $H_{c\bar{c}}^{(2)}$ is the second-order term in the $\as$ expansion of
$H_{c\bar{c}}$ defined in \eqn{eq:Hcdef} and $\hat{P}^{(0)}(z)$ is the
tree-level regularised Altarelli--Parisi splitting function.

The final steps in the derivation of the \minnlo algorithm for
$Q\bar{Q}$ production follow again the procedure described in
\citeres{Monni:2019whf,Monni:2020nks} for the colour-singlet
case. Thanks to the structure of \eqn{eq:master}, being a sum of
terms of the type~\eqref{eq:master1}, the derivation can be applied to
each of the terms in the sum individually. The more technical details
can therefore be found in \citeres{Monni:2019whf,Monni:2020nks},
while the main features of the derivation have been already described
in Section~\ref{sec:minnlo_colour_singlet}.  
To connect with the results in
Section~\ref{sec:minnlo_colour_singlet}, we now write
\eqn{eq:master} as
\begin{align}
\label{eq:master_sum}
  \frac{\mathd\sigma}{\mathd{} \pt\,\mathd{} \Phi_{Q\bar{Q}}}  = \frac{\mathd{}}{\mathd{}
  \pt}\Bigg\{\sum_{c}\,\Bigg[\sum_{i=1}^{n_c}{\cal C}^{[\gamma_i]}_\ccbar(\Phi_{Q\bar{Q}})e^{-\tilde{S}^{[\gamma_i]}_{\ccbar}(\pt)}\Bigg]{\cal L}_{c\bar{c}} (\PhiB,\pt)\Bigg\} + R_{\rm finite}(\pt) + {\cal O}(\alpha_s^5)\,,
\end{align}
where we have diagonalised the one-loop soft anomalous dimension in
${\mathbf V}_{\rm NLL}$ and used
\begin{equation}
  e^{-\tilde{S}_\ccbar (\pt) } \langle M_{c\bar{c}}^{(0)} | \left({\mathbf V}_{\rm
    NLL}\right)^\dagger{\mathbf V}_{\rm NLL} |M_{c\bar{c}}^{(0)}
\rangle =|M^{(0)}_{c\bar{c}}|^2\sum_{i=1}^{n_c} {\cal C}^{[\gamma_i]}_\ccbar(\Phi_{Q\bar{Q}})e^{-\tilde{S}^{[\gamma_i]}_{\ccbar}(\pt)}\,.
\end{equation}
We defined $\tilde{S}^{[\gamma_i]}_{\ccbar}$ as the Sudakov obtained from
\eqn{eq:master} via the replacement
\begin{equation}
\label{eq:B1tilde}
B_\ccbar^{(1)} \to \tilde{B}_\ccbar^{(1)}  \equiv B_\ccbar^{(1)} +
\gamma_{i}(\Phi_{Q\bar{Q}})\,,
\end{equation}
where the functions $\gamma_i\equiv \gamma_{\ccbar, i}$ (we omit the
$\ccbar$ subscript in the following) are obtained from the eigenvalues
of the operator ${\mathbf \Gamma}_t^{(1)}$.
Moreover, the coefficients
${\cal C}^{[\gamma_i]}_\ccbar\equiv {\cal C}^{[\gamma_i]}_\ccbar(\Phi_{Q\bar{Q}})$ define the linear combination in terms of the resulting
complex exponentials, and obey the constraint
\begin{equation}
  \sum_{i=1}^{n_c}{\cal C}^{[\gamma_i]}_\ccbar(\Phi_{Q\bar{Q}})=1\,.
\end{equation}
The number of terms $n_c$ in the sum depends on the $SU(3)$
representation of a given flavour configuration of the initial state,
and in particular one has $n_c=4$ for $q\bar{q}$ channels and $n_c=9$
for the $gg$ channel.
The expressions for the $\gamma_i$ and ${\cal C}^{[\gamma_i]}_\ccbar$
coefficients is reported in \app{app:resum}.
We also defined the luminosity factor in \eqn{eq:master_sum} as
\begin{equation}
\label{eq:lumi-tt}
{\cal L}_{c\bar{c}} (\PhiB,\pt)\equiv \frac{|M^{(0)}_{c\bar{c}}|^2}{2 m_{Q\bar{Q}}^2}\sum_{i,j}\left[\Tr(\tilde{\mathbf H}_{c\bar{c}}{\mathbf D})\, \,(\tilde{C}_{ci}\otimes f_i)
  \,(\tilde{C}_{\bar{c} j}\otimes f_j)\right]_\phi \,.
\end{equation}
We can now repeat the derivation of \eqn{eq:BtildeMinnlo} for each of
the terms in the sum over $i$ in \eqn{eq:master_sum}, and we obtain
the following definition of the \minnlo $\tilde B$ function
\begin{align}
  \label{eq:mastertt}
\tilde B_{Q\bar{Q}}&(\PhiBJ,\Phi_{\rm rad}) \equiv \sum_{c_{\rm  FJ}} \Bigg\{\sum_{i=1}^{n_{c_{\rm  F}\leftarrow c_{\rm  FJ}}}{\cal C}^{[\gamma_i]}_{c_{\rm  F}\leftarrow c_{\rm  FJ}}(\PhiB)\exp [-\tilde{S}_{c_{\rm  F}\leftarrow c_{\rm  FJ}}^{[\gamma_i]}(\PhiB,\pt)]  \notag\\
&\times  \Bigg[ B_{c_{\rm FJ}}(\PhiBJ) \left(1+\abarmu{\pt} [\tilde{S}_{c_{\rm F}\leftarrow c_{\rm FJ}}^{[\gamma_i]} (\PhiB,\pt)]^{(1)}\right)
 + V_{c_{\rm FJ}}(\PhiBJ) \Bigg]\Bigg\} \notag\\
 &+  \sum_{c_{\rm FJJ}} \Bigg\{\sum_{i=1}^{n_{c_{\rm  F}\leftarrow c_{\rm  FJJ}}}{\cal C}^{[\gamma_i]}_{c_{\rm  F}\leftarrow c_{\rm  FJJ}}(\PhiB)\exp [-\tilde{S}_{c_{\rm  F}\leftarrow c_{\rm  FJJ}}^{[\gamma_i]}(\PhiB,\pt)] \Bigg\} R_{c_{\rm FJJ}}(\PhiBJ,\Phi_{\rm   rad}) \notag\\
  &+\sum_{c_{\rm  FJ}}\Bigg\{\sum_{c_{\rm F}}\sum_{i=1}^{n_{c_{\rm  F}}}{\cal C}^{[\gamma_i]}_{c_{\rm  F}}(\PhiB)\exp[-\tilde{S}_{c_{\rm  F}}^{[\gamma_i]}(\PhiB,\pt)]  D_{c_{\rm  F}}^{[\gamma_i],\,{\scriptscriptstyle
    (\geq 3)}}(\PhiB,\pt)\Bigg\}{\Fcorr_{c_{\rm FJ}}(\PhiBJ)} \,.
\end{align}
In the above equation, the phase space $\PhiB$ is obtained with an
appropriate kinematic map from the $\PhiBJ$ phase space.
The projection can be obtained, for example (see
\citeres{Nason:2004rx,Frixione:2007vw,Alioli:2010xd}), with a
longitudinal boost such that the \F{}-system has zero rapidity,
followed by a transverse boost such that the \F{}-system has zero
transverse momentum, and a final longitudinal boost equal and opposite
to the first one.
Analogously to \eqn{eq:BtildeMinnlo} (and with a little abuse of
notation), the transverse momentum $\pt$ in each term of
\eqn{eq:mastertt} is calculated in the corresponding kinematics
(either $\PhiBJ$ or $\PhiBJJ$).

The sum over $c_{\rm FJ}$ in \eqn{eq:mastertt} runs over all possible
initial-state flavour compositions of the \FJ{} process (in this case
$c_{\rm FJ}=\{gg,gq, q\bar{q}\}$, and we omit for simplicity all
conjugate configurations in our notation).  The notation
$c_{\rm F}\leftarrow c_{\rm FJ}$ denotes instead the flavour
configuration of the final state \F{} corresponding to the flavour
projection of the \FJ{} final state. Analogously,
$c_{\rm F}\leftarrow c_{\rm FJJ}$ indicates a projection between the
flavour configuration of \FJJ{} and \F{}, which is obtained by first
deriving the flavour of \FJ{} via the \POWHEG{} flavour mapping and
then applying the $c_{\rm F}\leftarrow c_{\rm FJ}$ projection.
There are various ways to define such a projection, meaning that each
summand might be unphysical and only the total combination
(i.e.\ $\tilde B_{Q\bar{Q}}$) is physical.
We define the projection using the following procedure:
\begin{itemize}
\item If $c_{\rm FJ}=\{gg,q\bar{q}\}$ then there is no
  ambiguity and we set $c_{\rm F}= c_{\rm FJ}$;
\item If $c_{\rm FJ}=gq$ then we assign the flavour $c_{\rm F}$ as
  follows. In the partonic c.o.m. frame, we consider the rapidity
  $\eta_\ell$ of the light parton in the \FJ{} final state and we
  split the event into two hemispheres at $\eta_\ell=0$.
  If the parton shares the same hemisphere with an initial-state
  gluon, then we assign $c_{\rm F}=q\bar{q}$. Conversely, if the
  parton shares the same hemisphere with an initial-state quark (or
  anti-quark), then we assign $c_{\rm F}=gg$.
  This procedure guarantees that the flavour configuration is
  correctly assigned when the light parton is collinear to one of the
  initial-state legs (i.e.\ the collinear singularity is correctly
  reproduced by the Sudakov
  $\tilde{S}_{c_{\rm F}\leftarrow c_{\rm FJ}}^{[\gamma_i]}$).
\end{itemize}

\noindent The function
$D_{c_{\rm F}}^{[\gamma_i],\,{\scriptscriptstyle (\geq 3)}}$ is
defined by extending \eqn{eq:Ddef} as
\begin{align}
  \label{eq:Ddeftt}
  D_{c_{\rm F}}^{[\gamma_i],\,{\scriptscriptstyle (\geq 3)}}(\PhiB,\pt) =&\, D_{c_{\rm F}}^{[\gamma_i]}(\PhiB,\pt) -\abarmu{\pt} [D_{c_{\rm F}}^{[\gamma_i]}(\PhiB,\pt)]^{(1)}- \left(\abarmu{\pt}\right)^2[D_{c_{\rm F}}^{[\gamma_i]}(\PhiB,\pt)]^{(2)}\,,\notag\\
       D_{c_{\rm F}}^{[\gamma_i]}(\PhiB,\pt) =& -\frac{\mathd \tilde{S}_{c_{\rm F}}^{[\gamma_i]}(\PhiB,\pt)}{\mathd \pt} {\cal L}_{c_{\rm F}} (\PhiB,\pt)+\frac{\mathd {\cal L}_{c_{\rm F}} (\PhiB,\pt)}{\mathd \pt}\,.
\end{align}
The explicit expressions of the expansion coefficients
$[D_{c_{\rm F}}^{[\gamma_i]}(\PhiB,\pt)]^{(1)} $ and
$[D_{c_{\rm F}}^{[\gamma_i]}(\PhiB,\pt)]^{(2)}$ directly follow from
the colour-singlet formulae given in eqs.~(27), (28) of
\citere{Monni:2020nks}.
In \eqn{eq:mastertt} we have explicitly indicated the dependence of
the Sudakov $\tilde{S}$ and the luminosity factor ${\cal L}$ on the
phase space $\PhiB$.

\section{Computational aspects}
\label{sec:comput}
In this section we discuss some computational aspects of the
implementation of \eqn{eq:mastertt}. In particular, we derive the
dependence of \eqn{eq:mastertt} on the perturbative scales and
introduce their settings, whose variation is used to estimate the size
of subleading perturbative corrections. We further comment on the
treatment of the Landau pole in the strong coupling constant that is
encountered in the calculation of the Sudakov radiators
$\tilde{S}_\ccbar^{[\gamma_i]}$.
Finally, we also describe the implementation of the decay of the top quarks. 

\subsection{Renormalisation and factorisation scale dependence}
The fixed-order elements of \eqn{eq:mastertt} (i.e.\
$B_{c_{\rm FJ}}$,$V_{c_{\rm FJ}}$ and $R_{c_{\rm FJJ}}$) receive the
customary dependence on the renormalisation- and factorisation-scale
variations.
For collider reactions that are mediated by strong interactions
already at the Born level, such as the process considered here, the
$n_B$ powers of the strong coupling in the Born squared matrix element
are evaluated at a scale $\muRzero$ which is always of order
$m_{Q\bar{Q}}$, and we would like to maintain its dependence explicit
in the expressions that follow so that different choices for this
scale can be made by the user. In particular, we will set this scale
to a dynamical hard scale in the phenomenological study of this
article.
The remaining scales are defined with respect to the central value as
$\muR=\KR\, \mu_{0}$, $\muF=\KF\,\mu_0$, with $\mu_0\sim \pt$ in the
regime $\pt \ll \mQQ$.
For the time being we will consider $\mu_0=\pt$, but in
\sct{sec:scale} this will be consistently modified by means of a
prescription to turn off all resummation effects at large $\pt$.
The $\KR$ and $\KF$ dependence of the remaining terms in
\eqn{eq:mastertt} is obtained by following
\citeres{Monni:2019whf,Monni:2020nks} and, for completeness, it is
reported below.

We start from the Sudakov radiators
$\tilde{S}_\ccbar^{[\gamma_i]}(\pt)$, which now become
\begin{equation}
\label{eq:sud_final}
  \tilde{S}_\ccbar^{[\gamma_i]}\left(\pt,\KR\right) \equiv
  \int_{\pt^2}^{m_{Q\bar{Q}}^2}\frac{\mathd{}q^2}{q^2} \left[A_\ccbar(\alpha_s(\KR\,
    q),\KR)\ln\frac{m_{Q\bar{Q}}^2}{q^2}+\tilde{B}_\ccbar(\alpha_s(\KR \,q),\KR)\right]\,,
\end{equation}
where the $\KR$ dependent perturbative coefficients $A_\ccbar^{(i)}$ and
$\tilde{B}_\ccbar^{(i)}$ are obtained from those given in
\neqn{eq:sudakov}, \eqref{eq:B2hat}, \eqref{eq:B2tilde}, \eqref{eq:B1tilde} by
means of the relations \cite{Monni:2019whf}
\begin{align}
A_\ccbar^{(2)}(\KR) \equiv& \,A_\ccbar^{(2)} + (2\pi \beta_0) A_\ccbar^{(1)} \ln \KR^2,\notag\\ 
\tilde{B}_\ccbar^{(2)}(\KR) \equiv & \,\tilde{B}_\ccbar^{(2)}  + (2\pi \beta_0) \tilde{B}_\ccbar^{(1)} \ln \KR^2 +
                (2\pi\beta_0)^2\,n_B \ln \frac{(\muRzero)^2}{m_{Q\bar{Q}}^2}\,.
\end{align}
The factor $n_B$ (defined as the power of $\as$ of the Born squared
amplitude) is induced by the
presence of $H_{c\bar{c}}^{(1)}$ in the $\tilde{B}_\ccbar^{(2)}$ coefficient, which,
in turn, originates from evaluating the hard virtual corrections
$\tilde{\mathbf H}_{c\bar{c}}$ at $\pt$ in the factor ${\cal L}_{c\bar{c}}$ defined in
\eqn{eq:lumi-tt}, see also \eqn{eq:Dapprox}.
The integral in \eqn{eq:sud_final} is evaluated exactly without
additional approximations through a numerical integration using a
quadrature method, which allows us to adopt the strong coupling
constant used in the evolution of the parton densities, encoding the
effect of heavy-quark thresholds as taken from the parton-densities
provider.

We now continue with the scale dependence of the luminosity
factor~\eqref{eq:lumi-tt}.
All coupling constants are now evaluated at $\muR=\KR\,\pt$ and the
parton densities are calculated at $\muF=\KF\,\pt$.
The definition of the $[\dots]_{\phi}$ expectation value is given in
\eqn{eq:Dapprox}. The perturbative coefficients of the hard
factor $H_{c\bar{c}}$ acquire the following $\KR$ and $\KF$ dependence~\cite{Monni:2019whf}
\begin{align}
H_{c\bar{c}}^{(1)}(\KR) \equiv\,& H_{c\bar{c}}^{(1)} + (2\pi\beta_0) n_B \ln
                         \frac{(\muRzero)^2}{m_{Q\bar{Q}}^2}\,, \\
  \tilde{H}_{c\bar{c}}^{(2)}(\KR) \equiv\,& \tilde{H}_{c\bar{c}}^{(2)} - 2 \, n_B (1-n_B) \pi^2\beta_0^2 \ln^2 \frac{(\muRzero)^2}{m_{Q\bar{Q}}^2}
+4\pi^2 n_B \ln \frac{(\muRzero)^2}{m_{Q\bar{Q}}^2}\left(\beta_1 +
                                   \beta_0^2\ln \KR^2\right)\notag\\
                       & + 2 \pi \beta_0\left(n_B\ln \frac{(\muRzero)^2}{m_{Q\bar{Q}}^2}
                         +\ln \KR^2\right) H_{c\bar{c}}^{(1)}\,.
\end{align}
The coefficient functions $C_{ij}$ receive the following scale dependence \cite{Monni:2019whf}:
\begin{align}
C_{ij}^{(1)}(z,&\, \KF) \equiv\, C_{ij}^{(1)}(z) -
                                    \hat{P}_{ij}^{(0)}(z)\ln \KF^{2},\\
\tilde{C}_{ij}^{(2)}(z,&\, \KF, \KR) \equiv\, \tilde{C}_{ij}^{(2)}(z) +
                                    \pi\beta_0 \hat{P}_{ij}^{(0)}(z)\left(
                                    \ln^2\KF^{2} -
                                   2 \ln \KF^{2}
                                    \ln \KR^{2}\right) -
                                    \hat{P}_{ij}^{(1)}(z)\ln \KF^{2}\notag\\
& + \frac{1}{2}(\hat{P}^{(0)}\otimes \hat{P}^{(0)})_{ij}(z) \ln^2\KF^{2} -
  (\hat{P}^{(0)} \otimes C^{(1)})_{ij}(z) \ln \KF^{2} + 2\pi\beta_0
  C_{ij}^{(1)}(z) \ln \KR^{2}\,,\notag
\end{align}
while the part of the coefficient functions $G_{ij}$ as well as the
second term in the r.h.s.\ of \eqn{eq:Dapprox}, which together contain
all sources of azimuthal and colour correlations, remain unchanged
since they enter first at relative order ${\cal O}(\alpha_s^2)$.

We use the above coefficients in the evaluation of the \minnlo{}
coefficient
$D_{c_{\rm F}}^{[\gamma_i],\,{\scriptscriptstyle (\geq 3)}}$ defined
in \eqn{eq:Ddeftt}.
In addition to the scale dependence of the coefficients above (which
is understood in the following equations), the
$[D_\ccbar^{[\gamma_i]}(\PhiB,\pt)]^{(1)}$ and
$[D_\ccbar^{[\gamma_i]}(\PhiB,\pt)]^{(2)}$ terms acquire an additional
explicit dependence on the scale factors $\KR$ and $\KF$ due to
expanding the derivative of the luminosity in \eqn{eq:Ddeftt} in
powers of $\alpha_s(\KR\,\pt)$. This reads
\begin{align}
[D_\ccbar^{[\gamma_i]}(\PhiB,\pt)]^{(1)} (\KF,\KR) &\equiv [D_\ccbar^{[\gamma_i]}(\PhiB,\pt)]^{(1)}\,,\notag\\
[D_\ccbar^{[\gamma_i]}(\PhiB,\pt)]^{(2)} (\KF,\KR) &\equiv [D_\ccbar^{[\gamma_i]}(\PhiB,\pt)]^{(2)} - 2\beta_0 \pi \left[  \frac{\mathd {\cal L}_{c\bar{c}}(\PhiB,\pt)}{\mathd \pt}
  \right]^{(1)} \ln\frac{\KF^2}{\KR^2}\,.
\end{align}
On the other hand, the scale dependence in the
$D_{\ccbar}^{[\gamma_i],\,{\scriptscriptstyle (\geq 3)}}(\PhiB,\pt)$
originating from the luminosity factor ${\cal L}$ is retained
implicitly by performing the derivative of ${\cal L}$ numerically.

\subsection{Resummation scale dependence}
We also introduce some flexibility in the scale at which one turns off
the resummation effects in \eqn{eq:mastertt} in phase space
regions characterised by a large transverse momentum $\pt$ of the
heavy-quark pair.\footnote{Since the following formulas are reported here for the first time we 
stress that they apply also to the case of colour-singlet production.}
This is done by means of a resummation scale $Q\sim m_{Q\bar{Q}}$,
which is introduced in both the Sudakov radiators and the luminosity
factor. At the level of the Sudakov, the resummation scale $Q$
effectively replaces the upper bound of the integral, as we now
show. We start by re-writing the Sudakov factors in \eqn{eq:sud_final}
as (we define $\KQ\equiv Q/m_{Q\bar{Q}}$)
\begin{align}
  \label{eq:sud_Q}
  \tilde{S}_\ccbar^{[\gamma_i]}\left(\pt,\KR\right) =& 
  \int_{\pt^2}^{Q^2}\frac{\mathd{}q^2}{q^2} \left[A_\ccbar(\alpha_s(\KR\,
    q),\KR)\ln\frac{Q^2}{q^2}\right.\notag\\
  &\left.+\left(\tilde{B}_\ccbar(\alpha_s(\KR \,q),\KR)
    - A_\ccbar(\alpha_s(\KR\,
    q),\KR)\ln \KQ^2\right)\right] \notag\\
  &+\int_{Q^2}^{m_{Q\bar{Q}}^2}\frac{\mathd{}q^2}{q^2} \left[A_\ccbar(\alpha_s(\KR\,
    q),\KR)\ln\frac{Q^2}{q^2}\right.\notag\\
  &\left.+\left(\tilde{B}_\ccbar(\alpha_s(\KR \,q),\KR)
    - A_\ccbar(\alpha_s(\KR\,
    q),\KR)\ln \KQ^2\right)\right]\,.
\end{align}
Subsequently, for later convenience, we expand the second integral in
the r.h.s.\ of \eqn{eq:sud_Q} to second order in powers of
$\alpha_s(\KR/\KQ\, \pt)$, and absorb the terms proportional to
$\ln\frac{Q}{\pt}$ into the first integral neglecting subleading
corrections (specifically into the $\tilde{B}^{(2)}_\ccbar$
coefficient), namely
\begin{align}
\tilde{B}_\ccbar^{(2)}(\KR,\KQ) \equiv & \,\tilde{B}_\ccbar^{(2)}(\KR) -
                                    \pi\beta_0\ln
                                    \KQ^2\left(A_\ccbar^{(1)} \ln \KQ^2
                                    - 2 \tilde{B}_\ccbar^{(1)}\right)\,,
\end{align}
while the non-logarithmic terms are expanded out of the exponential
and absorbed into the $H_{c\bar{c}}$ hard factor.
Consistently, we then change the scales of the strong coupling and
parton densities in the hard factor $\tilde{H}_{c\bar{c}}$ and parton
distribution functions in the luminosity factor~\eqref{eq:lumi-tt} as
follows:
\begin{equation}
  \label{eq:newmu}
\muR = \KR \,\pt \to \muR = \frac{\KR}{\KQ}\pt\,,\quad \muF = \KF \,\pt \to \muF = \frac{\KF}{\KQ}\pt\,.
\end{equation}
The above procedure results in the following scale dependence in the
coefficients $\tilde{H}_{c\bar{c}}^{(i)}$ and $\tilde{C}_{ij}$
\begin{align}
  H_{c\bar{c}}^{(1)}(\KR,\KQ) &\equiv H_{c\bar{c}}^{(1)}(\KR) + 
                     \left(-\frac{A_\ccbar^{(1)}}{2} \ln
                     \KQ^2+\tilde{B}_\ccbar^{(1)}\right)\ln \KQ^2\,,
\end{align}
\begin{align}
  \tilde{H}_{c\bar{c}}^{(2)}(\KR,\KQ) &\equiv \tilde{H}_{c\bar{c}}^{(2)}(\KR) + \frac{(A_\ccbar^{(1)})^2}{8}\ln^4
                     \KQ^2 -
                     \left(\frac{A_\ccbar^{(1)}\tilde{B}_\ccbar^{(1)}}{2}+\pi\beta_0\frac{A_\ccbar^{(1)}}{3}\right)
                     \ln^3\KQ^2\notag\\
  & +
    \left(-\frac{A_\ccbar^{(2)}(\KR)}{2}+\frac{(\tilde{B}_\ccbar^{(1)})^2}{2}+\pi\beta_0
    \tilde{B}_\ccbar^{(1)} - n_B \pi \beta_0 A_\ccbar^{(1)}\ln \frac{(\muRzero)^2}{m_{Q\bar{Q}}^2}\right)
    \ln^2\KQ^2\notag\\
& + \left(\tilde{B}_\ccbar^{(2)}(\KR) + 2 n_B \pi\beta_0\tilde{B}^{(1)}_\ccbar
                         \ln
                         \frac{(\muRzero)^2}{m_{Q\bar{Q}}^2}\right)\ln
                         \KQ^2\notag\\
  & + \left(\tilde{B}_\ccbar^{(1)}\ln \KQ^2 -\frac{A_\ccbar^{(1)}}{2}\ln^2 \KQ^2-2\pi\beta_0\ln \KQ^2\right) H_{c\bar{c}}^{(1)}\,,
\end{align}
and
\begin{align}
  C_{ij}^{(1)}(z,\KF, \KQ) &\equiv  C_{ij}^{(1)}(z,\KF/\KQ)\,,\\
  \tilde{C}_{ij}^{(2)}(z,\KF, \KR, \KQ) &\equiv  \tilde{C}_{ij}^{(2)}(z,\KF/\KQ,\KR/\KQ)\,.
\end{align}

\subsection{Profiled logarithms and scale setting at large $\pt$}
\label{sec:scale}
In order to actually turn off all resummation effects at $\pt \sim Q$
we introduce a (new version of the) \textit{modified} logarithm by
means of the replacement
\begin{equation}
\label{eq:modlog}
\ln\frac{Q}{\pt} \to L\,
\end{equation}
with
\begin{equation}
\label{eq:modlogdef}
  L \equiv
  \begin{cases}
    \ln\frac{Q}{\pt} \,\, &{\rm if}\,\, \pt \leq \frac{Q}{2}\,,\\
    \ln\left( a_{0}+a_{1}\frac{\pt}{Q} + a_2\left(\frac{\pt}{Q}\right)^2 \right) \,\, &{\rm if}\,\, \frac{Q}{2} < \pt \leq Q\,,\\
    0 \,\, &{\rm if}\,\, \pt > Q\,,
  \end{cases}
\end{equation}
where $a_0=5$, $a_1=-8$, $a_2 = 4$.
The coefficients are chosen in such a way that $L$ and its derivative 
with respect to $\pt$ are continuous at $Q/2$ and that they both vanish at $Q$.   
The above parameterisation ensures that the resummation is exactly
turned off for $\pt \geq Q$, at variance with the modified logarithm
adopted in \citere{Monni:2020nks} which only vanishes
asymptotically for $\pt \gg Q$. 
Following \citere{Monni:2020nks}, the replacement in \eqn{eq:modlog}
can be implemented with the following simple procedure:
\begin{itemize}
\item We rewrite the scale setting in \eqn{eq:newmu} as
\begin{equation}
\label{eq:scales}
\muR = \KR \, m_{Q\bar{Q}} \,e^{-L}\,,\qquad \muF = \KF \, m_{Q\bar{Q}}\, e^{-L}\,.
\end{equation}
At small $\pt$ this reduces to \eqn{eq:newmu} (we recall that
$\KQ=Q/m_{Q\bar{Q}}$), while at large $\pt$ the scales tend to
$\KR\, \mQQ$ and $\KF\, \mQQ$, respectively.
\item We rewrite the lower integration bound of the first integral in
  the Sudakov~\eqref{eq:sud_final} (the integrand is unchanged) as
\begin{equation}
\pt \to Q e^{-L}.
\end{equation}
\item We multiply
  $D_{\ccbar}^{[\gamma_i],\,{\scriptscriptstyle (\geq 3)}}$ in
  \eqn{eq:mastertt} by the following Jacobian factor:
\begin{equation}
\label{eq:jacob}
D_{\ccbar}^{[\gamma_i],\,{\scriptscriptstyle (\geq 3)}}\to {\cal J}_Q D_{\ccbar}^{[\gamma_i],\,{\scriptscriptstyle (\geq 3)}},\qquad {\cal J}_Q =
\begin{cases}
    1 \,\, &{\rm if}\,\, \pt \leq \frac{Q}{2}\,,\\
    \frac{2 a_{2} \pt+a_{1} Q}{a_{0} Q^2+a_{1} Q \pt + a_2 \pt^2} \,\, &{\rm if}\,\, \frac{Q}{2} < \pt \leq Q\,,\\
    0 \,\, &{\rm if}\,\, \pt > Q\,.
  \end{cases}
\end{equation}
\end{itemize}
In order to avoid the Landau singularity in the calculation of the
Sudakov radiator and in the evolution of parton densities, we implement
the freezing procedure of \citere{Monni:2020nks} with a freezing
scale $Q_0=2$ GeV. This is easily implemented by further modifying the
scales in \eqn{eq:scales} as follows
\begin{equation}
\label{eq:scales_Q0}
\muR = \KR \, m_{Q\bar{Q}}\,\left(e^{-L}+\frac{Q_0}{Q} \,g(\pt)\right)\,,\qquad \muF = \KF\, m_{Q\bar{Q}}\,
\,\left(e^{-L} + \frac{Q_0}{Q} \,g(\pt)\right),
\end{equation}
where $g(\pt)$ is a damping function. A suitable choice (albeit
one has some freedom) is~\cite{Monni:2020nks}
\begin{equation}
\label{eq:gpt}
g(\pt) = \frac{1}{1+\frac{Q}{Q_0} e^{-L}}\,.
\end{equation}
This prescription is also consistently adopted in the Sudakov radiator
$\tilde{S}(\pt)$~\eqref{eq:sud_Q}, at the integrand level. The
integration is then still performed numerically.
Analogously to what has been discussed for the modified logarithms in
\eqn{eq:jacob}, the choice in \eqn{eq:scales_Q0} requires
the introduction of an additional factor ${\cal J}_{Q_0}$
\begin{equation}
{\cal  J}_{Q_0} \equiv Q \,e^{-L} \frac{1 -g^2(\pt)}{Q \,e^{-L} +
  Q_0 \, g(\pt)}\,,
\end{equation}
which multiplies \textit{only} the derivative of the
luminosity~\cite{Monni:2020nks} in the definition of
$D_{\ccbar}^{[\gamma_i],\,{\scriptscriptstyle (\geq
    3)}}$~\eqref{eq:Ddeftt}.
We stress that the freezing only affects running coupling effects, and
hence does not modify the double logarithmic structure of
\eqn{eq:mastertt} and the phase-space integration 
is performed down to vanishing $\pt{}$ including all power corrections at the
desired accuracy.
Therefore, the scale $Q_0$ does not play the
role of a slicing parameter in the \minnlo{} calculation.
  
A final important remark is that in all
formulae defined in this section we allow for the scale factors $\KR$
and $\KF$ to be set dynamically on an event-by-event basis. That is,
we can choose to perform the computation with reference
renormalisation and factorisation scales at large $\pt{}$ different from a multiple of
the top-quark pair invariant mass $m_{Q\bar{Q}}=m_{t\bar{t}}$, while their 
setting at small $\pt{}$ has to remain of order $\pt{}$ for consistency 
of the \minnlo{} procedure, as detailed before.
This has the advantage of allowing different scale settings in phase
space regions characterised by a large $\pt$, which better describe
the momentum transfer in the hard scattering.
Suitable candidates are for instance scales proportional to
\begin{equation}
\label{eq:Htt}
H_T^{t\bar{t}} = m_T^{t}+m_T^{\bar{t}}\,,\quad H_T^{t\bar{t}+{\rm\,
    jets}}= m_T^{t}+m_T^{\bar{t}}+\sum_{i\not \in t}p_{T,i}\,,
\end{equation}
with $m_T^i=\sqrt{m_i^2+p_{Ti}^2}$ being the transverse mass of
particle $i$.\footnote{A detailed discussion about suitable scale settings in top-quark pair
production is reported in \citeres{Czakon:2016dgf,Catani:2019hip,Caola:2021mhb}.}
In particular, such settings may be used for the fixed-order part of the 
calculation by selecting from the inputs of the \minnlo{} $t\bar{t}$ code
{\tt fixedscale\;0} and by choosing different values of the parameter {\tt largeptscales}.
Moreover, by setting the parameter {\tt whichscale} in the input file
such scale setting can be chosen for the powers of the coupling constant of the
Born squared amplitude, denoted by $\muRzero$ in the equations
given in this section.
In the phenomenological studies shown below, we use 
$\muRzero=\KR\,H_T^{t\bar{t}}/4$ ({\tt whichscale\;4}) 
for the scale of the two Born powers of the strong coupling, while 
not changing the setting of the other scales (keeping {\tt fixedscale\;0} and  {\tt largeptscales\;0}).

\subsection{Top-quark decays and spin correlations}
\label{sec:decay}
In this section we discuss the inclusion of the top-quark decays and
of spin correlations between the decay products.
Several event generators can achieve NLO accuracy for top-quark pair
production in both production and decay, with the exact inclusion of
off-shell (including non-resonant contributions) and spin correlation
effects~\cite{Frederix:2012ps,Hoeche:2014qda,Cormier:2018tog,%
  Jezo:2015aia,Jezo:2016ujg,Frederix:2016rdc}.

In fixed-order calculations up to NNLO QCD, the inclusion of 
top-quark decays and spin correlations has been
presented in \citeres{Behring:2019iiv,Czakon:2020qbd} in
the narrow-width approximation and in the double-resonant channel only.
Including such effects at this perturbative order in event generators
is currently still an open problem. In our study we simply consider
the inclusion of off-shell effects and spin correlations for the
double-resonant channels at the leading order.
This is motivated by the observation that the effect of spin
correlations in several important top-decay observables (such as the
azimuthal correlation between the two charged leptons in the fully
leptonic decay channel) is well described at this perturbative order
provided one includes NNLO QCD corrections to the production of the
top-quark pair (see e.g.\ \citere{Behring:2019iiv}).

Our implementation of the top-quark decays follows the approach of
\citere{Frixione:2007zp} that is already implemented in the generator
of \citere{Alioli:2011as} for the production of a top-quark pair and a
light jet, which we used as a starting point to build the \minlo{} and
\minnlo{} generators presented in this work.
In the results presented in the next section we consider top quarks and $W$ bosons to be
on-shell, but we give the user the possibility to include off-shell
effects according to the same algorithm used in the
\textsc{Powheg-hvq} process~\cite{Frixione:2007zp}, which we describe
in detail in \app{sec:tvirt}.
We apply the algorithm described in the articles mentioned above solely to the 
double-resonant contributions of the production of the decayed final
state, while keeping the top quarks (and the $W$ bosons in their decay chain) 
on their mass shell.
This simple procedure includes the top-quark decays at LO in the \FJ{}
and \FJJ{} configurations. Since in the limit of small $\pt$ the
entire decay process factorises from the structure of QCD radiation,
this procedure ensures a LO treatment of the top-quark decays (including  
spin correlations) also at the level of fully inclusive observables (i.e.\ those
related to the final state \F{}).
In the remainder of this paper we will refer to this implementation of
the top-quark decays as the \textit{native} implementation.

\begin{figure}[t!]
\begin{center}
\begin{tabular}{ccc}
\hspace{-.55cm}
\includegraphics[width=.34\textwidth,page=11]{plots/comparison_madspin_native/ATLAS_leptonic_with_tau.pdf}
&
\hspace{-0.6cm}
\includegraphics[width=.34\textwidth,page=1]{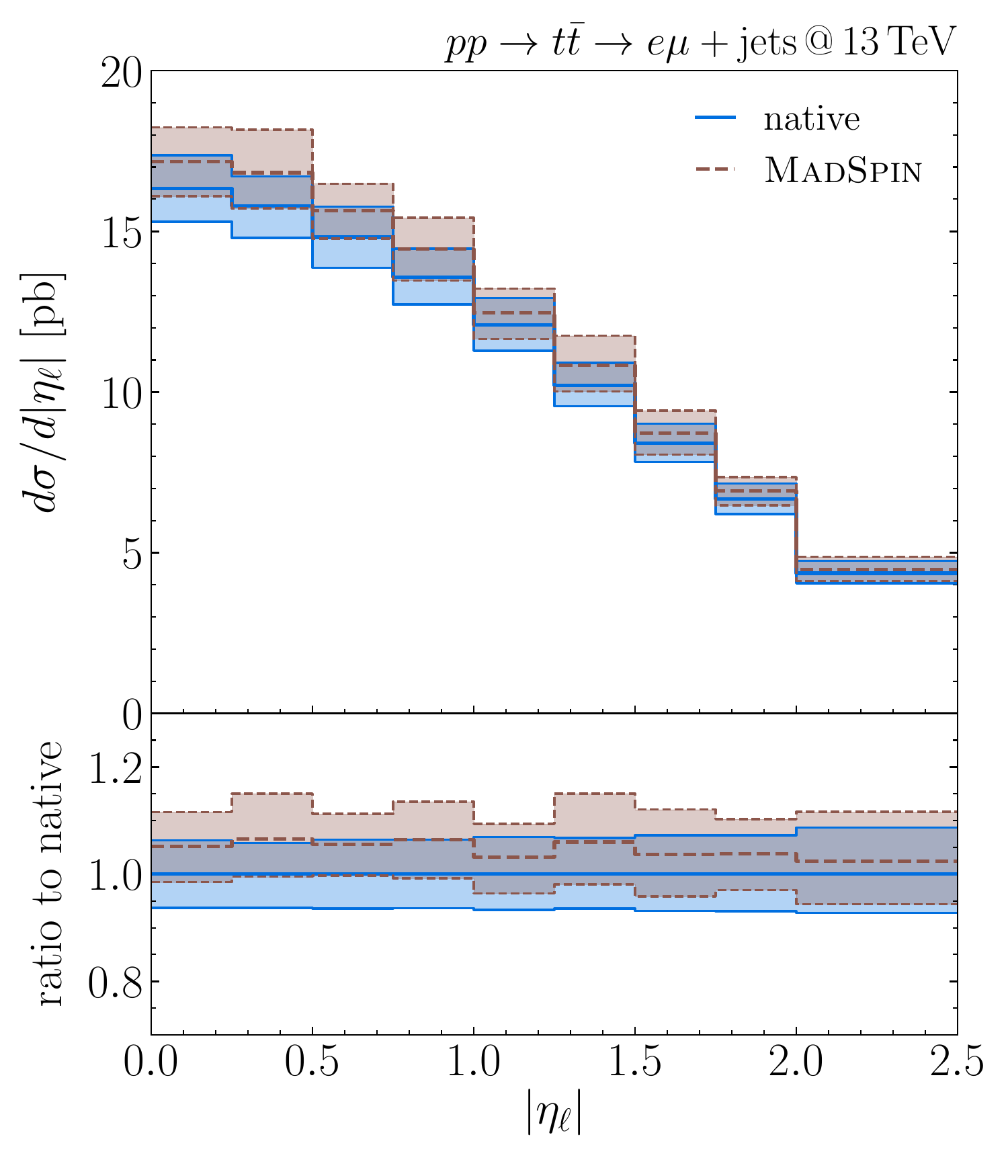} 
&
\hspace{-0.6cm}
\includegraphics[width=.34\textwidth,page=18]{plots/comparison_madspin_native/ATLAS_leptonic_with_tau.pdf}\\
\hspace{-.55cm}
\includegraphics[width=.34\textwidth,page=1]{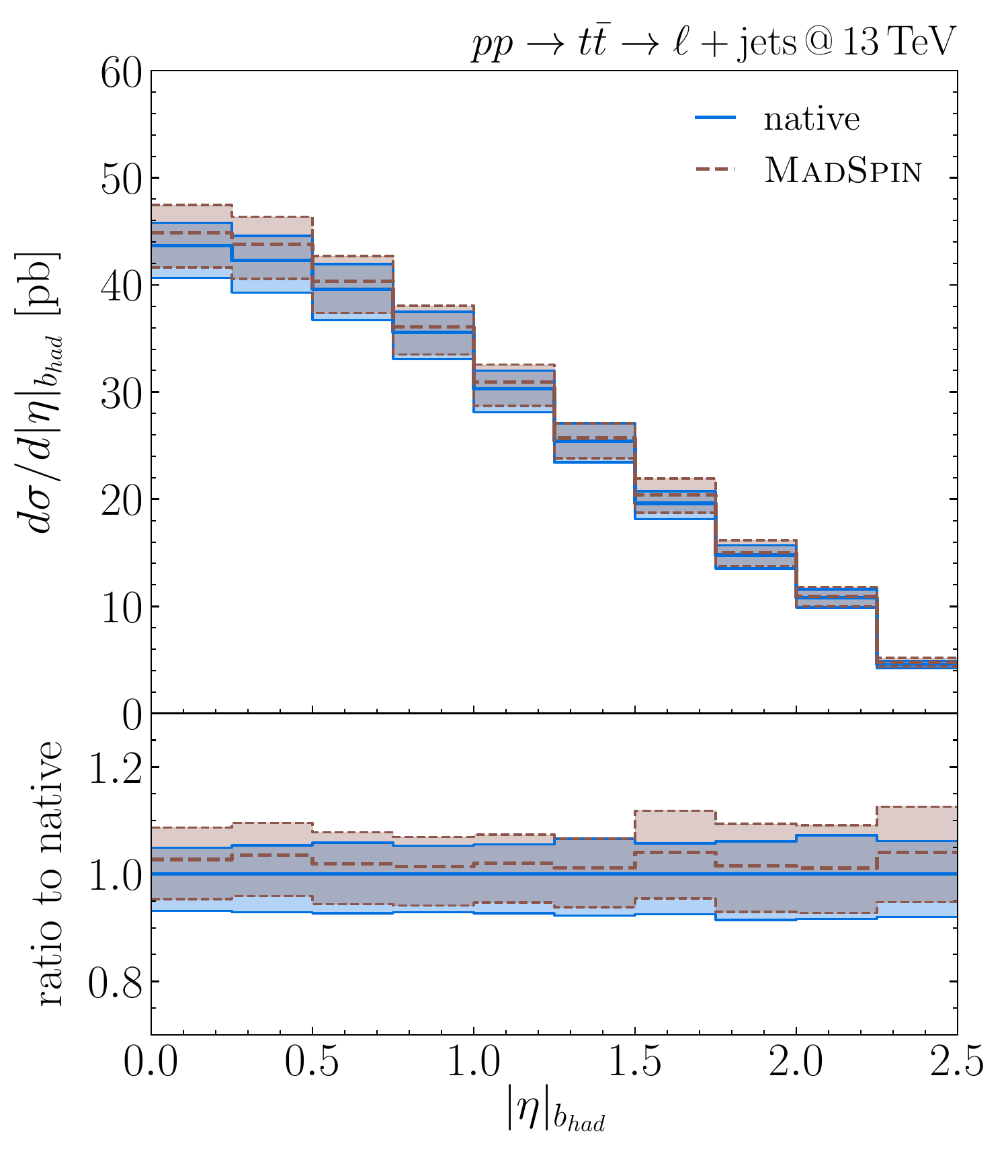}
&
\hspace{-0.6cm}
\includegraphics[width=.34\textwidth,page=3]{plots/comparison_madspin_native/CMS_semileptonic.pdf} 
&
\hspace{-0.6cm}
\includegraphics[width=.34\textwidth,page=11]{plots/comparison_madspin_native/CMS_semileptonic.pdf}
\end{tabular}
\caption{\label{fig:madspin}  Comparison of \minnlo{} predictions with native implementation of the top-quark decays with on-shell top quarks and $W$ bosons (blue, solid) against that of \MadSpin{} with 
off-shell top quarks and $W$ bosons (brown, dashed) in \setupleptonic{} (upper plots) and \setupsemi{} (lower plots). See \sct{sec:setup} for details on the setup.}
\end{center}
\end{figure}

In order to validate the native implementation of the top-quark decays
and spin correlations, we have produced corresponding results using
the well-tested and well-established \MadSpin{} code
\cite{Artoisenet:2012st}.  Those results have been obtained by passing
on-shell $t\bar{t}$ events at NNLO+PS to \MadSpin{}.  While both
approaches rely on the same basic concept, their implementation and
the choices made may differ in some details. In particular, we have
produced \MadSpin{} results where in the decayed matrix elements the
top quarks are off-shell. Moreover, \MadSpin{} always includes
off-shell effects for the $W$ bosons in the top-quark decays, while in
the native implementation they are on-shell when the top-quarks are
kept on-shell, as stated above.  Therefore, the comparison of the
native implementation to the \mbox{\MadSpin{}} results allows us to
quantify the differences (and uncertainties) that may be expected from
the approximation adopted for the top-quark decays.  In particular,
such comparison allows us to assess whether those differences,
although they are not expected to be described by scale-variation
uncertainties, could exceed the quoted perturbative uncertainties.

In \fig{fig:madspin} we show three observables in \setupleptonic{}, i.e.\ in the fully leptonic top-decay mode, and three 
observables in \setupsemi{}, i.e.\ in the semi-leptonic top-decay mode (see \sct{sec:setup}
for details on the setup). We compare \minnlo{} predictions with 
the native implementation of the top-quark decays in blue and solid against \minnlo{} predictions with the top-quark 
decays included through \MadSpin{} 
including off-shell effects in brown and dashed. By and large, both implementations lead to results
that are very close, regardless of whether or not off-shell top-quark effects are included, 
in several instances even barely different disregarding statistical fluctuations. In some cases,
like the $p_{T,\ell}$ spectrum, there can be differences ranging up to 10\%, which are not unexpected 
due the different approximations made for the top-quark decays, but in all cases 
those are fully covered by the quoted scale-uncertainty bands.
We have tested a large number of distributions in either of these two
top-decay modes, arriving to similar conclusions for all of them.

\section{Phenomenological results}
\label{sec:results}
In this section, we present phenomenological results for top-quark
pair production at \mbox{NNLO+PS} with and without decays of the top quarks.
After introducing our setup in \sct{sec:setup}, we perform a
comprehensive comparison against LHC data: In \sct{sec:incl} we
compare \minnlo{} predictions for on-shell top quarks against CMS data
extrapolated to the inclusive $t\bar t$ phase space.  We then
continue by considering top-quark decays including spin correlations
and we compare novel \minnlo{} predictions against data from both
ATLAS and CMS in various top-decay channels, including fully leptonic
(\sct{sec:leptonic}), semi-leptonic (\sct{sec:semi}) and hadronic
(\sct{sec:hadronic}) top-quark decays.

\subsection{Input parameters and fiducial cuts}
\label{sec:setup}

We present results for proton--proton collisions at the LHC with a
centre-of-mass energy of 13\,TeV. The top-quark pole mass is set to
$m_t = 173.3$\,GeV.  In the top-quark decays both the top quarks and
the $W$ bosons are treated in the narrow-width approximation (see
\sct{sec:decay} for details), i.e.\ $\Gamma_t = \Gamma_W = 0$\,GeV is
used, while assuming branching ratios for each charged lepton
$\ell\in\{e,\mu,\tau\}$ of ${\rm BR}(t\to b \,\ell \nu_\ell)=0.108$
and to all quarks ${\rm BR}(t\to b\, q \bar{q}')=0.676$.  For the
electroweak (EW) parameters, which are relevant for the top-quark
decays, we employ the $G_\mu$ scheme.  Thus, the EW coupling is
evaluated as
$\alpha_{G_\mu} = \sqrt{2}/\pi G_\mu m_W^2 \sin^2 \theta_W $ and the
EW mixing angle as $\cos^2 \theta_W = m_W^2 / m^2_Z$.  The EW inputs
are set to $G_F = 1.16637 \times 10^{-5}$\,GeV$^{-2}$,
$m_W = 80.399$\,GeV, 
$m_Z = 91.1876$\,GeV. 
We consider five massless quark flavours using the corresponding NNLO
set of the NNPDF31~\cite{Ball:2017nwa} with $\as=0.118$ via the
\textsc{lhapdf} interface~\cite{Buckley:2014ana} for all our
predictions.  In \minlo{} and \minnlo{} the PDFs are read by
\textsc{lhapdf}, but their evolution is performed internally through
\textsc{hoppet}~\cite{Salam:2008qg} together with all convolutions as
described in \citere{Monni:2019whf}.

The renormalisation and factorisation scale setting within \minlo{}
and \minnlo{} has been described in detail in \sct{sec:scale}. In
particular we use \eqn{eq:scales_Q0} with $Q_0=2$\,GeV and $\KR=\KF=1$
for the central scales.  To more appropriately describe the momentum
transfer in the hard scattering, the renormalisation scale for the two
powers of the strong coupling constant entering the Born cross section
is set to $\muRzero= \KR\,H_T^{t\bar{t}}/4$, with $H_T^{t\bar{t}}$
defined in \eqn{eq:Htt}. The value of $\KR$ here corresponds to that
used for the renormalisation scale within \minlo{} and \minnlo{} in
\eqn{eq:scales_Q0} with $m_{Q\bar{Q}}=m_{t\bar{t}}$, so that the scale
within all strong couplings is varied simultaneously.  The scale of
the modified logarithm in \eqn{eq:modlogdef} that turns off the
resummation effects in phase-space regions characterised by large
transverse momentum of the top-quark pair is set to
$Q=m_{t\bar{t}}/2$.  In our phenomenological study we also compare to
fixed-order NNLO results (obtained from the implementation of \citeres{Catani:2019iny,Catani:2019hip}), where the scales are set correspondingly as
$\muR=\KR\,m_{t\bar t}$, $\muF=\KF\,m_{t\bar t}$, and
$\muRzero= \KR\,H_T^{t\bar{t}}/4$.  Scale uncertainties in all cases
are estimated by varying $\KR$ and $\KF$ by a factor of two in each
direction, while keeping the minimal and maximal values within the
constraint $1/2\le \KR/\KF\le 2$.

All showered results have been obtained with
\PYTHIA{8}~\cite{Sjostrand:2014zea} with the Monash 2013
tune~\cite{Skands:2014pea} (specifically {\tt py8tune\,14} in the
input card). For the inclusive $t\bar{t}$ results, where no top-quark
decay is included, we have kept the shower purely at parton level
turning off all additional effects. By contrast, for the fiducial
results in the phase-space of the top-quark decay products we have
included hadronisation, multi-parton interactions as well as QED
showering effects. For the final states where these effects are
particularly important, like the fully hadronic top-decay mode, we
have checked explicitly that using a different tune, in particular the
A14 ATLAS tune \cite{TheATLAScollaboration:2014rfk} (specifically {\tt
  py8tune\,21} in the input card), leads to very similar results.

We have considered various setups of fiducial cuts for the different decay modes of the top quarks.
For brevity, we simply refer to the relevant experimental studies, where the respective fiducial cuts have been specified in detail.
In particular, we show results for a fully inclusive setup in the $t\bar{t}$ phase space of the on-shell top quarks that is dubbed \setupinclusive{} 
from now on, for the fully leptonic setup of \citere{ATLAS:2019hau} dubbed \setupleptonic{}, for the semi-leptonic setup of \citere{CMS:2018htd} dubbed \setupsemi{}, and 
for the hadronic setup of \citere{ATLAS:2020ccu} dubbed \setuphadronic{}.
\sloppy

\subsection[Comparison to data extrapolated to the inclusive $t\bar{t}$ phase space]{Comparison to data extrapolated to the inclusive \boldmath{$t\bar{t}$} phase space}
\label{sec:incl}

\begin{figure}[t!]
\begin{center}
\begin{tabular}{ccc}
\hspace{-.55cm}
\includegraphics[width=.34\textwidth,page=1]{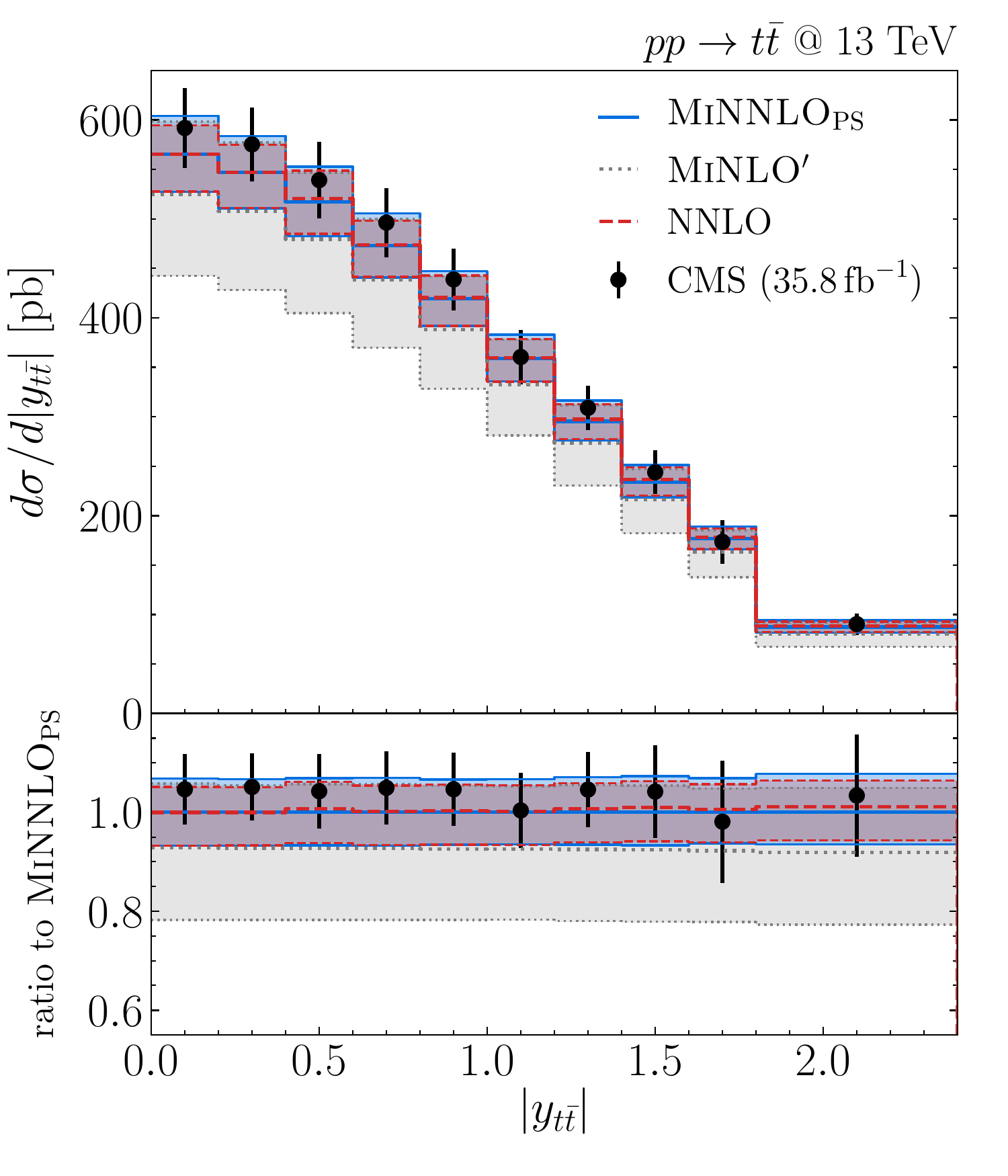}
&
\hspace{-0.6cm}
\includegraphics[width=.34\textwidth,page=2]{plots/comparison_data/all_plots_paper.pdf} 
&
\hspace{-0.6cm}
\includegraphics[width=.34\textwidth,page=3]{plots/comparison_data/all_plots_paper.pdf}
\end{tabular}
\begin{tabular}{cc}
\hspace{-.55cm}
\includegraphics[width=.34\textwidth,page=4]{plots/comparison_data/all_plots_paper.pdf}
&
\includegraphics[width=.34\textwidth,page=5]{plots/comparison_data/all_plots_paper.pdf} \\
\includegraphics[width=.34\textwidth,page=6]{plots/comparison_data/all_plots_paper.pdf}
&
\includegraphics[width=.34\textwidth,page=7]{plots/comparison_data/all_plots_paper.pdf} 
\end{tabular}
\caption{\label{fig:inclusive} Comparison of \minnlo{} (blue, solid), \minlo{} (black, dashed), and NNLO QCD (red, dashed) predictions with CMS data \cite{CMS:2018htd} (black points with errors) in \setupinclusive{}.}
\end{center}
\end{figure}

We start our presentation of phenomenological results by considering
distributions in the inclusive $t\bar{t}$ phase space.  In
\fig{fig:inclusive} we show \minnlo{} (blue, solid), \minlo{} (black,
dotted) and fixed-order NNLO (red, dashed) predictions compared to
data from CMS \cite{CMS:2018htd} (black points with errors) that has
been extrapolated from semi-leptonic top-quark decays to the inclusive
$t\bar{t}$ phase space. The first five distributions shown in that
figure, which include the rapidity ($y_{t\bar{t}}$), invariant mass
($m_{t\bar{t}}$), and transverse-momentum ($p_{T,t\bar{t}}$) of the
$t\bar{t}$ system as well as the rapidity ($y_{t_{\rm lep}}$) and
transverse momentum ($p_{T,t_{\rm lep}}$) of the leptonically decaying
top quark, have been considered already in our original publication in
\citere{Mazzitelli:2020jio}. We update that comparison here using the
newest settings, in terms of scales etc., considered in this paper. We
find excellent agreement between \minnlo{} and fixed-order NNLO
predictions for the rapidity distributions, both for the central
predictions and for the scale-uncertainty bands.  For the other
kinematical distributions some differences between \minnlo{} and NNLO
results can be observed, especially in terms of shape, which however
are largely covered by the respective scale uncertainties. With
respect to \minlo{} the \minnlo{} corrections lead to a significant
increase at the level of $+10$\% and a substantial reduction of the
scale uncertainties by more than a factor of two.  The agreement of
\minnlo{} predictions with data is quite remarkable. Only in the very
first bin of the $m_{t\bar t}$ distribution the data deviates beyond
the quoted uncertainties from the \minnlo{} result. Indeed, the
$m_{t\bar t}$ region close to the $2\,m_t$ threshold is strongly
affected by finite width effects of the top quarks, which are not
included due to the on-shell approximation of the top quarks that is
employed here.  Moreover, QED corrections through multiple-photon
radiation play an important role at $m_{t\bar t}$ values below the
$2\,m_t$ threshold, which have to be accounted for at the level of the
leptonic final states and not for the inclusive $t\bar{t}$ final
state.  This clearly shows that such extrapolations come with various
uncertainties and why comparisons in the fiducial phase space of the
top-quark decay products are advantageous.  Finally, it is interesting
to notice that for the shape of the $p_{T,t\bar{t}}$ distribution we
observe a better description of the data for \minnlo{} than at
fixed-order NNLO. This is not unexpected since the fixed-order result
suffers from large logarithmic contributions in the limit
$p_{T,t\bar{t}}\to 0$, which are resummed by the parton shower in the
\minnlo{} calculation.

The last two observables in \fig{fig:inclusive}, namely the transverse
momentum of the leading ($p_{T,t_1}$) and subleading ($p_{T,t_2}$) top
quark, have not been shown in \citere{Mazzitelli:2020jio}. These are
actually quite interesting as both are somewhat pathological at fixed
order \cite{Czakon:2019bcq,Catani:2019hip}: The first distribution is
affected by large logarithmic contributions at small $p_{T,t_1}$,
while the second one is affected by large logarithmic contributions
for $p_{T,t_2}\gtrsim m_t$, due to their sensitivity to soft-gluon
emissions.  Therefore, the matching to the parton shower becomes
particularly important.  We observe that the shapes of the $p_{T,t_1}$
distribution are only slightly different between \minnlo{} and
fixed-order NNLO predictions and both results are in agreement with
the experimental data within uncertainties. By contrast, substantial
shape effects can be observed for the $p_{T,t_2}$ distribution when
comparing the \minnlo{} and fixed-order NNLO curves. Here, the
parton-shower matched \minnlo{} prediction yields a substantially
improved description of the data. This is not unexpected for the
reasons discussed before and shows once more the relevance of matching
with the parton shower.

\subsection{Comparison to data  in the fully leptonic top-decay mode}
\label{sec:leptonic}

\begin{figure}[t!]
\begin{center}
\begin{tabular}{cc}
\hspace{-.55cm}
\includegraphics[width=.34\textwidth,page=18]{plots/comparison_data/ATLAS_leptonic_with_tau.pdf}
&
\includegraphics[width=.34\textwidth,page=17]{plots/comparison_data/ATLAS_leptonic_with_tau.pdf} 
\end{tabular}
\begin{tabular}{cccccc}
\hspace{-0.3cm}
\includegraphics[height=.305\textheight,page=1]{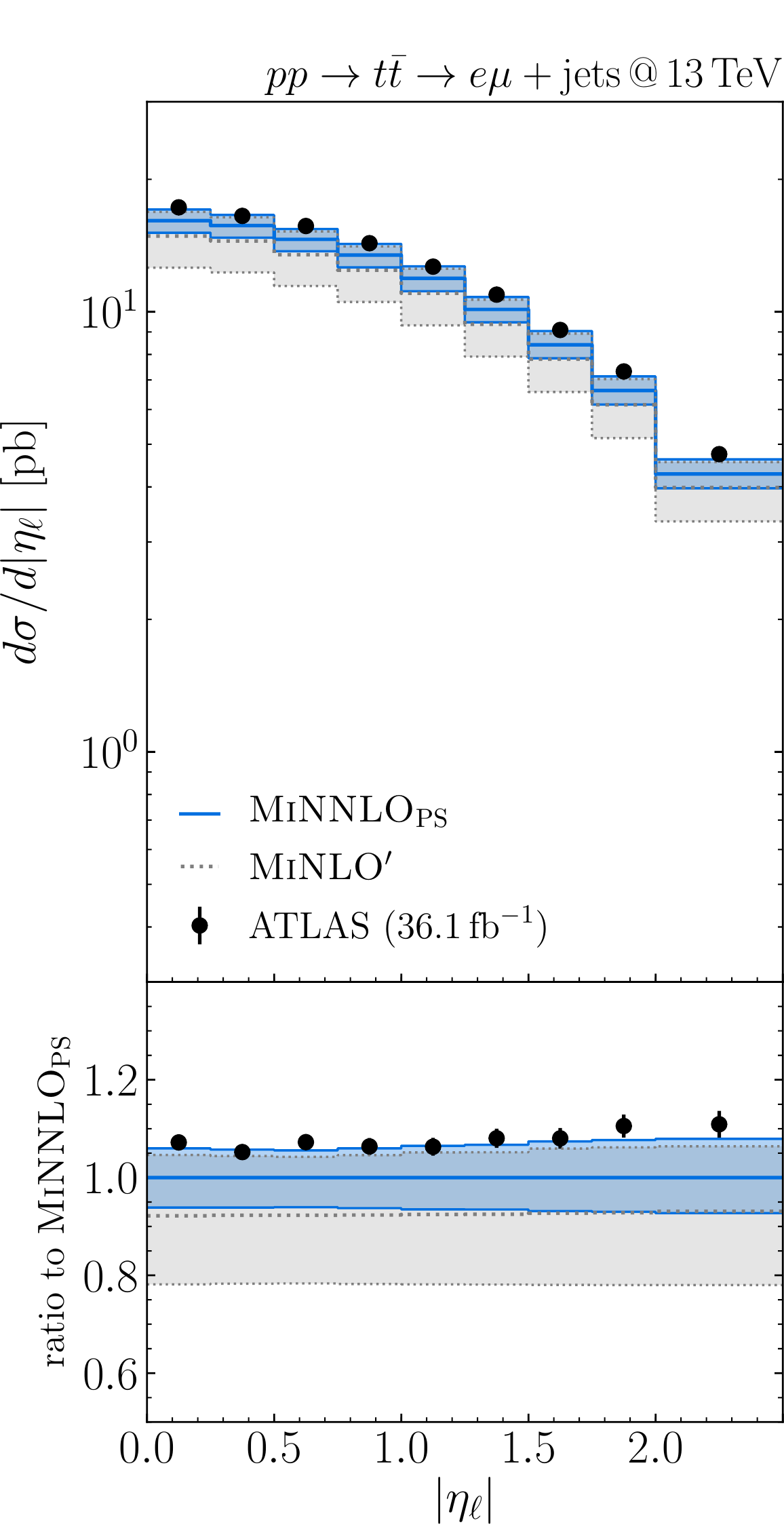}
&
\hspace{-0.56cm}
\includegraphics[height=.305\textheight,page=2]{plots/comparison_data/ATLAS_leptonic_with_tau.pdf}
&
\hspace{-0.56cm}
\includegraphics[height=.305\textheight,page=3]{plots/comparison_data/ATLAS_leptonic_with_tau.pdf} 
&
\hspace{-0.56cm}
\includegraphics[height=.305\textheight,page=4]{plots/comparison_data/ATLAS_leptonic_with_tau.pdf}
&
\hspace{-0.56cm}
\includegraphics[height=.305\textheight,page=5]{plots/comparison_data/ATLAS_leptonic_with_tau.pdf}\\
\hspace{-0.3cm}
\includegraphics[height=.305\textheight,page=11]{plots/comparison_data/ATLAS_leptonic_with_tau.pdf}
&
\hspace{-0.56cm}
\includegraphics[height=.305\textheight,page=12]{plots/comparison_data/ATLAS_leptonic_with_tau.pdf}
&
\hspace{-0.56cm}
\includegraphics[height=.305\textheight,page=13]{plots/comparison_data/ATLAS_leptonic_with_tau.pdf} 
&
\hspace{-0.56cm}
\includegraphics[height=.305\textheight,page=14]{plots/comparison_data/ATLAS_leptonic_with_tau.pdf} 
&
\hspace{-0.675cm}
\includegraphics[height=.305\textheight,page=15]{plots/comparison_data/ATLAS_leptonic_with_tau.pdf}\\
\end{tabular}\caption{\label{fig:leptonic}  Comparison of \minnlo{} (blue, solid) and \minlo{} (black, dashed) predictions with ATLAS data \cite{ATLAS:2019hau} (black points with errors) in \setupleptonic{}, including decays of $\tau$ leptons.}
\end{center}
\end{figure}

\afterpage{\clearpage}

We continue by considering \minnlo{} predictions in comparison to data for various distributions in the phase space of the top-decay products.
For simplicity, 
we only show the \minlo{} results as a reference prediction in the following.

We start with the decay channel where both top quarks decay
leptonically, requiring one electron and one muon in the final
signature.  Figure\,\ref{fig:leptonic} shows a selection of various
distributions that have been measured by ATLAS at $13$\,TeV in
\citere{ATLAS:2019hau} (black points with errors) in comparison to
\minnlo{} (blue, solid) and \minlo{} (black, dotted) predictions. In
the results considered here, the electrons and muons may also stem
from top-quark decays to $\tau$ leptons and their subsequent leptonic
decays.  Corresponding plots for all other observables considered in
the measurement of \citere{ATLAS:2019hau} with and without $\tau$
decays to electrons and muons can be found in \app{app:leptonictau}
and \app{app:leptonicnotau}, respectively.

The first two observables in Figure\,\ref{fig:leptonic} are the
transverse-momentum spectrum of the two leptons ($p_{T,\ell}$), binned
at each event for both the electron and the muon, and the invariant
mass of the electron--muon pair ($m_{e\mu}$). We find good agreement
within the respective uncertainties between \minnlo{} predictions and
data in all bins of the $p_{T,\ell}$ distribution. In particular the
$p_{T,\ell}$ shape is very well described, while there is a slight
off-set in the normalisation, with all data points being at the upper
edge of the \minnlo{} uncertainty band.  This is not unexpected, as
the \minnlo{} predictions do not include corrections beyond NNLO+PS,
such as the additional resummation of threshold logarithms
\cite{Czakon:2019txp}, that may slightly increase the
normalisation. Also for the $m_{e\mu}$ distribution we find that
\minnlo{} predictions describe the data well, with larger differences
only in the first two bins in $m_{e\mu}$.  These differences can be
traced back to two possible effects missing here. The first one is
that the top-quarks are treated in the on-shell approximation. Since
this leads to an underestimate of the cross section for
$m_{t\bar{t}}\gtrsim 2\,m_t$, as observed in the previous section,
this might induce an effect also at low $m_{e\mu}$.  Moreover, in this
region QED/EW corrections through multiple photon emissions are
relevant. While we include those through the QED shower, the
approximation of the QED shower might not yield a complete description
of such effects.

The lower two plots in Figure\,\ref{fig:leptonic} show double differential distributions. In particular the rapidity of the two leptons ($\eta_{\ell}$),
binned each event for both the electron and the muon, and the azimuthal difference between the electron and the muon ($\Delta\phi_{e\mu}$) 
are shown in slices of $m_{e\mu}$ in the following order: inclusive, $m_{e\mu}<80$\,GeV, $80$\,GeV$<m_{e\mu}<120$\,GeV, 
$120$\,GeV$<m_{e\mu}<200$\,GeV, and $200$\,GeV$<m_{e\mu}<500$\,GeV. For the $\eta_{\ell}$ distribution we observe a rather 
similar agreement between \minnlo{} predictions and data as for the $p_{T,\ell}$ distribution: The shapes are described very well and 
the data is at the upper edge of the uncertainty band, but overall still compatible. The same is found inclusively and in essentially all 
bins of $m_{e\mu}$. For $m_{e\mu}<80$\,GeV the differences become slightly larger at forward rapidities, which could have 
the same origin as the difference observed at low $m_{e\mu}$.
What is interesting to observe from the main frame is the fact that for the bins with larger $m_{e\mu}$ the leptons become successively more 
forward. In fact in the highest $m_{e\mu}$ bin there is even a small dip at central rapidity.

The $\Delta\phi_{e\mu}$ distribution is quite an important observable, as it is directly sensitive to spin correlations in the top-quark
decays. It therefore allows us to test our implementation in comparison to data. By and large, we find good agreement in
$\Delta\phi_{e\mu}$ between \minnlo{} predictions and data in all slices of $m_{e\mu}$. Only in the region where 
the electron and the muon are close in the azimuthal angle slight differences in shape appear.
Nevertheless, given the fact that spin correlations in the top-quark decays have been included only through an approximation at tree-level,
the observed agreement is quite remarkable. 
Another interesting feature can be observed in the main frame of the $\Delta\phi_{e\mu}$ distributions. For $m_{e\mu}<80$\GeV the shape of the 
distribution is reversed with respect to the other slices in $m_{e\mu}$.

For all observables in Figure\,\ref{fig:leptonic} we find that the corrections induced by \minnlo{}  with respect to \minlo{} 
are at the order of $10$\% and generally quite flat in phase space. The only case where some
differences in shape between \minnlo{} and \minlo{} can be appreciated are the $\Delta\phi_{e\mu}$
distributions at high $m_{e\mu}$. As far as scale uncertainties are concerned, the \minnlo{} predictions 
feature significantly smaller uncertainty bands than those of \minlo{}, with a reduction of roughly a factor of two.
We remark again that we have considered a large number of observables, comparing \minnlo{}, \minlo{} and ATLAS data
in \setupleptonic{}, and similar conclusions as those pointed out for the distributions shown in this section are found for all of them,
as the reader can appreciate from the additional figures in \app{app:leptonictau} and \app{app:leptonicnotau}.

\subsection{Comparison to data in the semi-leptonic top-decay mode}
\label{sec:semi}
\begin{figure}[t!]
\begin{center}
\begin{tabular}{ccc}
\hspace{-.55cm}
\includegraphics[width=.34\textwidth,page=7]{plots/comparison_data/CMS_semileptonic.pdf}
&
\hspace{-0.6cm}
\includegraphics[width=.34\textwidth,page=8]{plots/comparison_data/CMS_semileptonic.pdf} 
&
\hspace{-0.6cm}
\includegraphics[width=.34\textwidth,page=15]{plots/comparison_data/CMS_semileptonic.pdf}\\
\hspace{-.55cm}
\includegraphics[width=.34\textwidth,page=3]{plots/comparison_data/CMS_semileptonic.pdf}
&
\hspace{-0.6cm}
\includegraphics[width=.34\textwidth,page=1]{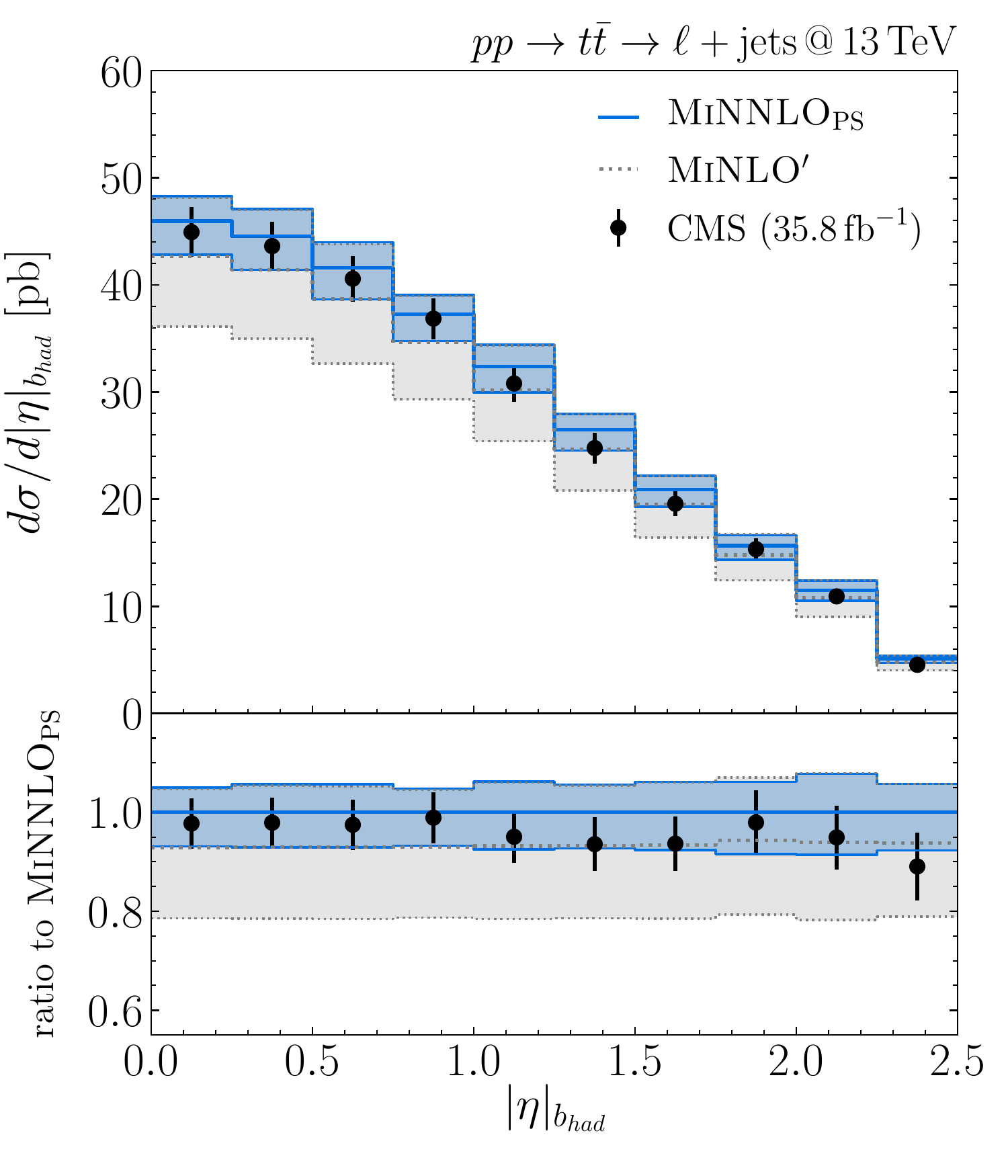} 
&
\hspace{-0.6cm}
\includegraphics[width=.34\textwidth,page=2]{plots/comparison_data/CMS_semileptonic.pdf}\\
\hspace{-.55cm}
\includegraphics[width=.34\textwidth,page=11]{plots/comparison_data/CMS_semileptonic.pdf}
&
\hspace{-0.6cm}
\includegraphics[width=.34\textwidth,page=9]{plots/comparison_data/CMS_semileptonic.pdf} 
&
\hspace{-0.6cm}
\includegraphics[width=.34\textwidth,page=10]{plots/comparison_data/CMS_semileptonic.pdf}
\end{tabular}\caption{\label{fig:semi}  Comparison of \minnlo{} (blue, solid) and \minlo{} (black, dashed) predictions with CMS data \cite{CMS:2018htd} (black points with errors) in \setupsemi{}.}
\end{center}
\end{figure}

We now turn to results for semi-leptonic top-quark decays, where one of the $W$ bosons stemming from the top quark decays to leptons and the other to hadrons.
Figure\,\ref{fig:semi} shows a selection of various distributions measured by CMS at $13$\,TeV in \citere{CMS:2018htd} (black points with errors) 
compared to \minnlo{} (blue, solid) and \minlo{} (black, dotted) predictions. Other distributions measured in this analysis can be found 
in \app{app:semi}.

We start with the rapidity ($y_{t\bar t}$), invariant mass  ($m_{t\bar t}$), and transverse momentum  ($p_{T,t\bar t}$) distributions
of the two reconstructed top quarks. 
As for the corresponding distributions extrapolated to the inclusive $t\bar t$ phase space studied in \fig{fig:inclusive} of \sct{sec:incl},
the agreement with data is very good for the \minnlo{} predictions, with none of the data points 
deviating beyond the quoted uncertainties. It is interesting to notice that, contrary to \fig{fig:inclusive}, also 
the first bin in the $m_{t\bar t}$ distribution is well described by \minnlo{}. This indicates that QED effects on the leptonic final states
included through the QED shower here, which could not be applied in \sct{sec:incl}, 
may indeed be important in that region. Moreover, this shows the importance of performing such comparison directly in the fiducial phase space
of the top-quark decay products.
Note, however, that there is still a slight difference 
in terms of shape between the \minnlo{} curve and the data points at low $m_{t\bar t}$ that is likely due to the missing finite width effects 
in the applied on-shell approximation of the top quarks.

The other plots shown in \fig{fig:semi} are the rapidity and transverse-momentum distributions of the leading jet coming from the hadronic $W$ decay 
($\eta_{j_{W1}}$ and $p_{T,j_{W1}}$), of the bottom-flavoured jet coming from the hadronically decaying
top quark ($\eta_{b_{\rm had}}$ and $p_{T,{b_{\rm had}}}$) and of the bottom-flavoured jet coming from the leptonically decaying
top quark ($\eta_{b_{\rm lep}}$ and $p_{T,{b_{\rm lep}}}$). \minnlo{} predictions and data agree perfectly for all rapidity 
distributions, both in terms of shape and in terms of normalisation.
As far as the transverse-momentum spectra are concerned, the same level of agreement as for the observables just discussed is found also for $p_{T,j_{W1}}$.
Also for the $p_{T,{b_{\rm had}}}$ distribution \minnlo{} provides a reasonable description of data with slight shape differences 
towards large $p_{T,{b_{\rm had}}}$. Among all the distributions we have considered throughout this paper, it is only $p_{T,{b_{\rm lep}}}$
that shows a significant deviation of the data with respect to our  \minnlo{} prediction: There is a clear difference in the shape, such that
towards large $p_{T,{b_{\rm lep}}}$ the discrepancy between \minnlo{} and data increases. Given that the \minnlo{} prediction describes 
the measured $p_{T,{b_{\rm had}}}$ spectrum reasonably well it is surprising that such discrepancy is found for the 
$p_{T,{b_{\rm lep}}}$ spectrum, since one would expect the latter actually to be cleaner both in terms of its theoretical 
modeling and in terms of its measurement. At the moment, we cannot provide any deeper explanation for the 
observed behaviour and it will be interesting to see whether this trend persists in future measurements in the
semi-leptonic top-decay mode.

\subsection{Comparison to data in the fully hadronic top-decay mode}
\label{sec:hadronic}

\begin{figure}[t!]
\begin{center}
\begin{tabular}{ccc}
\hspace{-.55cm}
\includegraphics[width=.34\textwidth,page=30]{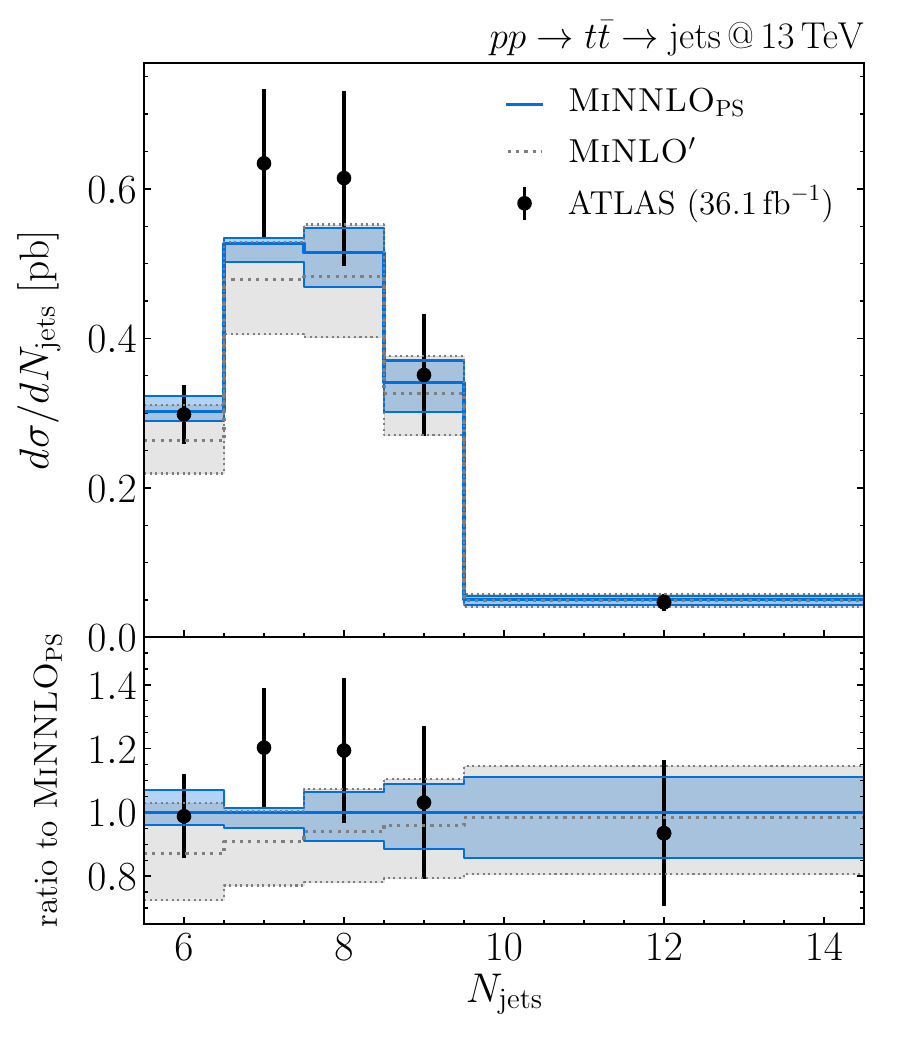}
&
\hspace{-0.6cm}
\includegraphics[width=.34\textwidth,page=9]{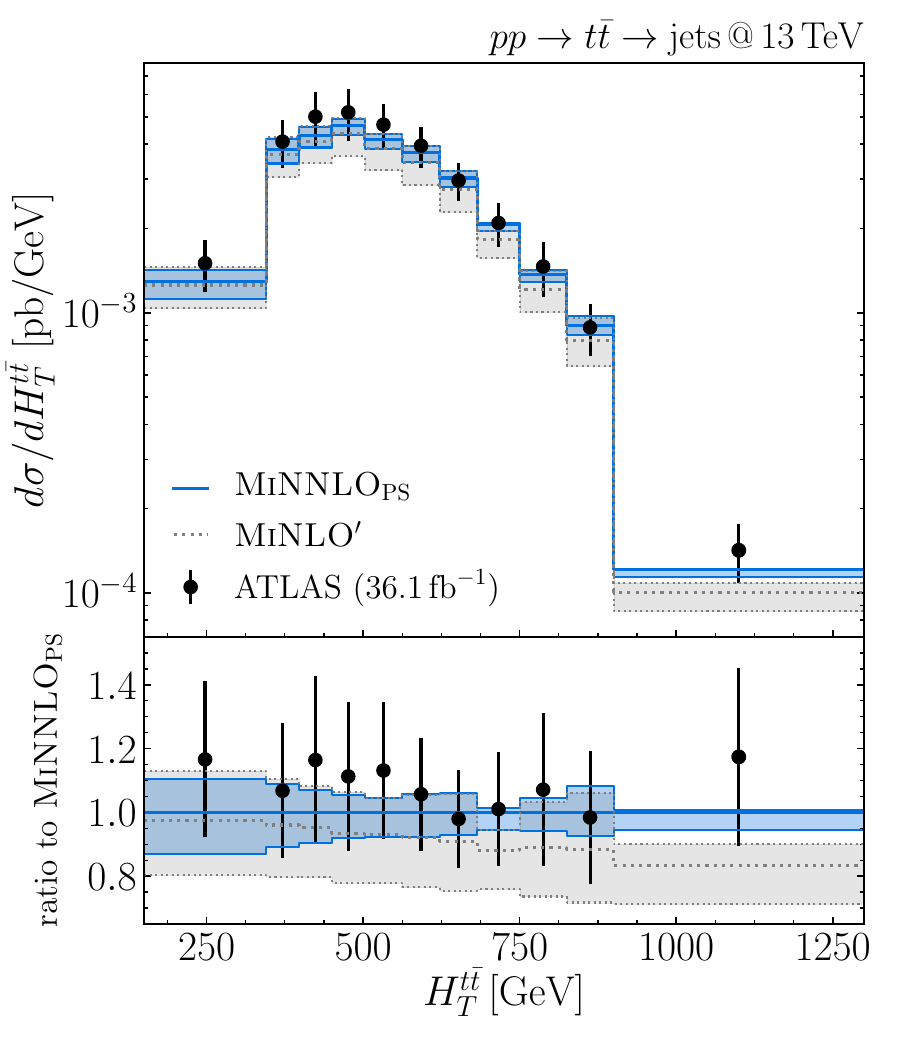} 
&
\hspace{-0.6cm}
\includegraphics[width=.34\textwidth,page=12]{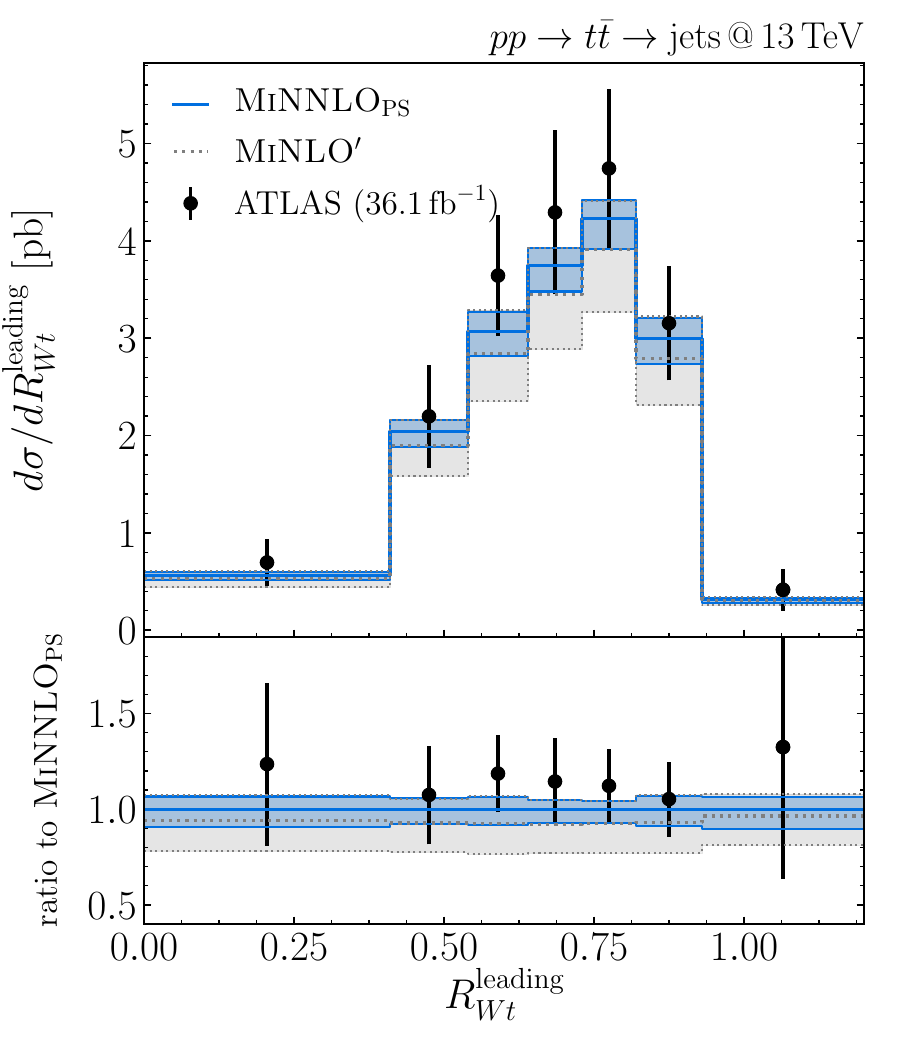}\\
\hspace{-.55cm}
\includegraphics[width=.34\textwidth,page=31]{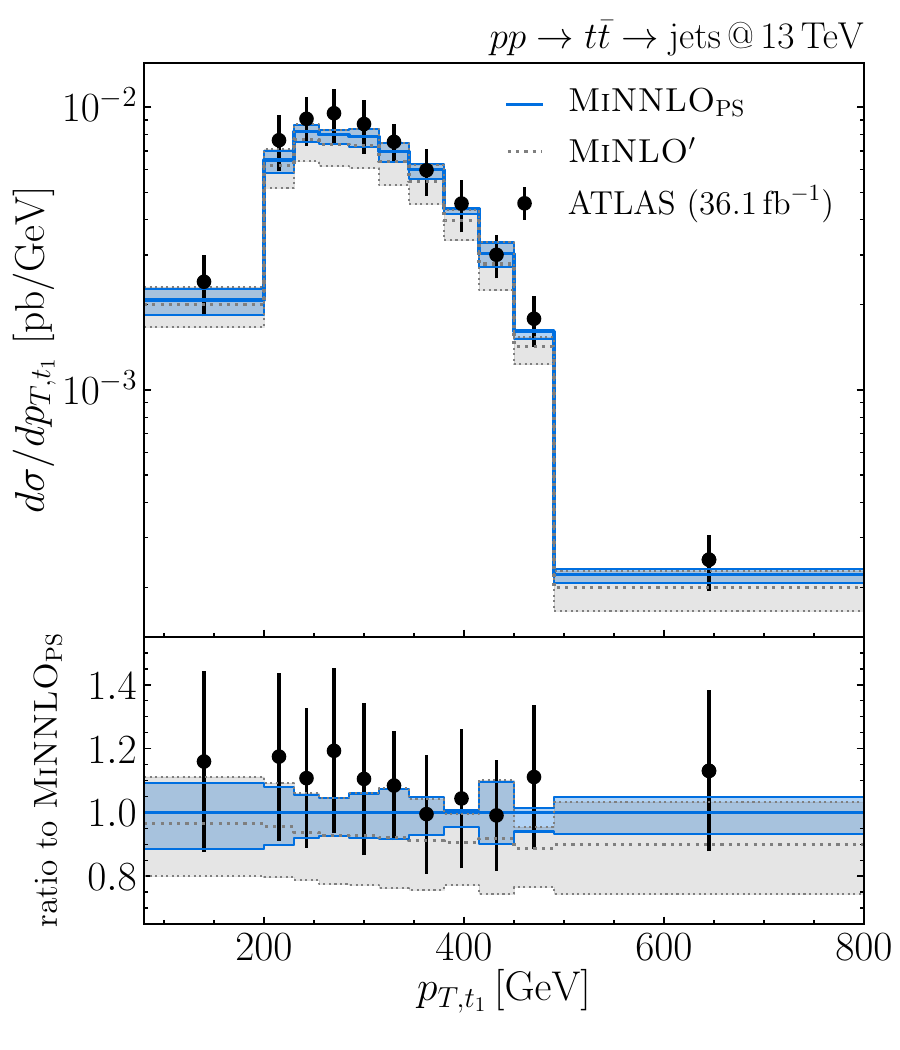}
&
\hspace{-0.6cm}
\includegraphics[width=.34\textwidth,page=32]{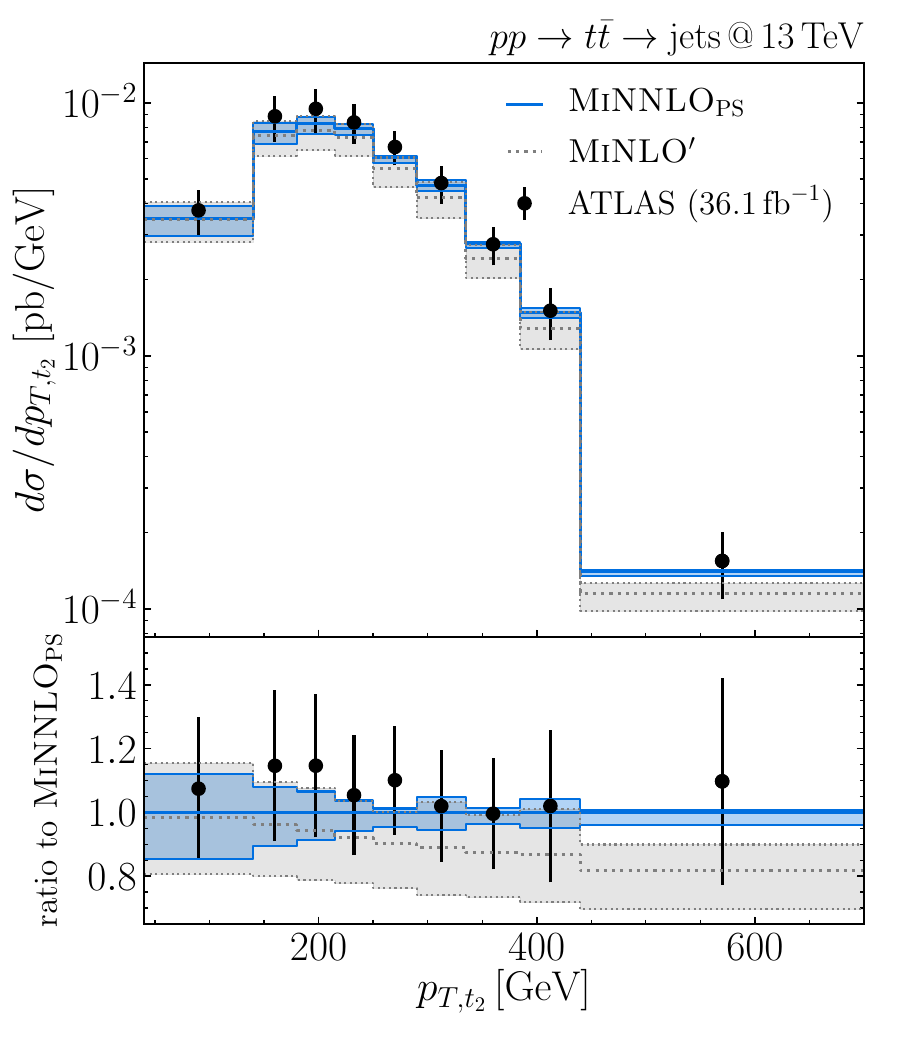} 
&
\hspace{-0.6cm}
\includegraphics[width=.34\textwidth,page=23]{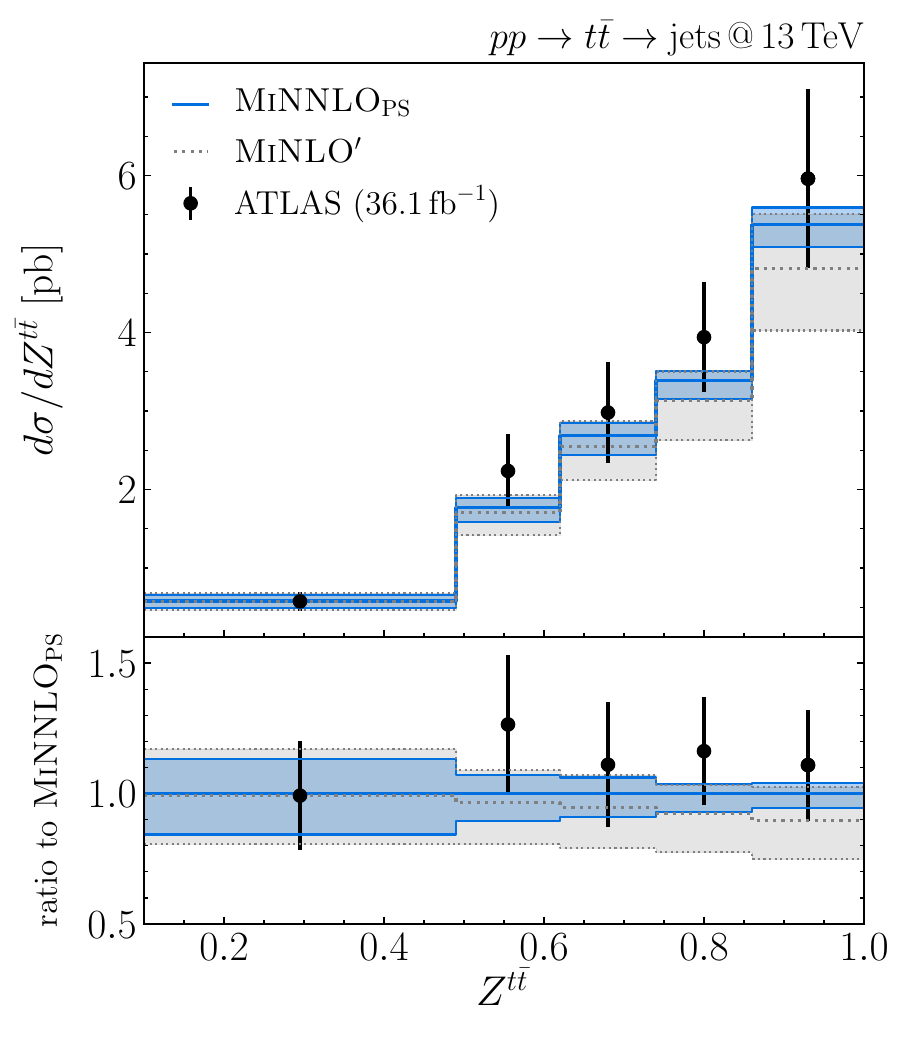}
\end{tabular}\caption{\label{fig:hadronic}  Comparison of \minnlo{} (blue, solid) and \minlo{} (black, dashed) predictions with ATLAS data \cite{ATLAS:2020ccu} (black points with errors) in \setuphadronic{}.}
\end{center}
\end{figure}

Finally, we have also considered the fully hadronic decay mode of the top quarks. In this case the  
measurements are plagued by relatively large experimental uncertainties due to the substantial QCD backgrounds.
In \fig{fig:semi} we present a selection of various distributions measured by ATLAS at $13$\,TeV in \citere{ATLAS:2020ccu} (black points with errors) 
compared to \minnlo{} (blue, solid) and \minlo{} (black, dotted) predictions. Other distributions measured in this analysis can be found 
in \app{app:hadronic}.

The first observable shown in \fig{fig:semi} is the cross section as a
function of the number of jets $N_{\rm jets}$. Since we consider
hadronically decaying top quarks only there are already six partons in
the tree-level process, including the two bottom quarks.  Therefore
\minnlo{} predictions are NNLO+PS accurate for $N_{\rm jets}=6$,
NLO+PS accurate for $N_{\rm jets}=7$, LO+PS accurate for
$N_{\rm jets}=8$, and all multiplicities $N_{\rm jets}\ge 9$ are
described purely by the parton shower. Indeed, the \minnlo{} scale
uncertainties increase towards larger jet multiplicities. Similarly,
the corrections induced by \minnlo{} with respect to \minlo{} and the
ensuing reduction of the scale uncertainties become successively more
significant for smaller jet multiplicities. Taking into account the
decreased accuracy at higher multiplicities, \minnlo{} is doing a good
job in describing the data for the different jet multiplicities. In
particular the first NNLO-accurate $N_{\rm jets}=6$ bin is in
excellent agreement.

The second plot in \fig{fig:semi} shows the distribution in $H_T^{t\bar t}$, as defined in \eqn{eq:Htt}. Since this observables 
determines our preferred scale choice for $\muRzero$, the scale of the two powers of the strong coupling constant entering the Born cross section,
it is reassuring to see that the agreement with data for this observable is excellent, both in terms of shape and in terms of normalisation.
Moreover, as one can see, the \minlo{} calculation predicts a quite different shape compared to \minnlo{}, leading to a
worse agreement in the tail of the $H_T^{t\bar t}$ distribution.

The third observable shown in \fig{fig:semi} is the cross section as a
function of the ratio ($R^{\rm leading}_{W\,t}$) of the transverse
momentum of the $W$ boson originating from the leading top quark with
respect to the transverse momentum of the leading top quark.  As can
be seen, the cross section peaks around the value where $W$ boson
takes 75\% of the top-quark transverse momentum, while it is rather
unlikely that the $W$ boson has a transverse momentum which is larger
than or less than half of that of its parent (leading) top quark. Also
here \minnlo{} predictions are in good agreement with the data. It is
reassuring to see especially the shape to be extremely well described,
given that the definition of this ratio originates entirely from the
hadronic top-quark decays that are included only at tree level in our
calculation. In fact, \citere{ATLAS:2020ccu} has defined and measured
various such ratios and we have found our \minnlo{} predictions to be
in excellent agreement with the data in all cases, as can be seen from
the additional observables provided in \app{app:hadronic}.

The last three observables in \fig{fig:semi} concern the transverse momenta of the top quarks. More precisely the one of the 
leading ($p_{T,t_1}$) and the subleading ($p_{T,t_2}$) top quark is shown as well as the ratio of the latter to the former  ($Z^{t\bar t}=p_{T,t_2}/p_{T,t_1}$).
As discussed before, these observables cannot be described appropriately by fixed-order calculations as they are subject to 
large logarithmic corrections due to soft-gluon effects. By contrast, our \minnlo{} calculation is perfectly able to describe the 
shapes and normalisation of all three measured distributions in all bins within the quoted uncertainties. 
In particular for $p_{T,t_2}$ we find that the agreement 
is better than compared to the \minlo{} result, which features quite a different shape than \minnlo{} towards large $p_{T,t_2}$.
The distribution in $Z^{t\bar t}$ peaks around one, which is not unexpected since the top quarks are back-to-back at LO. 
The distribution for $Z^{t\bar t}<1$ is filled only upon inclusion of addition real QCD emissions starting from NLO, which is
also indicated by the enlarged uncertainties towards small $Z^{t\bar t}$.

\section{Summary}
\label{sec:summary}


The experimental precision of LHC measurements
is challenging the accuracy of theoretical predictions, which need to keep 
up with the steadily decreasing uncertainties of the measured observables.
In this context, the accuracy of full-fledged Monte Carlo event 
generation tools is particularly important, and it becomes indispensable to advance the description 
of LHC interactions through the inclusion of perturbative information at 
highest-possible order.

In this paper, we have presented a novel formulation of the \minnlo{} method 
to achieve the matching of NNLO QCD corrections to parton showers
for processes with a pair of heavy quarks in the final state.
The goal was to keep the various positive features of the \minnlo{} approach, while
making it applicable to a new class of collider processes. In particular, 
\minnlo{} preserves the (LL) accuracy of any parton shower 
order in the transverse momentum, it does not introduce any new scale or 
slicing parameter to separate between different jet multiplicities, 
and it directly includes the NNLO corrections on-the-fly in the event generation.
The latter two render the implementation particularly efficient as no large 
(non-local) cancellations appear and no post-processing of the events is required.

The new formulation of \minnlo{} for a heavy quark pair is based on the observation that the transverse-momentum
resummation formula of the pair, despite being considerably more 
complicated, can be rewritten in a form very similar to that for a colour singlet
to the accuracy we are interested in. After achieving this, the NNLO 
matching to the parton shower can be obtained with steps analogous to
those of the \minnlo{} procedure for colour singlets. A detailed derivation
of the final matching formula has been provided and all technical and computational
aspects have been discussed, including the renormalisation, factorisation and resummation
scale dependence, introduction of a new modified logarithm and its profiling,
as well as the respective scale settings.

We have applied the formalism to the production of a top-quark pair,
one of the most relevant and best studied processes at LHC. The decay
of the (on-shell) top-quarks has been consistently included via the
tree-level amplitudes, fully accounting for spin correlations of the
top quarks and $W$ bosons in the decay chain.  A comprehensive
phenomenological study of our \minnlo{} predictions has been performed
in comparison to a fixed-order NNLO calculation for on-shell top-quark
pair production without decay, to our less accurate \minlo{} results,
and to experimental data in all relevant top-decay modes.  For the
case of on-shell top-quarks without their decays we found excellent
agreement of \minnlo{} for fully inclusive observables with the
fixed-order NNLO predictions, while observables sensitive to
soft-gluon effects demonstrated the advantages of the \minnlo{}
matching to the parton shower, with (the expected) improved agreement
with data (extrapolated to the inclusive $t\bar{t}$ phase space)
compared to fixed order. When comparing to less accurate \minlo{}
results we find that the \minnlo{} procedure induces corrections of
typically ${\cal O}(+10\%)$ that are generally, but for a few
exceptions, quite flat in phase space, as well as substantially
smaller scale uncertainties, typically by a factor of two or more.  We
note that also the \minlo{} prediction stems from a genuinely new
calculation emerging from the developed formalism.  Compared to
standard NLO+PS calculations the improvements can be expected to be
even more prominent.

As far as the top-quark decays are concerned, we have compared two different approaches to include the 
tree-level decays of the top quarks, both through the native implementation in the $t\bar{t}$+jet calculation 
and through \MadSpin{}. We have shown the ensuing results to be in good  agreement, fully covered by the 
residual scale uncertainties.
We have then considered a significant number of observables in 
the leptonic, semi-leptonic and hadronic top-decay modes using the native implementation and compared
\minnlo{} predictions with LHC data. Additional results were reported in the appendices.
In conclusion, \minnlo{} describes data in all three top-decay modes remarkably well, also for observables 
sensitive to the top-quark decays and the respective spin correlations. Among all results we have found tension
with data only in a single observable, namely for the transverse-momentum spectrum of the bottom-flavoured jet
from the leptonically decaying top quark in the semi-leptonic top decay mode, whose source we can not 
identify at the moment.

We reckon that this calculation will be very useful especially for the experimental analyses based on top-quark
final states at the LHC.
Our \minnlo{} $t\bar{t}$ generator is implemented in the \POWHEGBOXVTWO{} framework. It is publicly available 
and can be downloaded within \POWHEGBOXVTWO{} as any standard (NLO+PS) process as instructed on the 
corresponding webpage \url{http://powhegbox.mib.infn.it/}.

\section*{Acknowledgments}
We are grateful to Simone Amoroso for several fruitful discussions and
for help with setting up the top-quark decays through \MadSpin{}. We
also would like to thank Rikkert Frederix for help with requiring
on-shell top quarks within \MadSpin{}, and Massimiliano Grazzini for
comments on the manuscript. This research was supported in part by the
National Science Foundation under Grant No. NSF PHY-1748958 in the
context of the KITP programme ``New Physics from Precision at High
Energies''. P.N. acknowledges support from Fondazione
Cariplo and Regione Lombardia, grant 2017-2070, and from INFN.

\appendix

\section{Resummation ingredients}
\label{app:resum}
We provide here the necessary ingredients for implementing the \minnlo
method for top-quark pair production following
\eqn{eq:mastertt}, that is, the explicit results for
$\gamma_{i}$ and ${\cal C}^{[\gamma_i]}_\ccbar$.

Instead of considering $\gamma_i$ with $i=1,\dots,n_c$ we will use the
alternative notation $\gamma_{(k,l)}$ with
$k,l=1,\dots,\sqrt{n_c}$. We will accordingly consider
${\cal C}^{[\gamma_{(k,l)}]}_\ccbar$. Using this notation, we can
write the following compact expressions:
\begin{eqnarray}
\gamma_{(k,l)} \equiv \gamma_{(k,l)}^\ccbar&=& 2[\lambda_k^\ccbar + (\lambda_l^\ccbar)^*]\,, \label{eq:gamma_i}\\
{\cal C}^{[\gamma_{(k,l)}]}_\ccbar &=&
\bar{\mathbf{C}}_{(l,k)}^\ccbar \bar{\mathbf{H}}_{(k,l)}^\ccbar\,. \label{eq:C_i}
\end{eqnarray}
In the above equations, we denoted by $\lambda^{\ccbar}_k$ the $k$-th
eigenvalue of the one-loop soft anomalous dimension, working in a
colour basis given by the matrix ${\mathbf{C}}^\ccbar$ and where the
matrix $\mathbf{H}^\ccbar$ is given by the colour-decomposed LO
hard-scattering amplitude. The matrices $\bar{\mathbf{C}}^\ccbar$ and
$\bar{\mathbf{H}}^\ccbar$ are obtained from ${\mathbf{C}}^\ccbar$ and
${\mathbf{H}}^\ccbar$ by performing a rotation to the basis in which
$\mathbf{\Gamma}_t^{(1)}$ is diagonal.

Explicit results for the colour matrices and eigenvalues needed to
compute \neqn{eq:gamma_i} and \eqref{eq:C_i} can be found in
\citere{hayk}. For the sake of completeness, they are given below.
The colour matrices ${\mathbf{C}}^\ccbar$ for the quark- and gluon-initiated
channels can be chosen in the following way,
\begin{equation}
{\mathbf{C}}^{q\bar{q}} = \frac{C_A}{2}
\begin{pmatrix}
2 C_A & 0 \\ 0 & C_F
\end{pmatrix}
\,,\;\;\;
{\mathbf{C}}^{gg} = C_F
\begin{pmatrix}
2 C_A^2 & 0 & 0 \\ 0 & C_A^2 & 0 \\ 0 & 0 & C_A^2 - 4
\end{pmatrix}\,.
\end{equation}
The two eigenvalues of $\mathbf{\Gamma}_t^{(1)}$ in the
quark-initiated channel are given by
\begin{eqnarray}
\lambda_1^{q\bar{q}} &=& \frac{1}{2} \left(
C_A (\Gamma_s + \Gamma_t) +
\sqrt{C_A^2(\Gamma_t-\Gamma_s)^2+4\Gamma_s\Gamma_t}
\right) + \Gamma_{\rm I}^{q\bar{q}}\,,\\
\lambda_2^{q\bar{q}} &=& \frac{1}{2} \left(
C_A (\Gamma_s + \Gamma_t) -
\sqrt{C_A^2(\Gamma_t-\Gamma_s)^2+4\Gamma_s\Gamma_t}
\right) + \Gamma_{\rm I}^{q\bar{q}}\,,
\end{eqnarray}
while the three eigenvalues corresponding to the
gluon-initiated channel can be written in the following way:
\begin{eqnarray}
\lambda_1^{gg} &=& \frac{1}{6} \left[
X^{1/3} + Y + 2(A_{11}+2A_{22})
\right] + \Gamma_{\rm I}^{gg}\,,\\
\lambda_2^{gg} &=& \frac{1}{12} \left[
-X^{1/3}-Y+\sqrt{3} i (X^{1/3}-Y)+4(A_{11}+2A_{22})
\right] + \Gamma_{\rm I}^{gg}\,,\\
\lambda_3^{gg} &=& \frac{1}{12} \left[
-X^{1/3}-Y-\sqrt{3} i (X^{1/3}-Y)+4(A_{11}+2A_{22})
                   \right] + \Gamma_{\rm I}^{gg}\,,
\end{eqnarray}
where we have defined the following quantities
\begin{eqnarray}
\Gamma_t &=& -\frac{1}{4} \ln \left(\frac{p_1\cdot p_3\, p_2\cdot p_4}{p_1\cdot p_4\, p_2\cdot p_3}\right)\,,\\
\Gamma_s &=& -\frac{1}{4}\left[
\frac{1}{2v} \ln\left(
\frac{1+v}{1-v}
\right) -i\pi\left(\frac{1}{v}+1\right)
-\ln\left(
\frac{4 p_1\cdot p_4 \, p_2\cdot p_3}{s\, m_t^2}
\right)
\right] \,,\\
\Gamma_{\rm I}^{q\bar{q}} &=& -\frac{C_F}{2}\left(
1-\frac{1}{2v}\left[
\ln\left(\frac{1+v}{1-v}\right)-2i\pi
\right] 
- \ln \left(\frac{p_1\cdot p_3\, p_2\cdot p_4}{p_1\cdot p_4\, p_2\cdot p_3}\right)
\right)\,,\\
\Gamma_{\rm I}^{gg} &=& -\frac{C_F}{2}\left(
1-\frac{1}{2v}\left[
\ln\left(\frac{1+v}{1-v}\right)-2i\pi
\right] \right)
+\frac{C_A+C_F}{4} \ln \left(\frac{p_1\cdot p_3\, p_2\cdot p_4}{p_1\cdot p_4\, p_2\cdot p_3}\right)\,,
\end{eqnarray}
and the following variables specific to the gluon channel,
\begin{eqnarray}X &=& -18A_{12}^2(C_A^2-8)(A_{11}-A_{22})+8(A_{11}-A_{22})^3\,,\\
&+& \Big\{
4(A_{11}-A_{22})^2 \left[
4(A_{11}-A_{22})^2-9A_{12}^2(C_A^2-8)
\right]^2 \nonumber\,,\\
&-&\left[
4(A_{11}-A_{22})^2 + 3A_{12}^2(C_A^2+4)
\right]^3
\Big\}^{1/2}\nonumber\,,\\
Y &=& \left[
4(A_{11}-A_{22})^2 + 3A_{12}^2(C_A^2+4)
\right] X^{-1/3}\,,\\
A_{11} &=& \Gamma_t (C_A+C_F)\,,\\
A_{22} &=& \Gamma_t (C_F+C_A/2) + \Gamma_s C_A\,,\\
A_{12} &=& -\Gamma_t\,.
\end{eqnarray}
In the present colour basis, the LO matrix $\mathbf{H}^\ccbar$ takes the
following form
\begin{equation}
{\mathbf{H}}^{q\bar{q}} = \frac{2}{C_A C_F}
\begin{pmatrix}
0 & 0 \\ 0 & 1
\end{pmatrix}
\,,\;\;\;
{\mathbf{H}}^{gg} = \frac{1}{C_F [C_A^2 (1+r^2) -2]}
\begin{pmatrix}
1/C_A^2 & r/C_A & 1/C_A \\ r/C_A & r^2 & r \\ 1/C_A & r & 1
\end{pmatrix}\,,
\end{equation}
with $r=(t-u)/s$.
Finally, $\bar{\mathbf{C}}^\ccbar$ and $\bar{\mathbf{H}}^\ccbar$ are obtained by applying
the following change of basis,
\begin{equation}
\bar{\mathbf{C}}^\ccbar = ({\mathbf{R}^\ccbar})^\dagger {\mathbf{C}^\ccbar}\, {\mathbf{R}^\ccbar}
\,,\;\;\;
\bar{\mathbf{H}}^\ccbar = ({\mathbf{R}^\ccbar})^{-1} {\mathbf{H}^\ccbar}\, (({\mathbf{R}^\ccbar})^{-1})^\dagger
\end{equation}
where the rotation matrices to go into the basis in which $\mathbf{\Gamma}_t^{(1)}$
is diagonal are given by
\begin{equation}
{\mathbf{R}}^{q\bar{q}} =
\begin{pmatrix}
\frac{1}{2\Gamma_t}\left[
\lambda_2^{q\bar{q}} - \Gamma_{\rm I}^{q\bar{q}} - 2C_F\Gamma_t
\right] &
\frac{1}{2\Gamma_t}\left[
\lambda_1^{q\bar{q}} - \Gamma_{\rm I}^{q\bar{q}} - 2C_F\Gamma_t
\right] \\
1&1
\end{pmatrix}\,,
\end{equation}
for the quark-antiquark channel, and
\begin{equation}
{\mathbf{R}}^{gg} =
\begin{pmatrix}
\frac{A_{12}}{\lambda_1^{gg}-\Gamma_{\rm I}^{gg}-A_{11}} &
\frac{A_{12}}{\lambda_2^{gg}-\Gamma_{\rm I}^{gg}-A_{11}} &
\frac{A_{12}}{\lambda_3^{gg}-\Gamma_{\rm I}^{gg}-A_{11}} \\
1 & 1 & 1 \\
\frac{C_A A_{12}/2}{\lambda_1^{gg}-\Gamma_{\rm I}^{gg}-A_{22}} &
\frac{C_A A_{12}/2}{\lambda_2^{gg}-\Gamma_{\rm I}^{gg}-A_{22}} &
\frac{C_A A_{12}/2}{\lambda_3^{gg}-\Gamma_{\rm I}^{gg}-A_{22}}
\end{pmatrix}\,,
\end{equation}
for the gluon-initiated channel.

\section{Generation of virtualities in top-quark decays} 
\label{sec:tvirt}
We now discuss how we generate off-shell effects
  (when the option {\tt zerowidth 0} is used)
in $t_{} \bar{t}$ decays. For clarity, in the following we indicate with $Q$ and $\bar{Q}$ the $t$ and
$\bar{t}$ quarks. The way the
virtualities are chosen will be described later.
We start by describing the mapping from the kinematics configurations with on-shell $t\bar{t}$ quarks
in the final state to the configuration where the $t\bar{t}$ have virtualities $M_Q$ and $M_{\bar Q}$.
Quantities before/after the mapping will be written in lower/upper case.
We will consider two reference frames: the partonic centre-of-mass (CM) frame (i.e.\ the rest frame of the
incoming partons), denoted with $S_{\tmop{CM}}$, and the centre-of-mass frame of the $Q\bar{Q}$ system
denoted by $S_{Q\bar{Q}}$. Frame dependent quantities in the  $S_{Q\bar{Q}}$ frame will be accompanied by
a prime, to avoid confusion. We proceed as follows:
\begin{itemizedot}
  \item We start from the $S_{\tmop{CM}}$ frame. We boost the heavy
  quark and antiquark system to their rest frame $S_{Q \bar{Q}}$, with the boost
  velocity
\begin{equation}
  \vec{\beta} = - \frac{\vec{p}_Q + \vec{p}_{\bar{Q}}}{p^0_Q +
    p^0_{\bar{Q}}}\,,
  \end{equation}
  where $p_{Q / \bar{Q}}$ denote the heavy-quark four-momenta, and
  $\vec{p}_{Q / \bar{Q}}$ and $p^0_{Q/\bar{Q}}$ are their vector components
  and energies in the $S_{\tmop{CM}}$ frame.
  
  \item In the $S_{Q \bar{Q}}$ frame, we change the energy of the quark and
  antiquark maintaining their three-momenta fixed in such a way that, denoting
  with $P_{Q / \bar{Q}}$ the four-momenta after the mapping, we have
\begin{equation}
  P_Q^2 = M_Q^2,\qquad  P_{\bar{Q}}^2 = M_{\bar{Q}}^2,\qquad \vec{P}^{\,\prime}_{Q / \bar{Q}} =
  \vec{p}^{\,\prime}_{Q / \bar{Q}}\,,
  \end{equation}
  where $M_{Q/\bar{Q}}$ are the generated virtualities of the heavy quarks, which may be 
  different from $m_{Q/\bar Q}$, and
  $\vec{P}_{Q / \bar{Q}}^{\,\prime} $ and $\vec{p}_{Q / \bar{Q}}^{\,\prime}$ are the
  three-momenta of the corresponding four-vectors evaluated in the $S_{Q
    \bar{Q}}$ frame.

  \item The momenta $P_Q$ and $P_{\bar{Q}}$ are boosted back to the
  $S_{\tmop{CM}}$ frame with boost velocity
    \begin{equation}
      \vec{\beta}^{\,\prime} = \frac{\vec{p}_Q + \vec{p}_{\bar{Q}}}{\sqrt{(P_Q +
          P_{\bar{Q}})^2 + (\vec{p}_Q + \vec{p}_{\bar{Q}})^2}}\,.
      \end{equation}
We thus have, in the $S_{\tmop{CM}}$ frame,
  $ \vec{p}_Q + \vec{p}_{\bar{Q}} = \vec{P}_Q + \vec{P}_{\bar{Q}}$\,,
  but in general $\vec{p}_{Q / \bar{Q}} \ne \vec{P}_{Q / \bar{Q}}$.
  
  \item The four-momenta of any associated light partons are left unchanged by this
  procedure. Thus, calling $e_{\tmop{CM}}$ and $E_{\tmop{CM}}$ the total
  initial partonic energy in the $S_{\tmop{CM}}$ frame before and after the
  mapping procedure, we have
  \begin{equation}
    E_{\tmop{CM}} = e_{\tmop{CM}} + E_{Q \bar{Q}} - e_{Q \bar{Q}}\,,
    \end{equation}
    while the total three-momentum in the $S_{\tmop{CM}}$ frame
    remains zero.  Thus, the rapidity of the partonic CM frame is not
    changed by our procedure. Denoting with $x_{1 / 2}$ and
    $X_{1 / 2}$ the incoming parton momentum fractions before and
    after the mapping, we must have $X_1/X_2=x_1/x_2$, and
  \begin{equation}
    X_{1 / 2} = x_{1 / 2} \times \frac{E_{\tmop{CM}}}{e_{\tmop{CM}}}\,.
    \end{equation}
\end{itemizedot}
The above procedure provides the off-shell $t\bar t$ kinematics for given virtualities (invariant masses) $M_{Q/\bar{Q}}$
of the top quarks starting from on-shell $t\bar t$ momenta. The values of $M_{Q/\bar{Q}}$ for each event are 
generated according to a Breit-Wigner distribution
\begin{equation}
  M^2 = m^2\, +\,  \Gamma  m \tan\left( \frac{r \pi}{2}\right)\,,
  \end{equation}
where $r$ is a random number uniformly distributed between $-1$ and 1. This
formula is used not only to generate $M_{Q/\bar{Q}}$, it is also used to generate the 
virtualities of the $W^{\pm}$ bosons $M_{W^+/W^-}$, using the top and $W$-boson masses 
and widths. The generated values are subject to the following restrictions:
\begin{itemizedot}
  \item Values of $M^2_{Q / \bar{Q}} < m_W^2$ or $M^2_{Q / \bar{Q}} > 2 m_t^2$
  are rejected.
  \item Values of the virtualities that prevent the construction of a
  consistent decay chain are rejected.
\end{itemizedot}
Finally, the weight of the event is multiplied by the following
correction factor, accounting for the enlarged off-shell phase space,
changed kinematics, PDFs, etc.:
\begin{eqnarray}
  w_c & = & \frac{E_{\tmop{CM}}}{e_{\tmop{CM}}} \times \frac{\sqrt{m_Q^2 +
  |\vec{p}^{\, \prime}_Q|^2}}{\sqrt{M_Q^2 + |\vec{p}^{\, \prime}_Q|^2}} \times \frac{\sqrt{m_{\bar{Q}}^2 +
  |\vec{p}^{\, \prime}_{\bar Q}|^2}}{\sqrt{M_{\bar{Q}}^2 + |\vec{p}^{\, \prime}_{\bar Q}|^2}} \times \frac{e_{Q \bar{Q}} M_{Q
  \bar{Q}}}{E_{Q \bar{Q}} m_{Q \bar{Q}}}\notag\\
  & \times & \frac{f_1 (X_1)}{f_1 (x_1)} \times \frac{f_2 (X_2)}{f_2 (x_2)} \notag\\
  &  \times & \frac{\frac{|\vec{K}_b|}{M^2_Q}}{\frac{|\vec{k}_b|}{m^2_Q}}\times
  \frac{\frac{|\vec{K}_{\bar{b}}|}{M^2_{\bar{Q}}}}{\frac{|\vec{k}_{\bar{b}}|}{m^2_{\bar{Q}}}}
  \times \frac{\frac{|\vec{K}_{l^+}|}{M^2_{W^+}}}{\frac{|\vec{k}_{l^+}|}{m^2_{W^+}}} \times
  \frac{\frac{|\vec{K}_{l^-}|}{M^2_{W^-}}}{\frac{|\vec{k}_{l^-}|}{m^2_{W^-}}}\,,
\label{eq:cweight}
\end{eqnarray}
where $|\vec{p}^{\, \prime}_Q| = |\vec{p}^{\, \prime}_{\bar Q}|$ by
definition, and $\vec{k}$ and $\vec{K}$ denote the three-momentum of a
decay product in the \textit{resonance} rest frame (i.e. both for top
and $W$ decays).
The second line corrects for the fact that the parton luminosities are
evaluated at different momentum fractions in the off-shell $t\bar{t}$
kinematics.

It is obtained by writing the full phase space as
\begin{eqnarray}
  \int \mathd \Phi & = & \int \mathd y_{\tmop{CM}} \mathd \tau_X \mathd^4 P_{Q
  \bar{Q}} \mathd \Phi_k \delta^4 (X_1 p_1 + X_2 p_2 - P_{Q \bar{Q}} - k)\notag\\
  & \times & \int \frac{\mathd^3 \vec{P}'_Q}{P_Q^{0\prime}}  \frac{\mathd^3
  \vec{P}'_{\bar{Q}}}{P_{\bar{Q}}^{0\prime}} \delta^4 (P_Q + P_{\bar{Q}} -
  P_{Q \bar{Q}})\notag\\
  & = & \int \mathd y_{\tmop{CM}} \mathd \tau_X \mathd^3 P_{Q \bar{Q}} \mathd
  \Phi_k \delta^3 (- \vec{P}_{Q \bar{Q}} - \vec{k}) \delta (X_1 p^0_1 + X_2
  p^0_2 - P^0_{Q \bar{Q}} - k^0)\notag\\
  &  & \mathd P^0_{Q \bar{Q}} \mathd M_{Q \bar{Q}}^2 \delta (P_{Q \bar{Q}}^2
  - M_{Q \bar{Q}}^2) \int \frac{\mathd^3 \vec{P}'_Q}{P_Q^{0\prime}
  P_{\bar{Q}}^{0\prime}} \delta (P^{\prime 0}_Q + P^{\prime 0}_{\bar{Q}} - M_{Q \bar{Q}})\notag\\
  & = &  \int \mathd y_{\tmop{CM}}  \frac{2 \sqrt{\tau_X} \mathd
  y_{\tmop{CM}}}{2 E_{\tmop{beam}}} \mathd \Phi_k  \frac{M_{Q \bar{Q}}}{P^0_{Q
  \bar{Q}}}  \int \frac{\mathd^3 \vec{P}'_Q}{P_Q^{0\prime}
  P_{\bar{Q}}^{0\prime}},
\end{eqnarray}
where all frame-dependent quantities in the outer integral (including $p_{1 /
2}^0$) are evaluated in the $S_{\tmop{CM}}$ frame, while those in the inner
integral are evaluated in the $S_{Q \bar{Q}}$ frame. We have used
\begin{eqnarray}
X_1 &=& \exp (y_{\tmop{CM}})  \sqrt{\tau_X}\,,\qquad X_2 = \exp (- y_{\tmop{CM}}) 
\sqrt{\tau_X}\,,\notag \\
p_1^0 &=& E_{\tmop{beam}} \exp (-y_{\tmop{CM}})\,,\qquad p_2^0 =
E_{\tmop{beam}} \exp (y_{\tmop{CM}})\,,
\end{eqnarray}
where $\tau_X=X_1X_2$ and $y_{\tmop{CM}}$ is the 
rapidity of the partonic CM system.
The phase space for the  light partons in the final state is represented by $\mathd
\Phi_k$. It is clear that the only difference with respect to the phase space
at fixed masses is given by the factor
\begin{equation}
  \frac{\sqrt{\tau_X}}{\sqrt{\tau_x}} \times \frac{\frac{M_{Q
   \bar{Q}}}{P^0_{Q \bar{Q}}} }{\frac{m_{Q \bar{Q}}}{p^0_{Q \bar{Q}}}} \times
   \frac{\frac{1}{P_Q^{0\prime} P_{\bar{Q}}^{0\prime}}}{\frac{1}{p_Q^{0\prime}
   p_{\bar{Q}}^{0\prime}}} = \frac{E_{\tmop{CM}}}{e_{\tmop{CM}}} \times
   \frac{p^0_{Q \bar{Q}} M_{Q \bar{Q}}}{P^0_{Q \bar{Q}} m_{Q \bar{Q}}} \times
   \frac{p_Q^{0\prime} p_{\bar{Q}}^{0\prime}}{P_Q^{0\prime}
     P_{\bar{Q}}^{0\prime}}\,.
   \end{equation} 

As far as the correction factors to the resonant decays are concerned, they are included through the third line 
in \eqn{eq:cweight}. The phase space together with the normalisation of the width is given by
$ \frac{| \vec{k} |}{M^2}$.
Thus a factor
\begin{equation}
  \frac{\frac{| \vec{K} |}{M^2}}{\frac{| \vec{k} |}{m^2}}
  \end{equation}
is provided for each decaying resonance.

For the correction factor in \eqn{eq:cweight} an upper bound is determined, which is 
implemented using a standard hit-and-miss technique.
In our implementation, the hit-and-miss technique is used for each
on-shell event that is produced. Strictly speaking, it would be more correct
to regenerate the on-shell event in case of a miss. Whether the improved
accuracy would be worth the loss of efficiency is a matter yet to be studied.

\afterpage{\clearpage}
\newpage

\section{Additional comparisons to data}
\label{app:additionalplots}
In this appendix we show the comparison of \minnlo{} and \minlo{}
predictions to data for additional observables. Since similar comments to those made about the plots shown in the main
text generally apply here as well, we refrain from commenting on these results any further.

\subsection{Fully leptonic top-decay mode including $\tau$ decays to electrons/muons}
\label{app:leptonictau}

\begin{figure}[h!]
\begin{center}
\begin{tabular}{ccc}
\hspace{-.55cm}
\includegraphics[width=.34\textwidth,page=16]{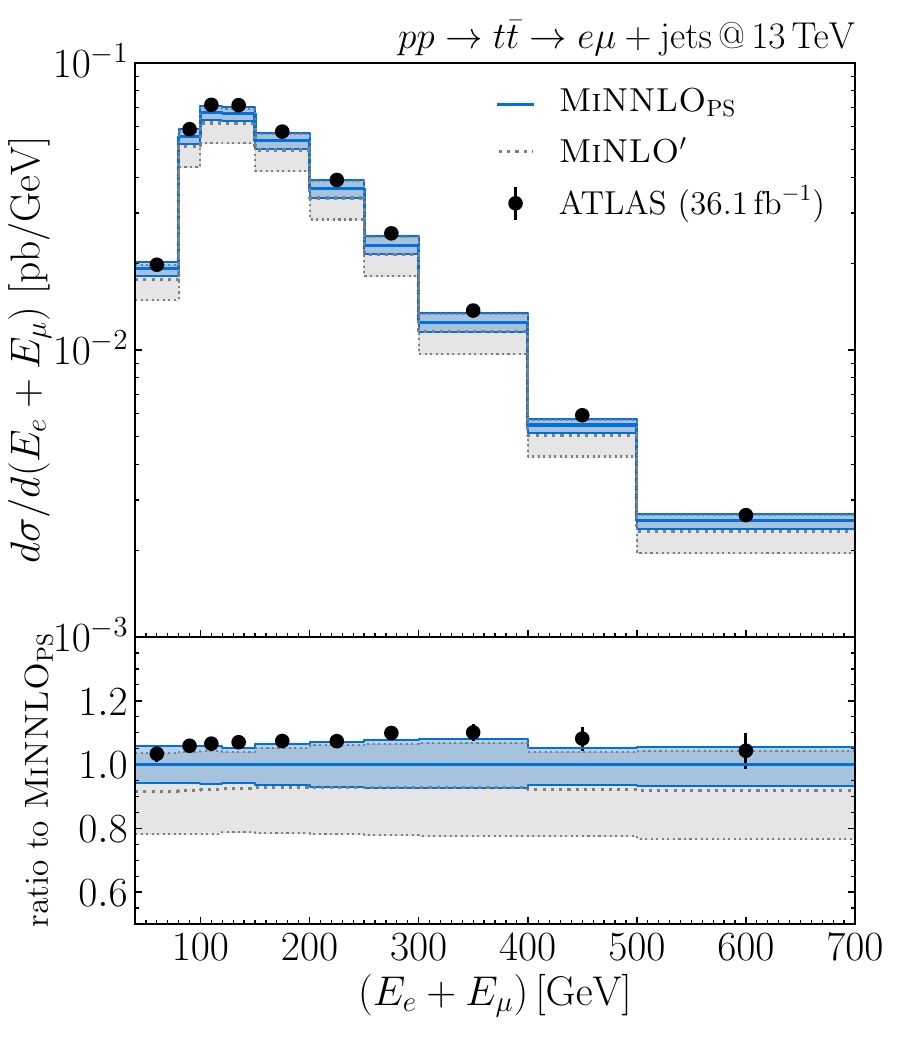}
&
\hspace{-0.6cm}
\includegraphics[width=.34\textwidth,page=19]{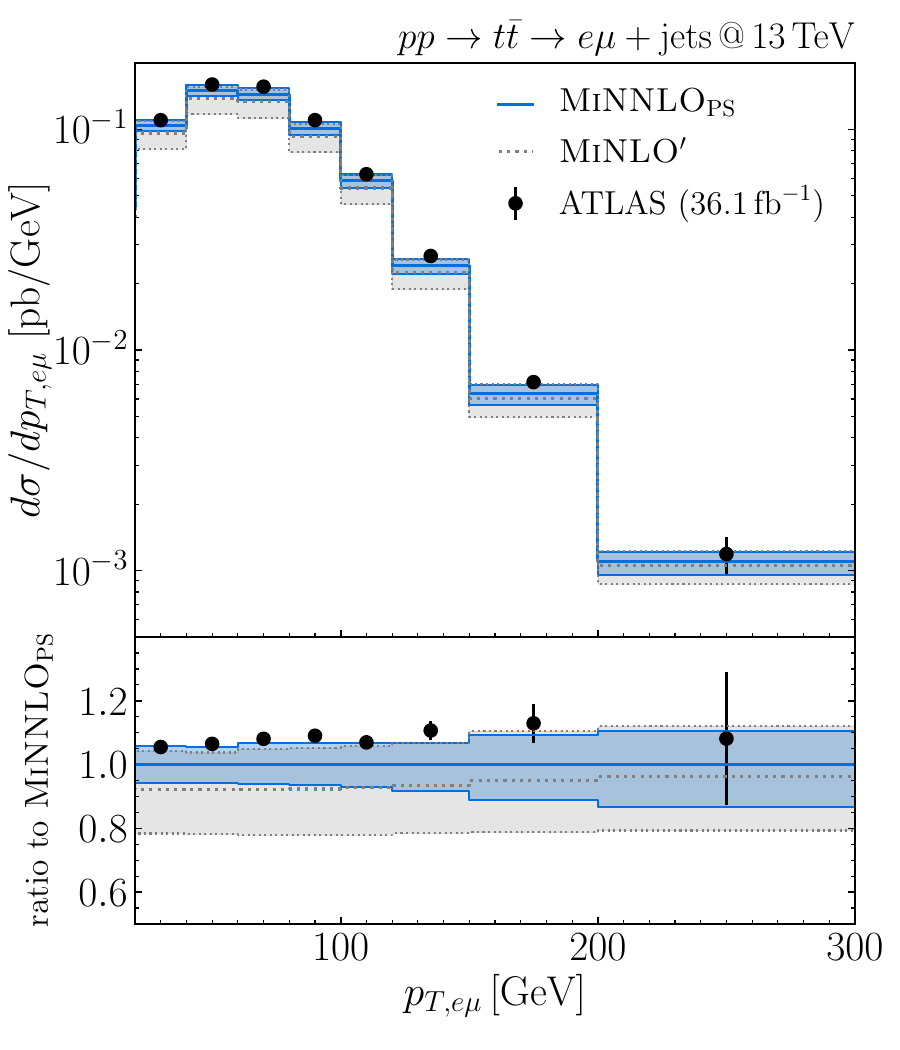} 
&
\hspace{-0.6cm}
\includegraphics[width=.34\textwidth,page=20]{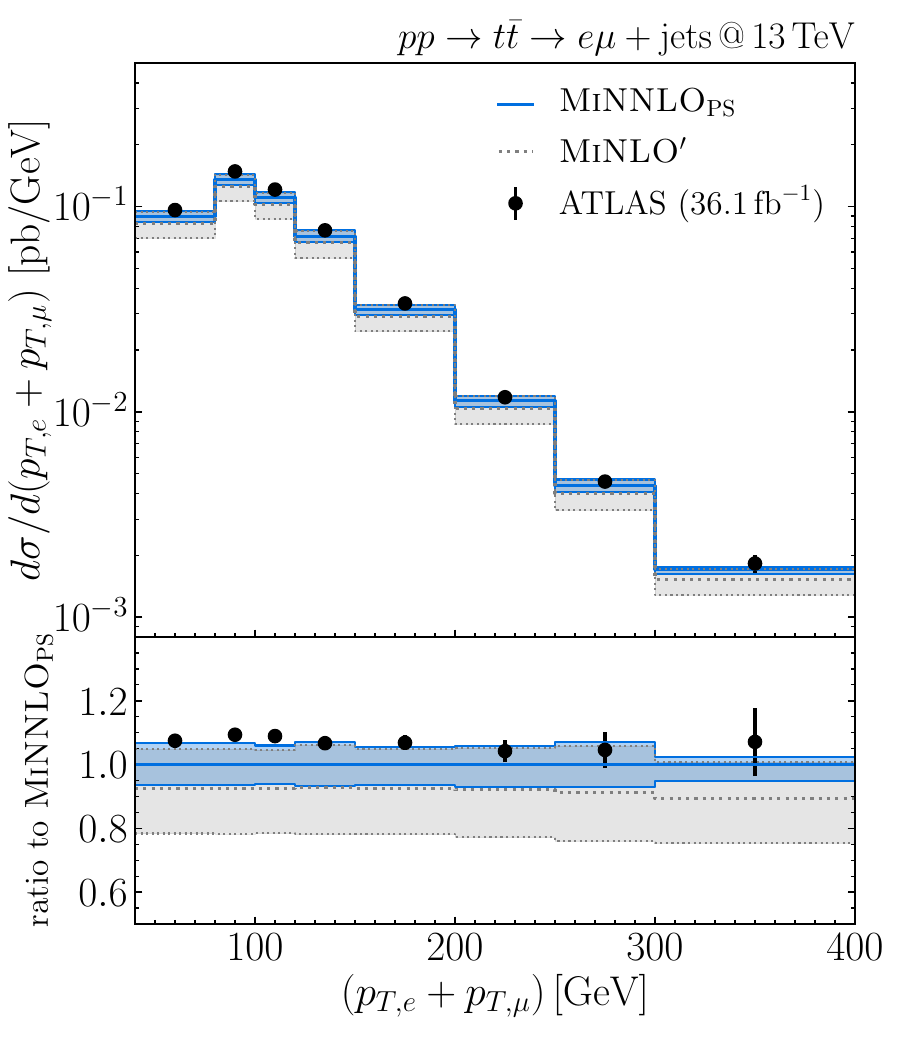}
\end{tabular}
\begin{tabular}{cccccc}
\hspace{-0.3cm}
\includegraphics[height=.305\textheight,page=6]{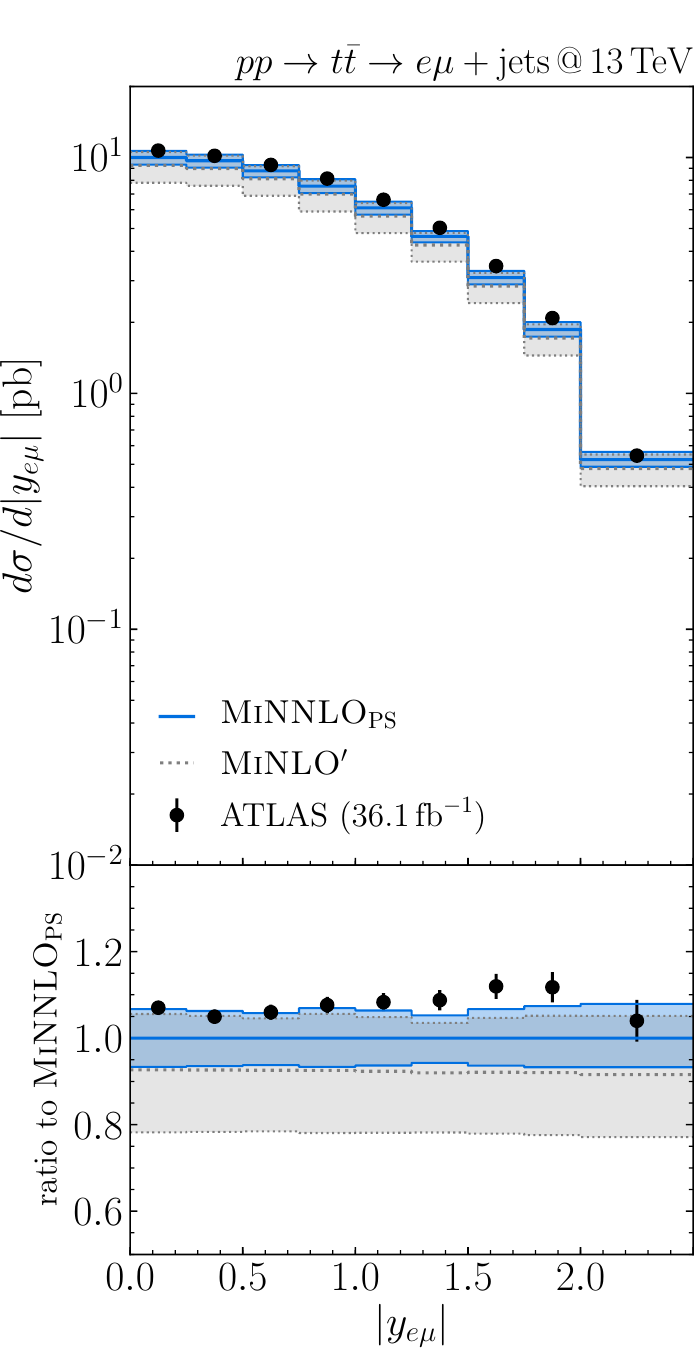}
&
\hspace{-0.56cm}
\includegraphics[height=.305\textheight,page=7]{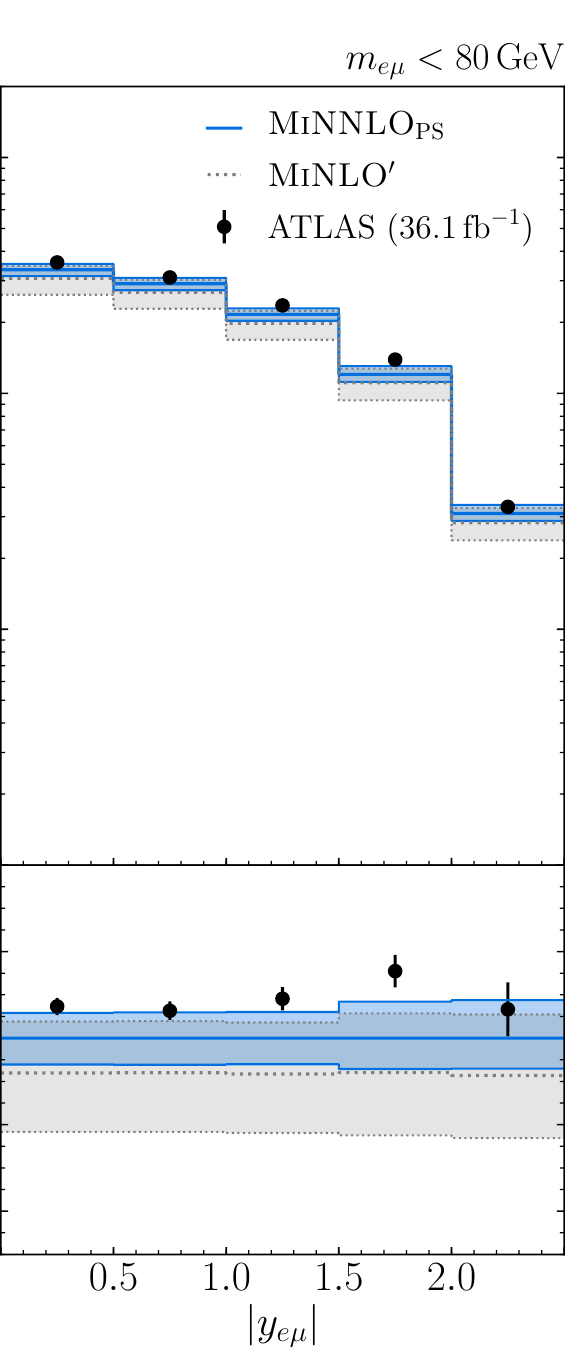}
&
\hspace{-0.56cm}
\includegraphics[height=.305\textheight,page=8]{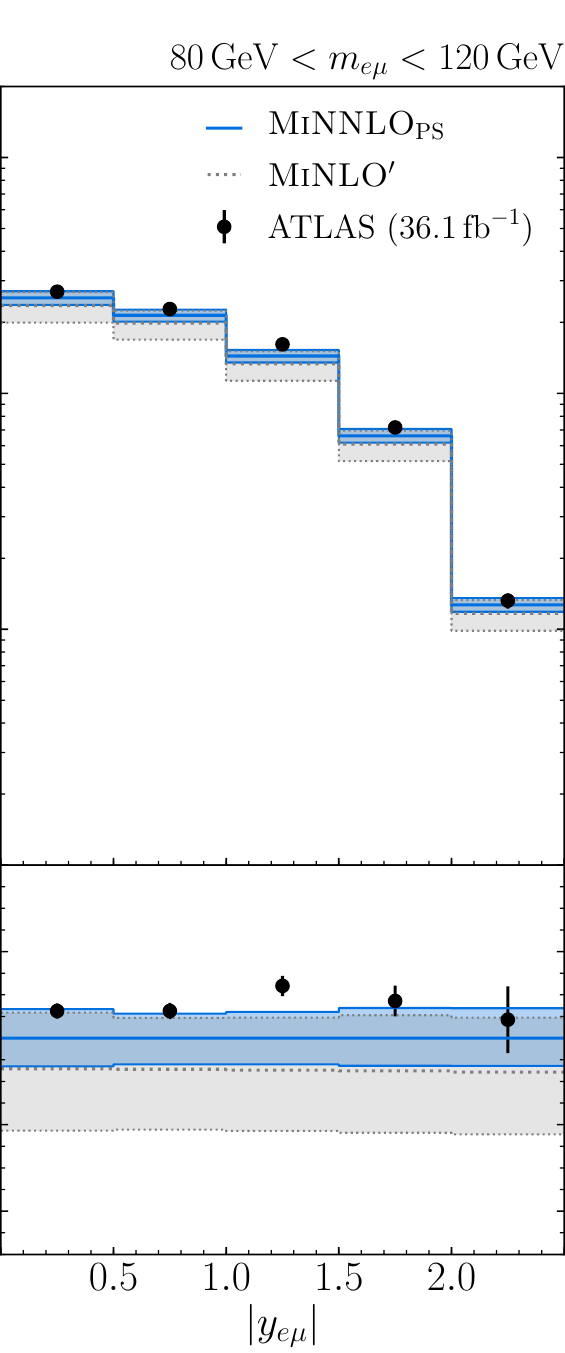} 
&
\hspace{-0.56cm}
\includegraphics[height=.305\textheight,page=9]{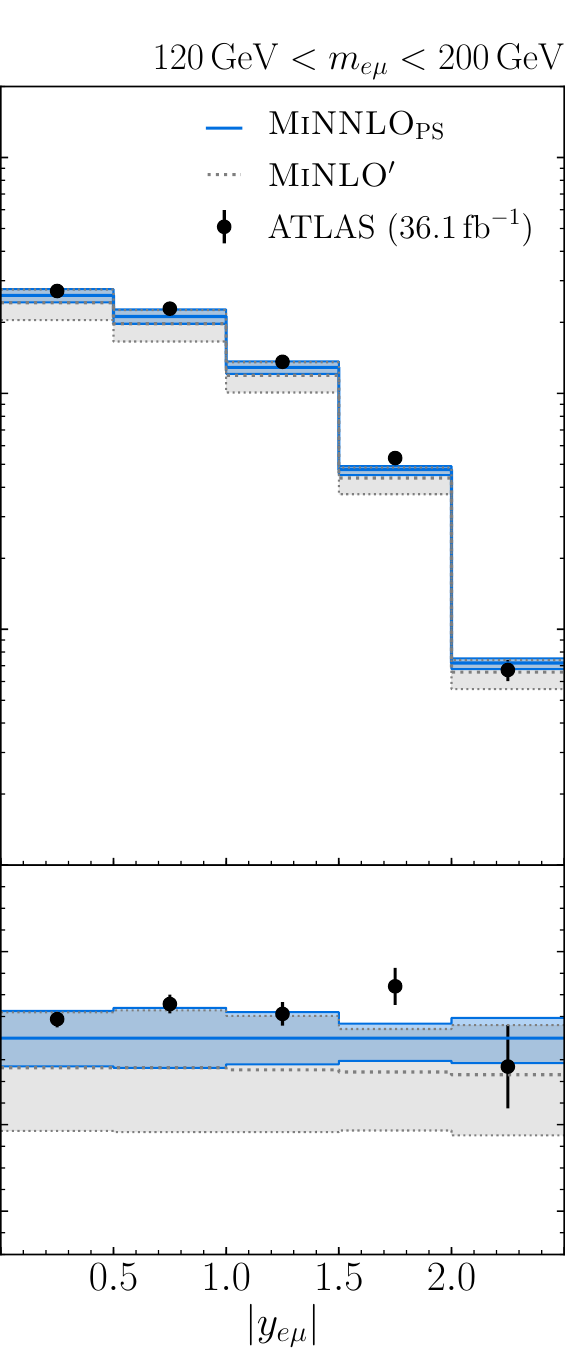}
&
\hspace{-0.56cm}
\includegraphics[height=.305\textheight,page=10]{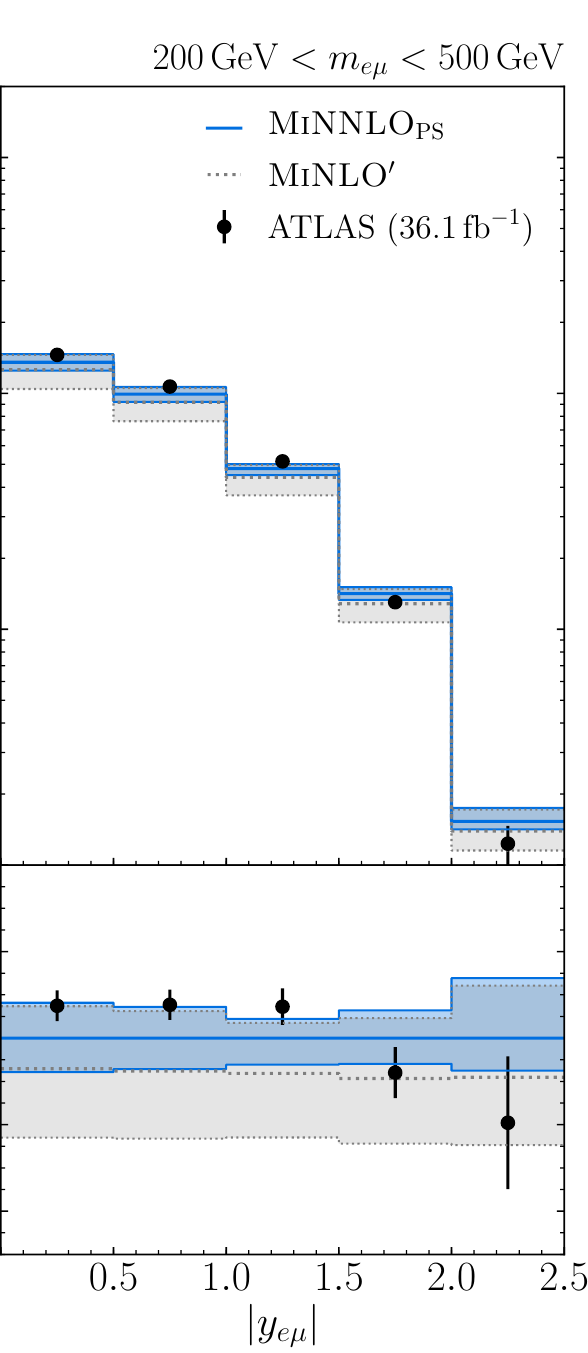}\\
\end{tabular}\caption{\label{fig:leptonicapp}  Comparison of \minnlo{} (blue, solid) and \minlo{} (black, dashed) predictions with ATLAS data \cite{ATLAS:2019hau} (black points with errors) in \setupleptonic{}, including decays of $\tau$ leptons.}
\end{center}
\end{figure}

\afterpage{\clearpage}
\newpage

\subsection{Fully leptonic top-decay mode excluding $\tau$ decays to electrons/muons}
\label{app:leptonicnotau}

\begin{figure}[h!]
\begin{center}
\begin{tabular}{cccccc}
\hspace{-0.3cm}
\includegraphics[height=.305\textheight,page=1]{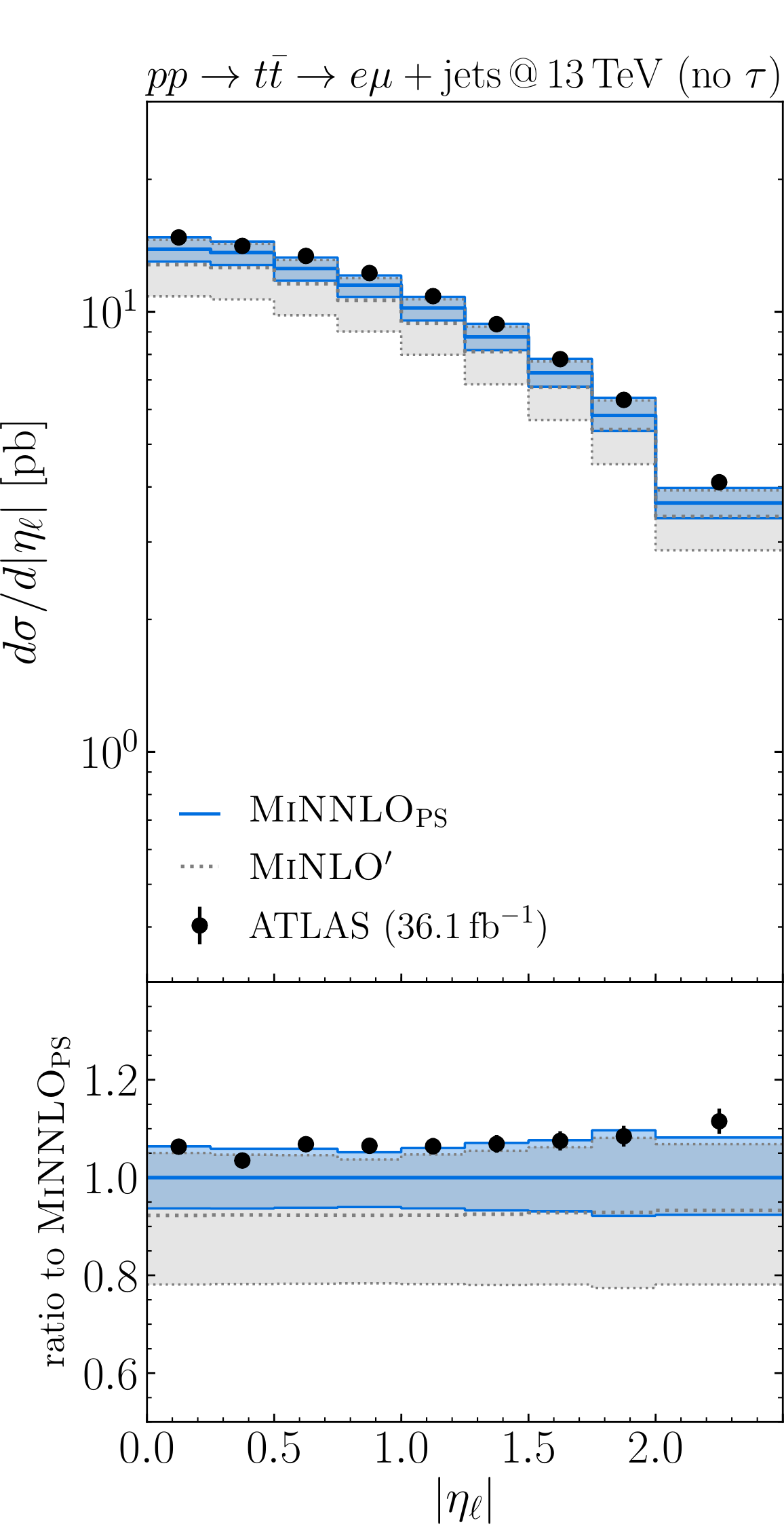}
&
\hspace{-0.56cm}
\includegraphics[height=.305\textheight,page=2]{plots/comparison_data/ATLAS_leptonic_no_tau.pdf}
&
\hspace{-0.56cm}
\includegraphics[height=.305\textheight,page=3]{plots/comparison_data/ATLAS_leptonic_no_tau.pdf} 
&
\hspace{-0.56cm}
\includegraphics[height=.305\textheight,page=4]{plots/comparison_data/ATLAS_leptonic_no_tau.pdf}
&
\hspace{-0.56cm}
\includegraphics[height=.305\textheight,page=5]{plots/comparison_data/ATLAS_leptonic_no_tau.pdf}\\
\hspace{-0.3cm}
\includegraphics[height=.305\textheight,page=6]{plots/comparison_data/ATLAS_leptonic_no_tau.pdf}
&
\hspace{-0.56cm}
\includegraphics[height=.305\textheight,page=7]{plots/comparison_data/ATLAS_leptonic_no_tau.pdf}
&
\hspace{-0.56cm}
\includegraphics[height=.305\textheight,page=8]{plots/comparison_data/ATLAS_leptonic_no_tau.pdf} 
&
\hspace{-0.56cm}
\includegraphics[height=.305\textheight,page=9]{plots/comparison_data/ATLAS_leptonic_no_tau.pdf}
&
\hspace{-0.56cm}
\includegraphics[height=.305\textheight,page=10]{plots/comparison_data/ATLAS_leptonic_no_tau.pdf}\\
\hspace{-0.3cm}
\includegraphics[height=.305\textheight,page=11]{plots/comparison_data/ATLAS_leptonic_no_tau.pdf}
&
\hspace{-0.56cm}
\includegraphics[height=.305\textheight,page=12]{plots/comparison_data/ATLAS_leptonic_no_tau.pdf}
&
\hspace{-0.56cm}
\includegraphics[height=.305\textheight,page=13]{plots/comparison_data/ATLAS_leptonic_no_tau.pdf} 
&
\hspace{-0.56cm}
\includegraphics[height=.305\textheight,page=14]{plots/comparison_data/ATLAS_leptonic_no_tau.pdf} 
&
\hspace{-0.675cm}
\includegraphics[height=.305\textheight,page=15]{plots/comparison_data/ATLAS_leptonic_no_tau.pdf}\\
\end{tabular}\caption{\label{fig:leptonicbis}  Comparison of \minnlo{} (blue, solid) and \minlo{} (black, dashed) predictions with ATLAS data \cite{ATLAS:2019hau} (black points with errors) in \setupleptonic{}, excluding decays of $\tau$ leptons.}
\vspace{-2cm}
\end{center}
\end{figure}

\afterpage{\clearpage}
\newpage

\begin{figure}[t!]
\begin{center}
\begin{tabular}{ccc}
\hspace{-.55cm}
\includegraphics[width=.34\textwidth,page=16]{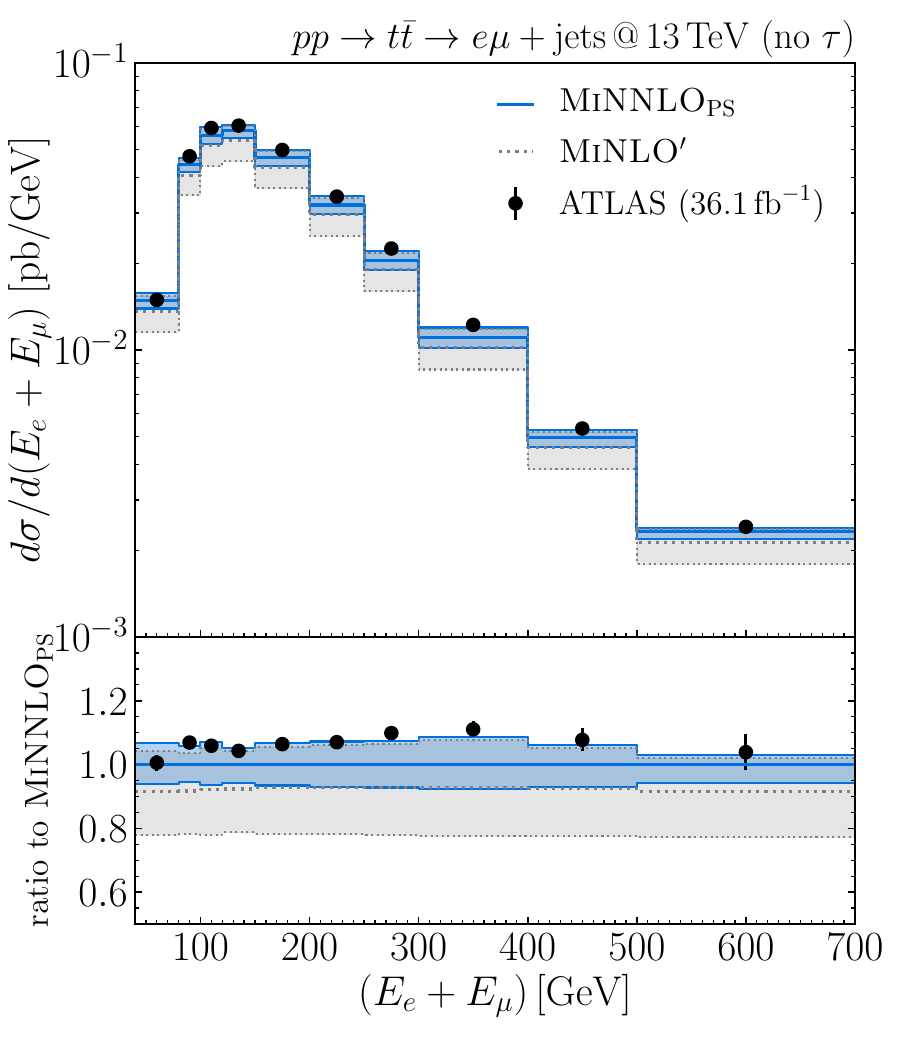}
&
\hspace{-0.6cm}
\includegraphics[width=.34\textwidth,page=17]{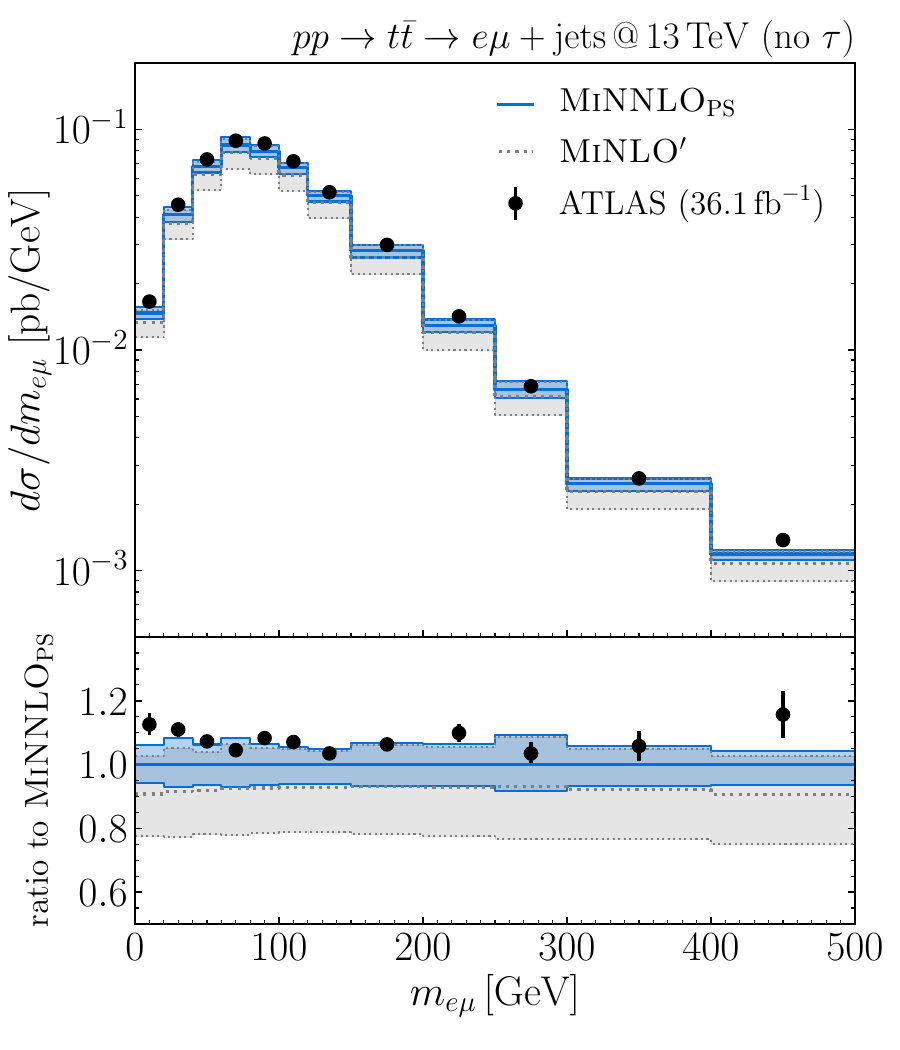} 
&
\hspace{-0.6cm}
\includegraphics[width=.34\textwidth,page=18]{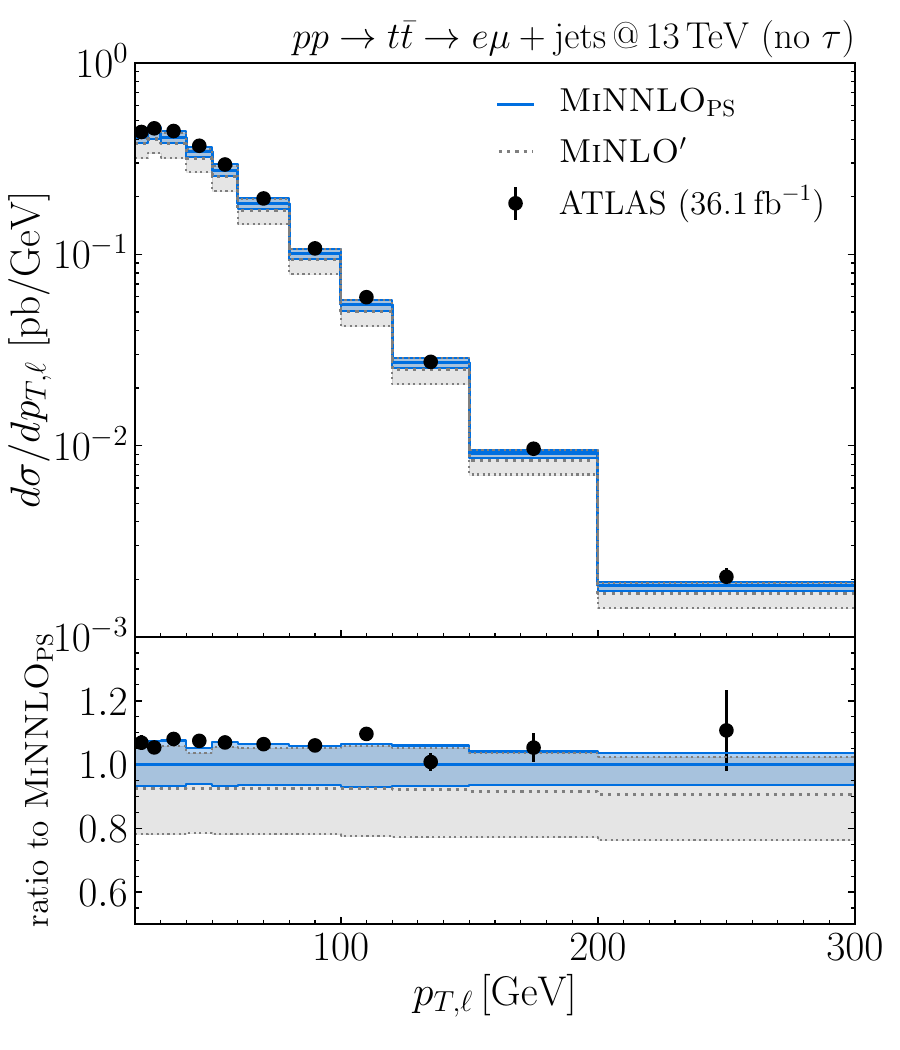}
\end{tabular}
\begin{tabular}{ccc}
\hspace{-.55cm}
\includegraphics[width=.34\textwidth,page=19]{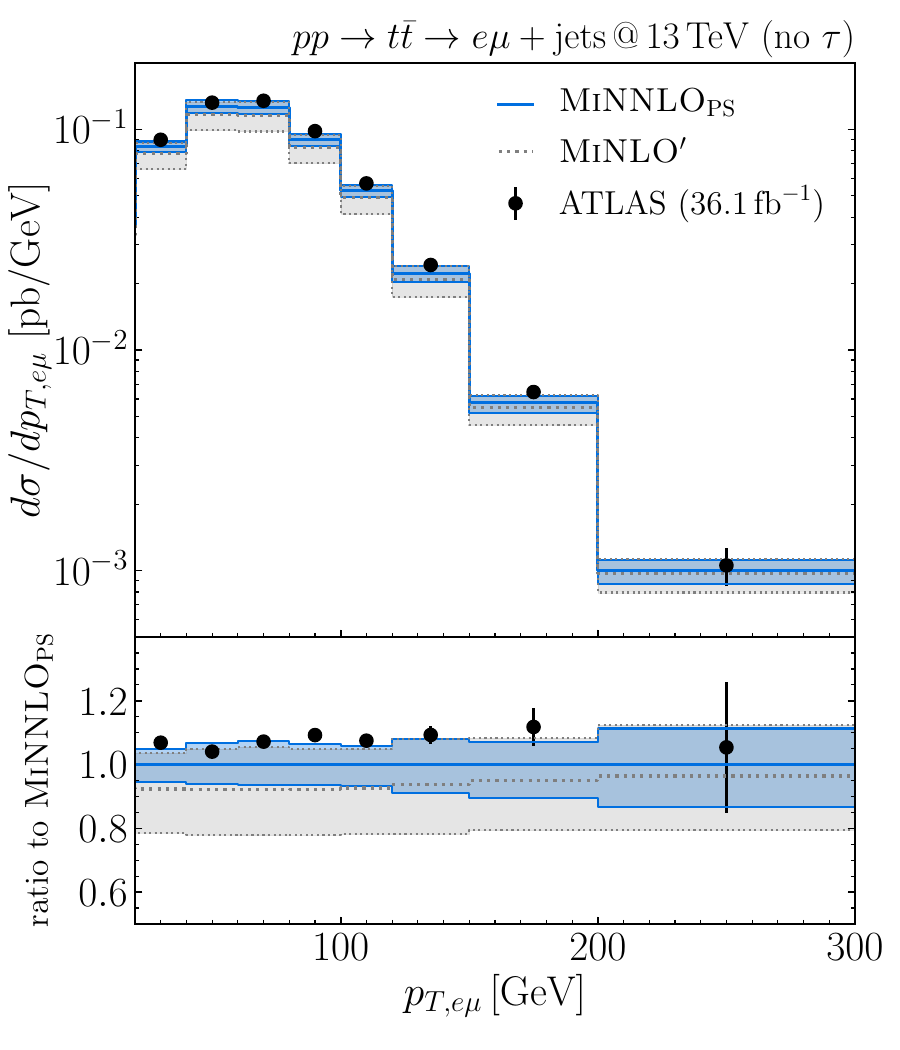}
&
\includegraphics[width=.34\textwidth,page=20]{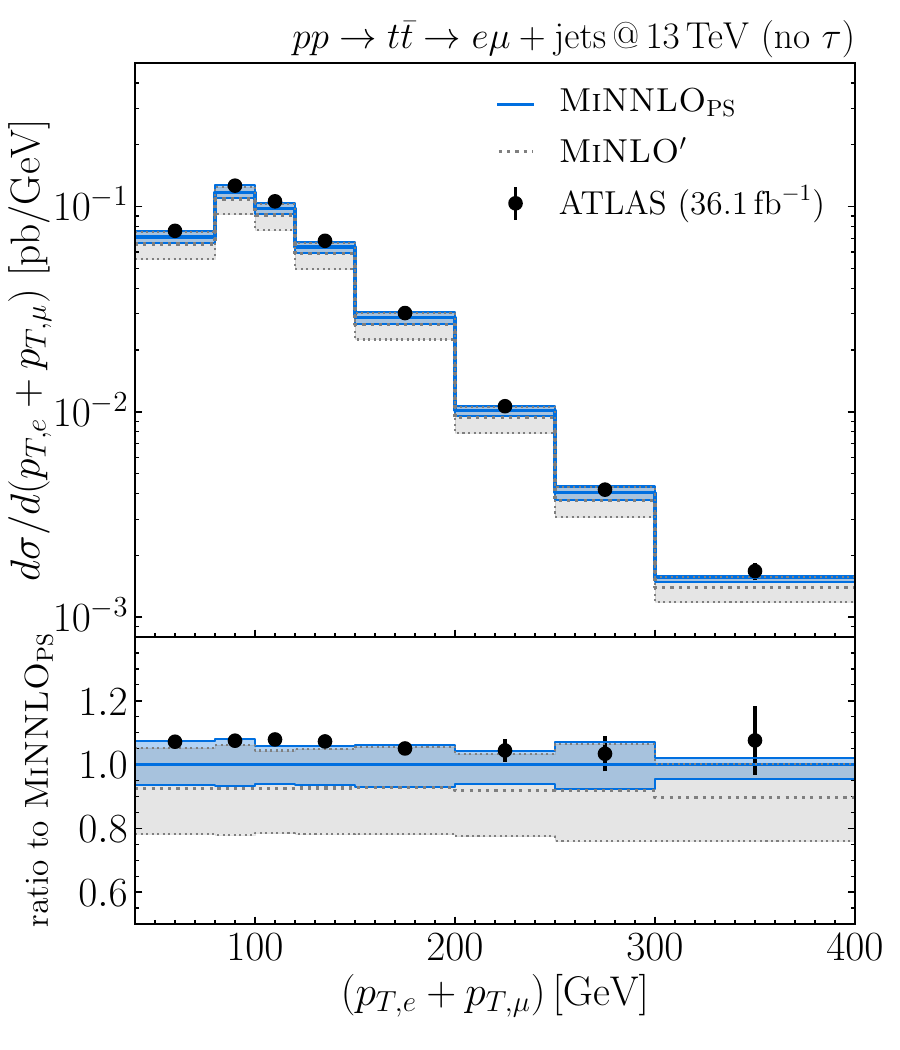} 
\end{tabular}
\caption{\label{fig:leptonic_appendix}  Comparison of \minnlo{} (blue, solid) and \minlo{} (black, dashed) predictions with ATLAS data \cite{ATLAS:2019hau} (black points with errors) in \setupleptonic{}, excluding decays of $\tau$ leptons.}
\end{center}
\end{figure}

\clearpage

\subsection{Semi-leptonic top-quark decay mode}
\label{app:semi}

\begin{figure}[h!]
\begin{center}
\begin{tabular}{ccc}
\hspace{-.55cm}
\includegraphics[width=.34\textwidth,page=4]{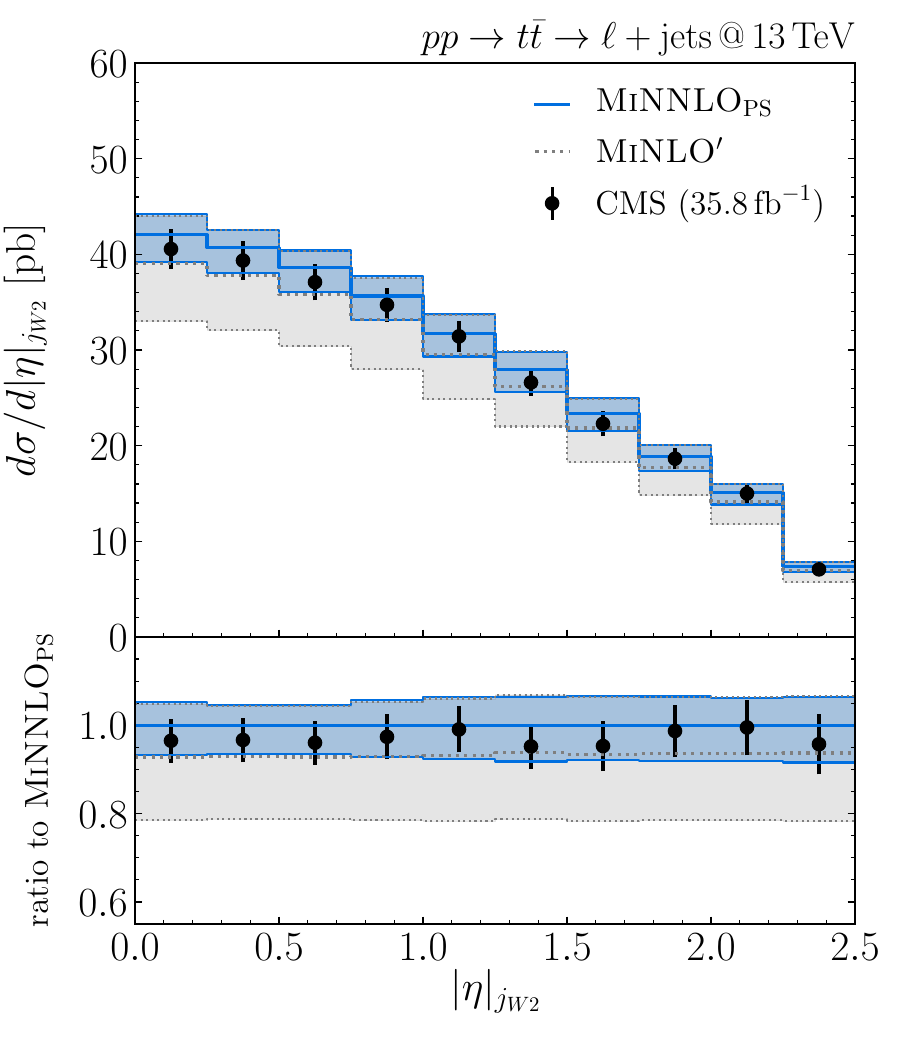}
&
\hspace{-0.6cm}
\includegraphics[width=.34\textwidth,page=5]{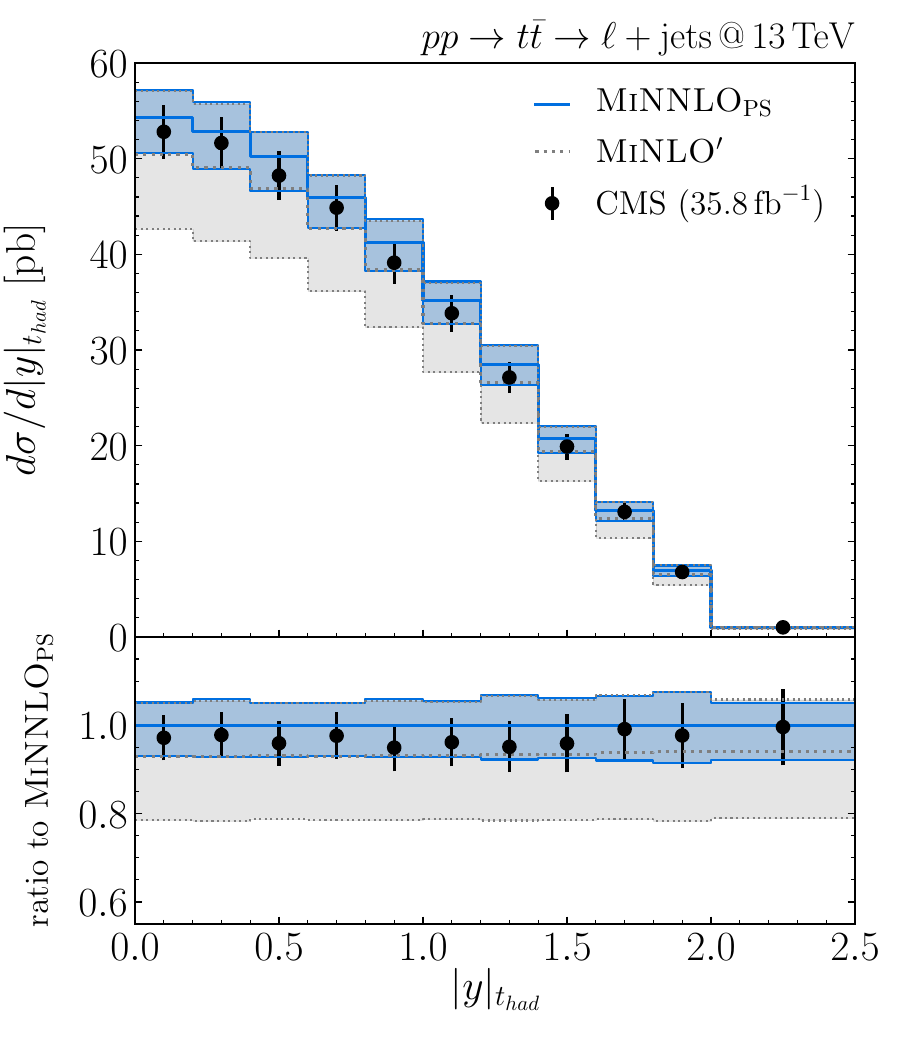} 
&
\hspace{-0.6cm}
\includegraphics[width=.34\textwidth,page=6]{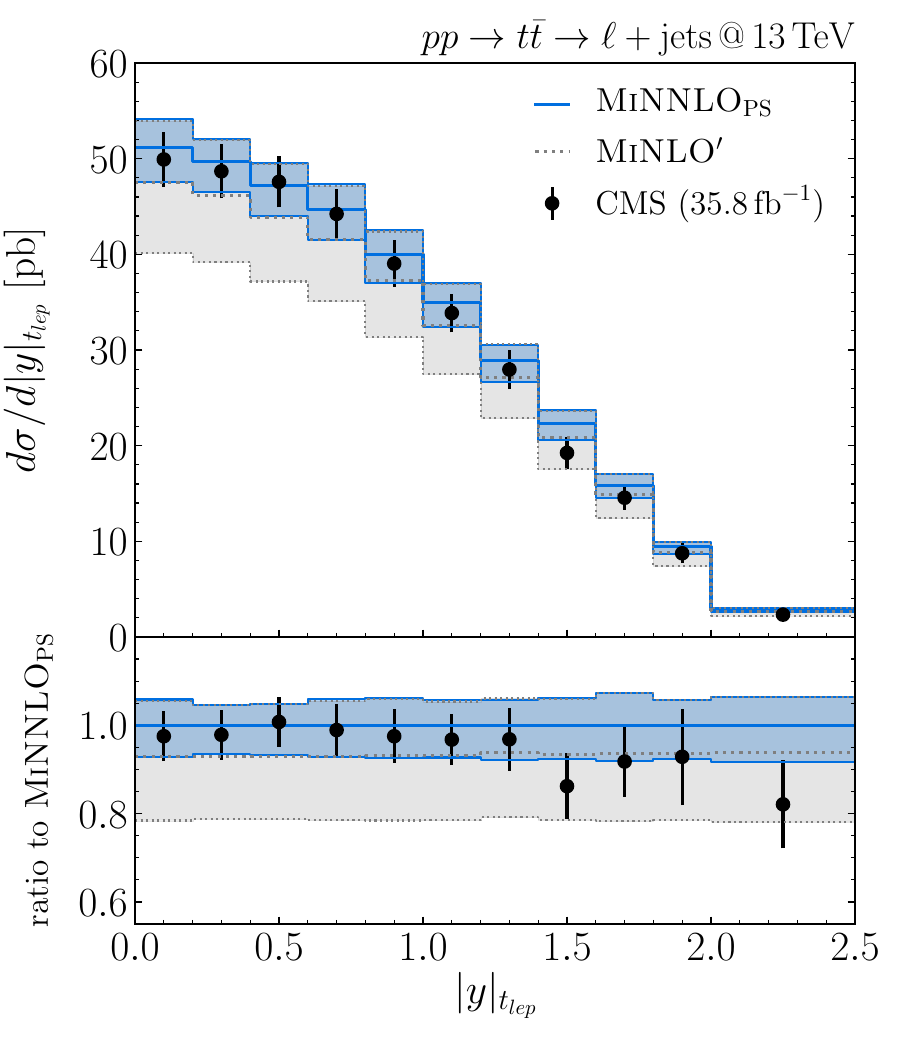}\\
\hspace{-.55cm}
\includegraphics[width=.34\textwidth,page=12]{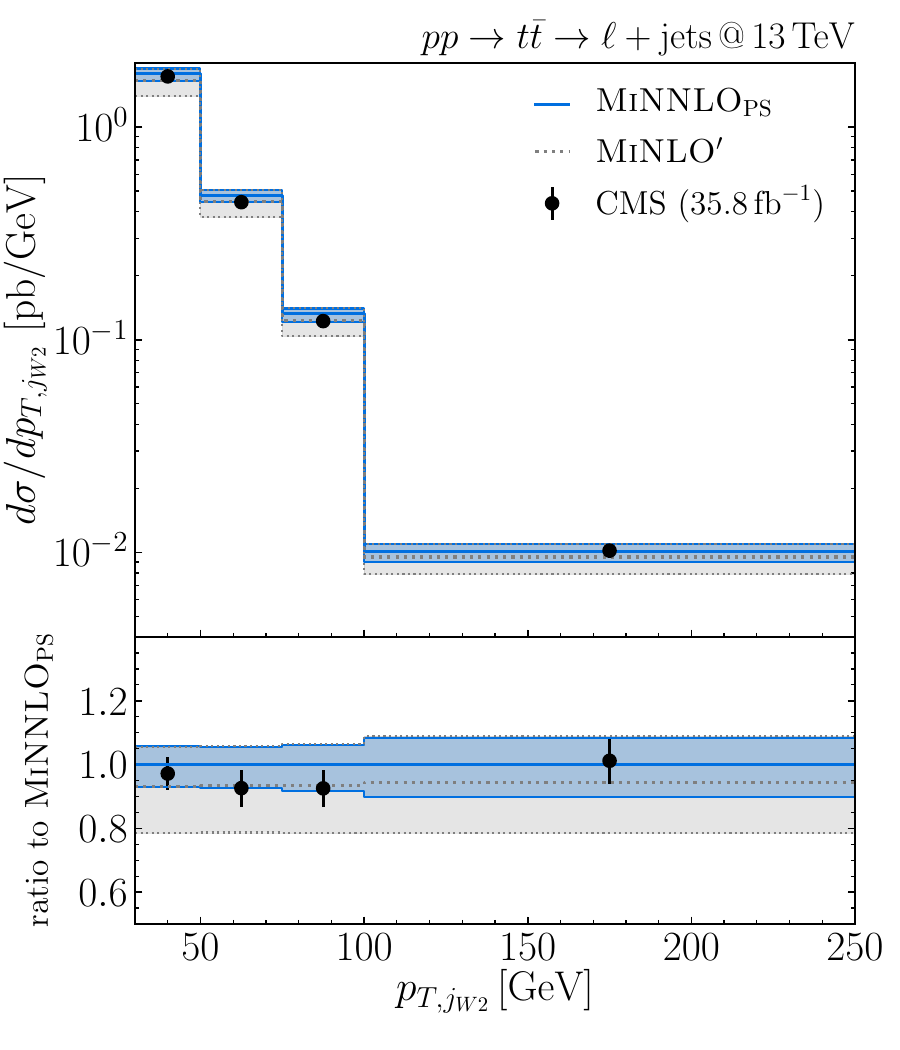}
&
\hspace{-0.6cm}
\includegraphics[width=.34\textwidth,page=13]{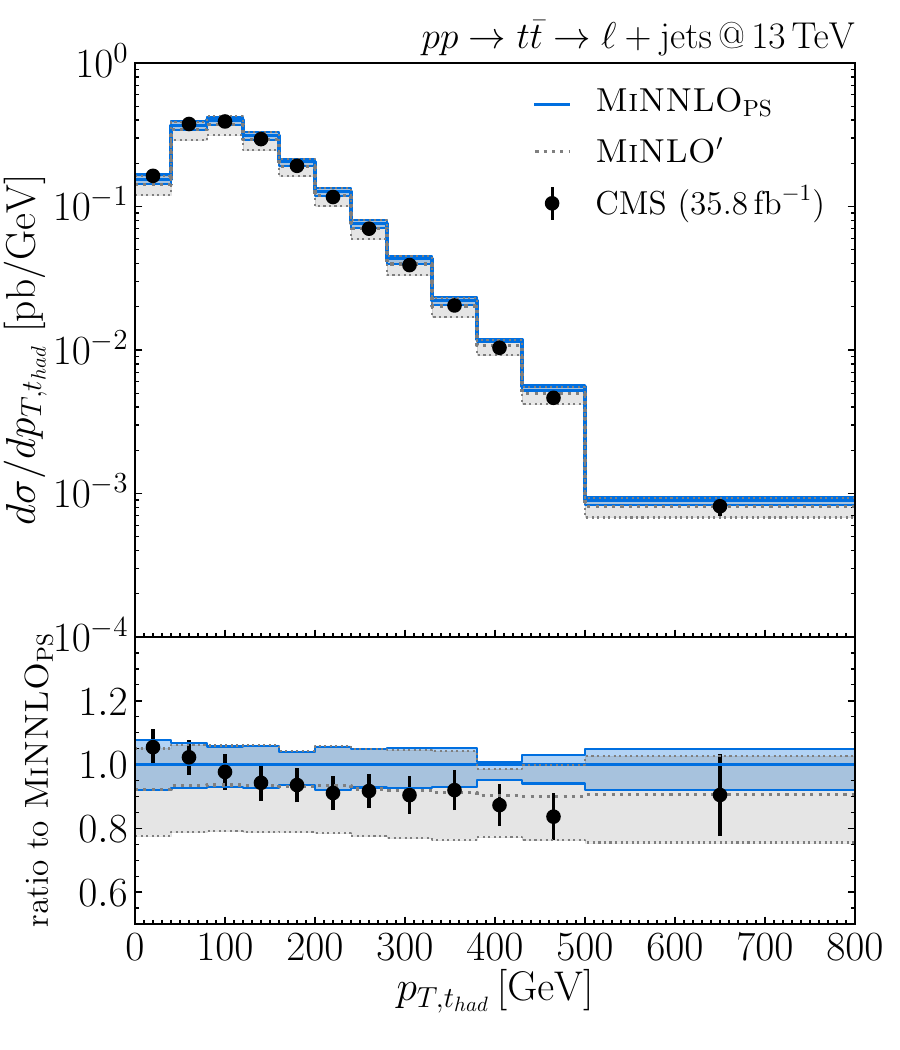} 
&
\hspace{-0.6cm}
\includegraphics[width=.34\textwidth,page=14]{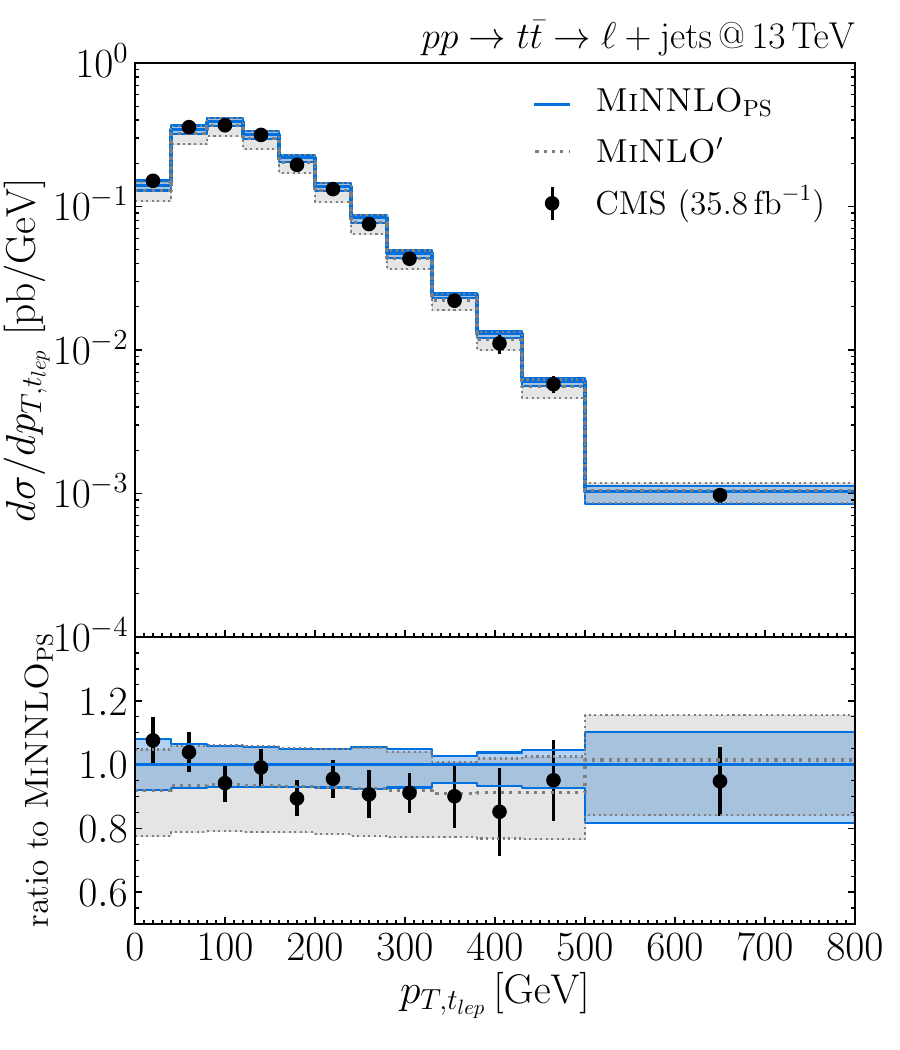}\\
\end{tabular}\caption{\label{fig:semileptonic_appendix}  Comparison of \minnlo{} (blue, solid) and \minlo{} (black, dashed) predictions with CMS data \cite{CMS:2018htd} (black points with errors) in \setupsemi{}.}
\end{center}
\end{figure}

\clearpage
\subsection{Fully hadronic top-quark decay mode}
\label{app:hadronic}

\begin{figure}[h!]
\begin{center}
\begin{tabular}{ccc}
\hspace{-.55cm}
\includegraphics[width=.34\textwidth,page=1]{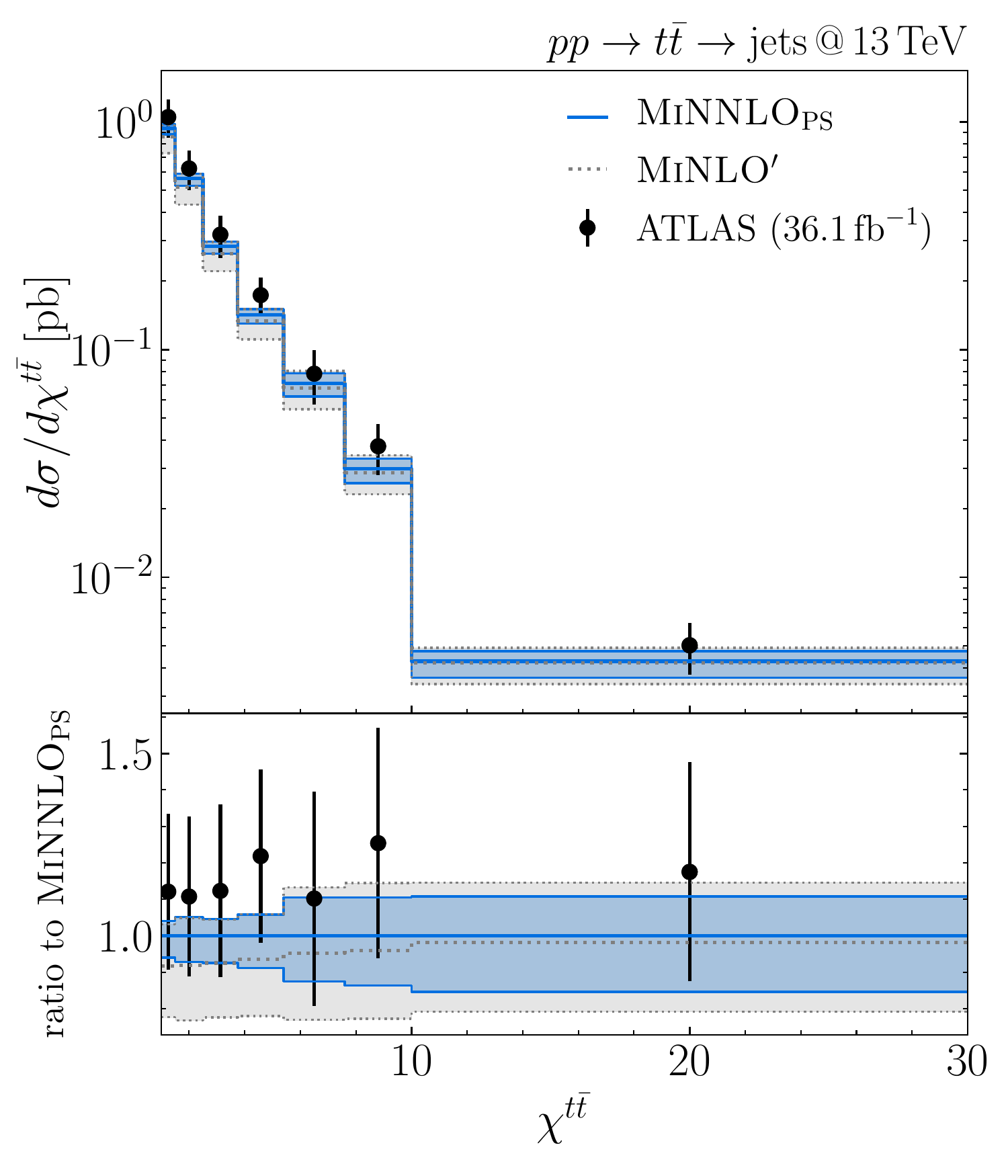}
&
\hspace{-0.6cm}
\includegraphics[width=.34\textwidth,page=2]{plots/comparison_data/ATLAS_hadronic.pdf} 
&
\hspace{-0.6cm}
\includegraphics[width=.34\textwidth,page=3]{plots/comparison_data/ATLAS_hadronic.pdf}\\
\hspace{-.55cm}
\includegraphics[width=.34\textwidth,page=4]{plots/comparison_data/ATLAS_hadronic.pdf}
&
\hspace{-0.6cm}
\includegraphics[width=.34\textwidth,page=5]{plots/comparison_data/ATLAS_hadronic.pdf} 
&
\hspace{-0.6cm}
\includegraphics[width=.34\textwidth,page=6]{plots/comparison_data/ATLAS_hadronic.pdf}\\
\hspace{-.55cm}
\includegraphics[width=.34\textwidth,page=7]{plots/comparison_data/ATLAS_hadronic.pdf}
&
\hspace{-0.6cm}
\includegraphics[width=.34\textwidth,page=8]{plots/comparison_data/ATLAS_hadronic.pdf} 
&
\hspace{-0.6cm}
\includegraphics[width=.34\textwidth,page=10]{plots/comparison_data/ATLAS_hadronic.pdf}
\end{tabular}\caption{\label{fig:hadronic_appendix_1}  Comparison of \minnlo{} (blue, solid) and \minlo{} (black, dashed) predictions with ATLAS data \cite{ATLAS:2020ccu} (black points with errors) in \setuphadronic{}.}
\end{center}
\end{figure}

\begin{figure}[t!]
\begin{center}
\begin{tabular}{ccc}
\hspace{-.55cm}
\includegraphics[width=.34\textwidth,page=11]{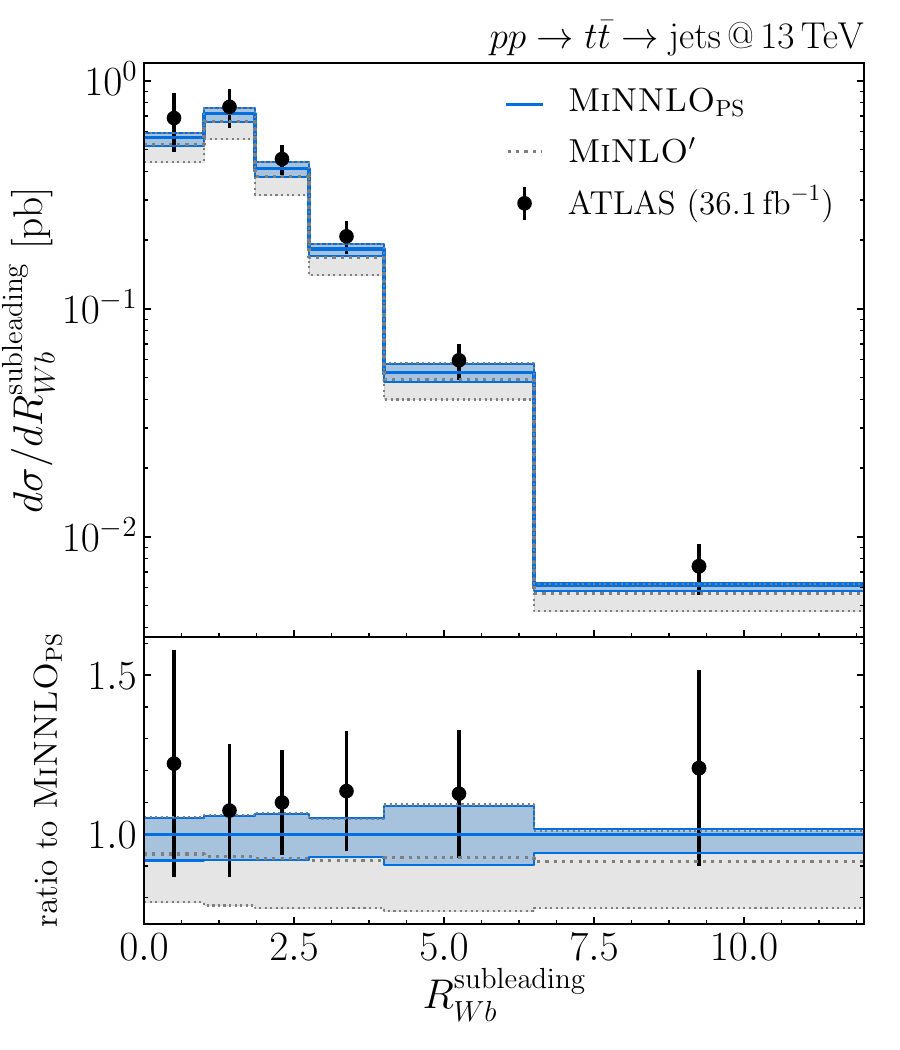}
&
\hspace{-0.6cm}
\includegraphics[width=.34\textwidth,page=13]{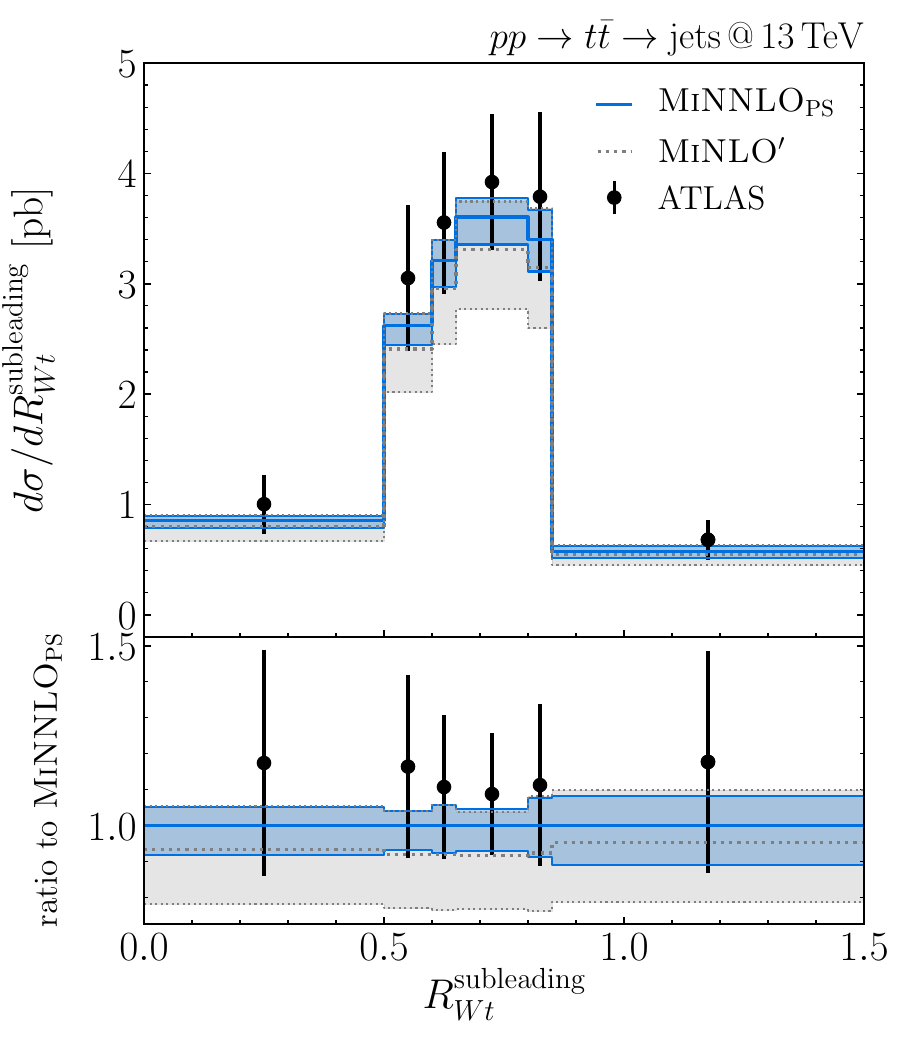} 
&
\hspace{-0.6cm}
\includegraphics[width=.34\textwidth,page=14]{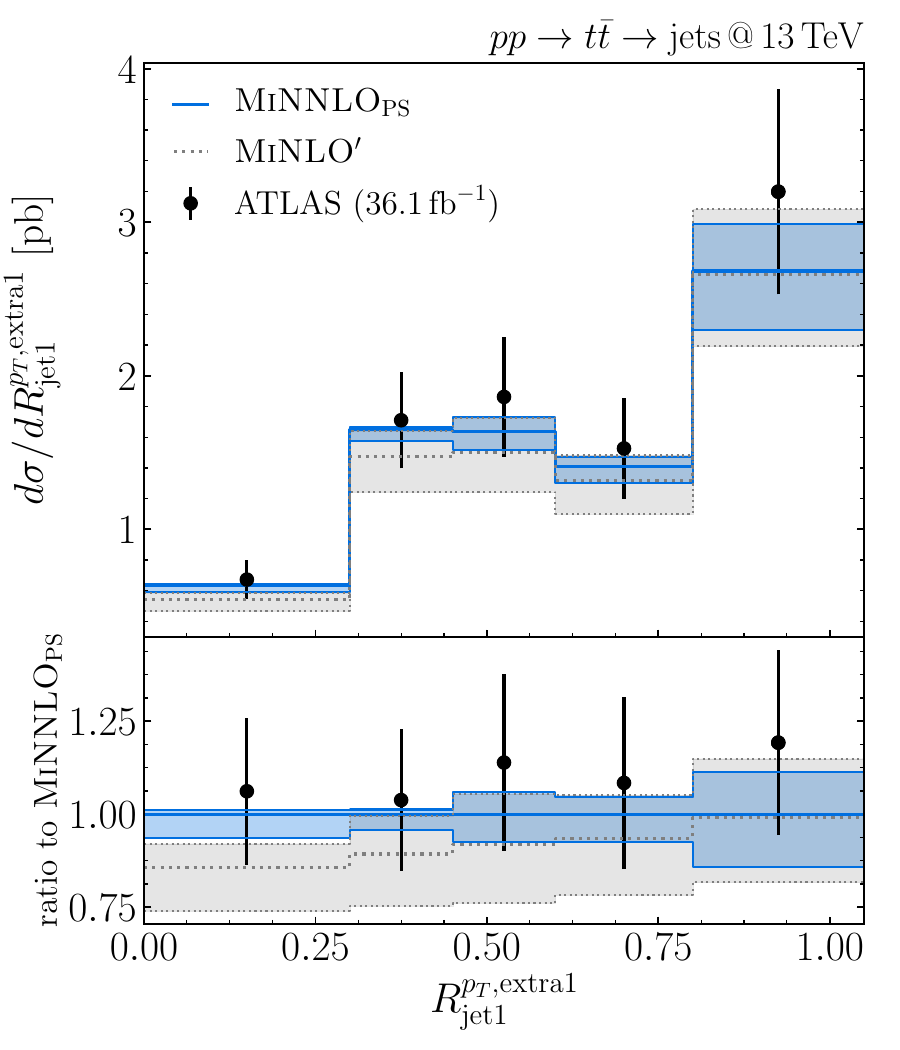}\\
\hspace{-.55cm}
\includegraphics[width=.34\textwidth,page=15]{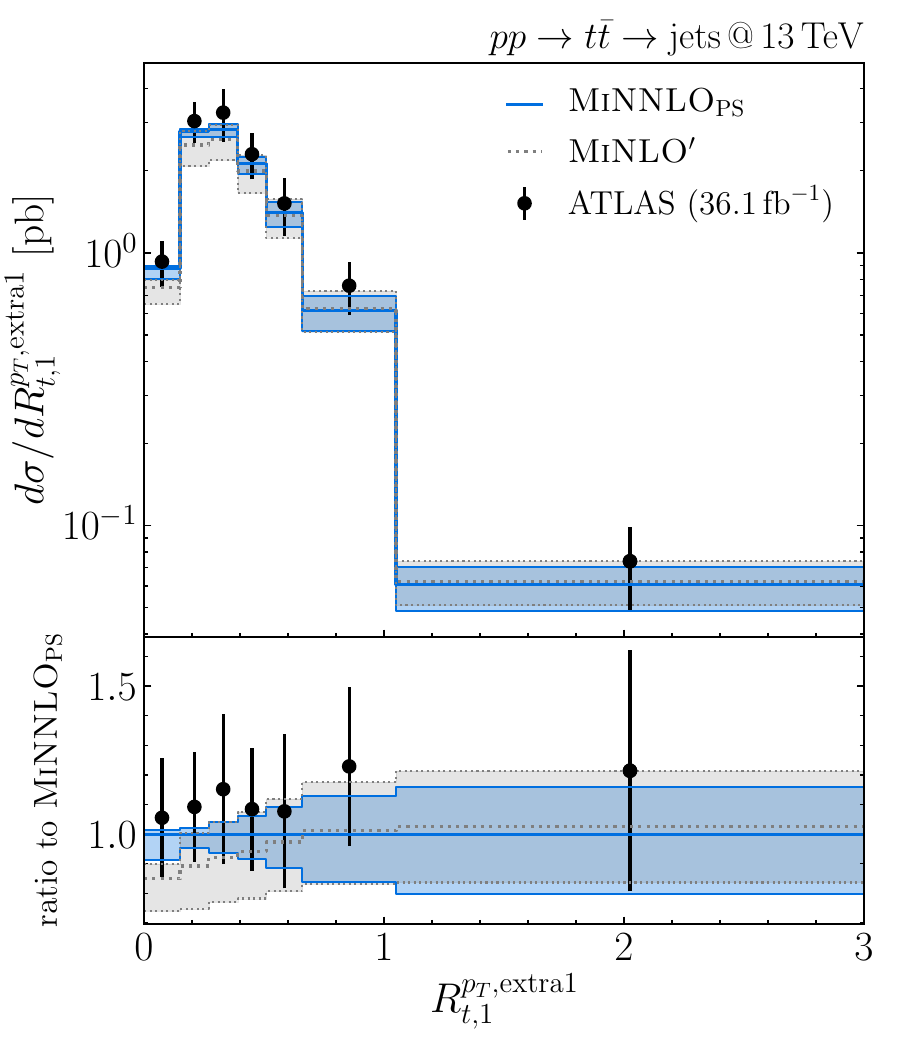}
&
\hspace{-0.6cm}
\includegraphics[width=.34\textwidth,page=16]{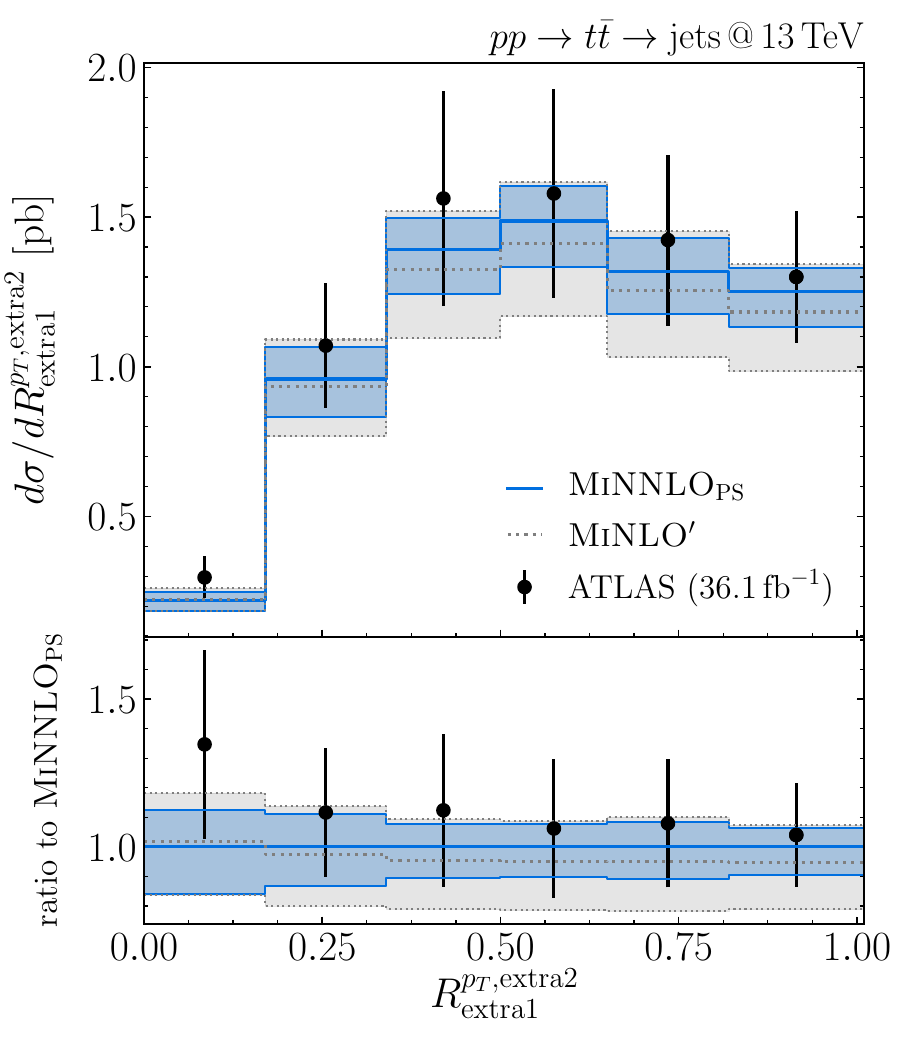} 
&
\hspace{-0.6cm}
\includegraphics[width=.34\textwidth,page=17]{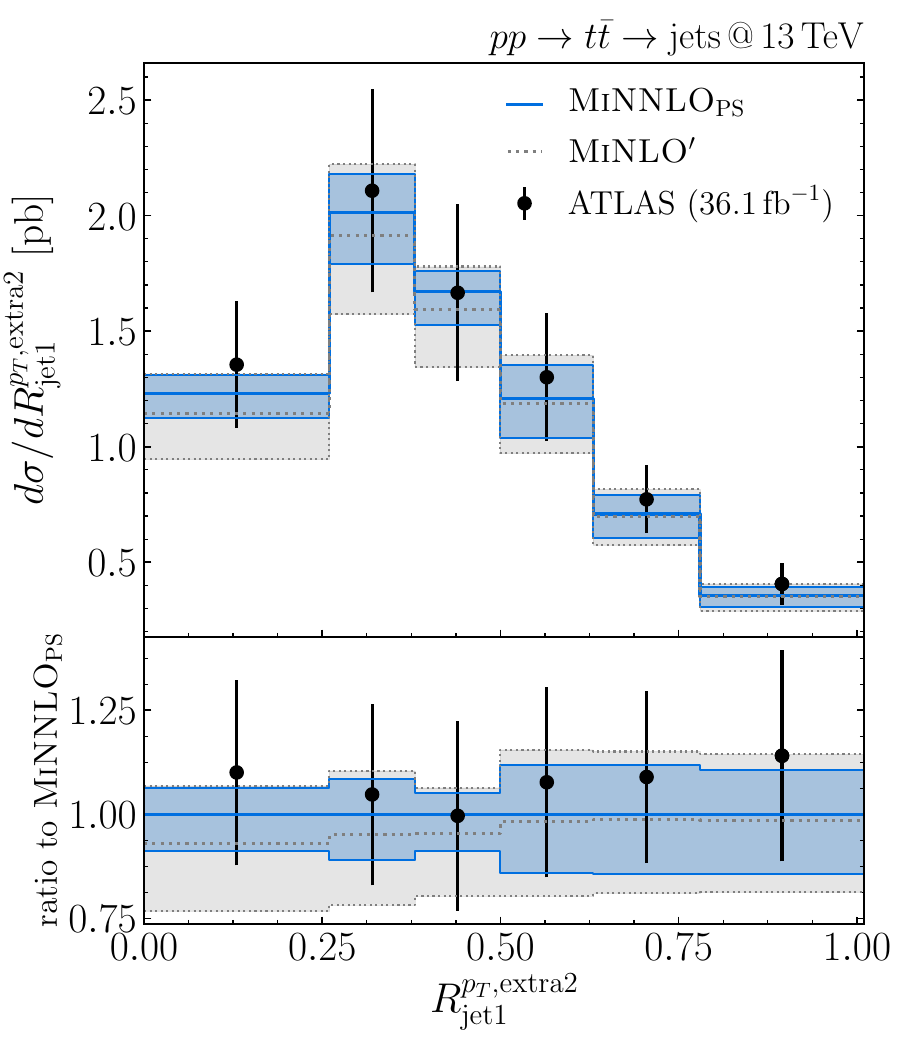}\\
\hspace{-.55cm}
\includegraphics[width=.34\textwidth,page=18]{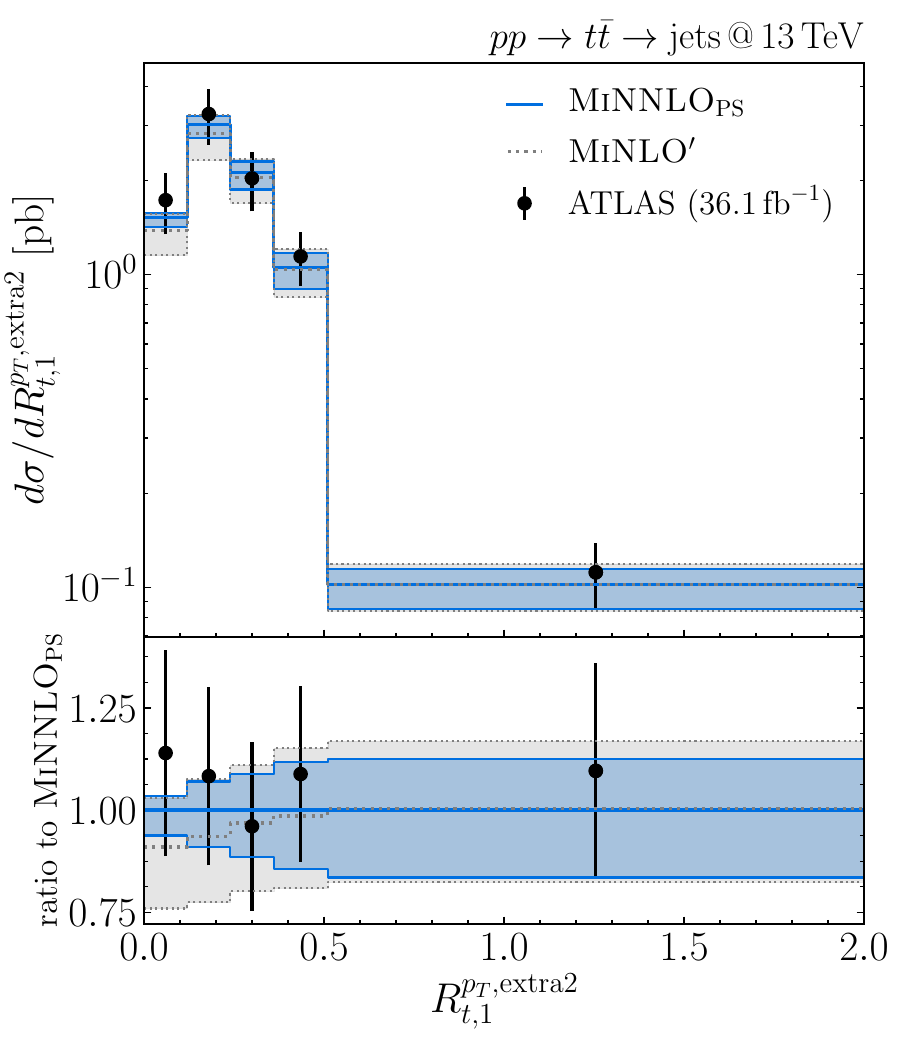}
&
\hspace{-0.6cm}
\includegraphics[width=.34\textwidth,page=19]{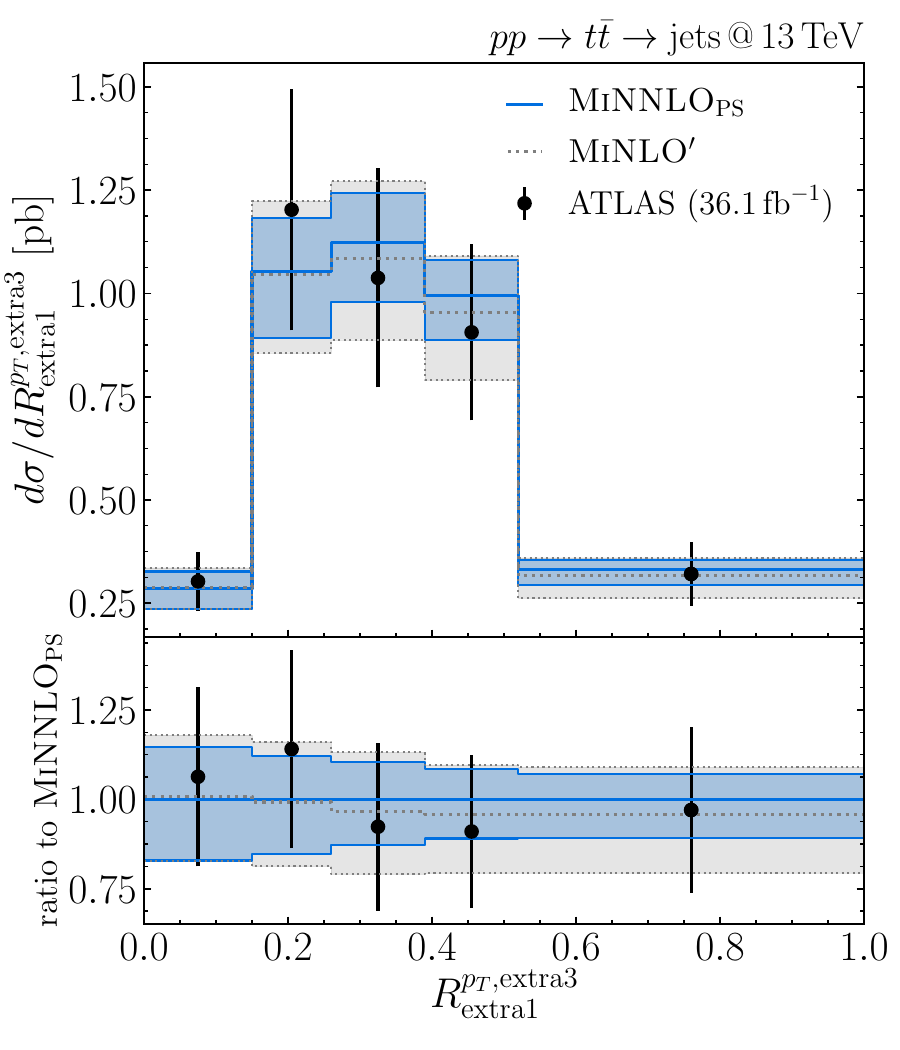} 
&
\hspace{-0.6cm}
\includegraphics[width=.34\textwidth,page=20]{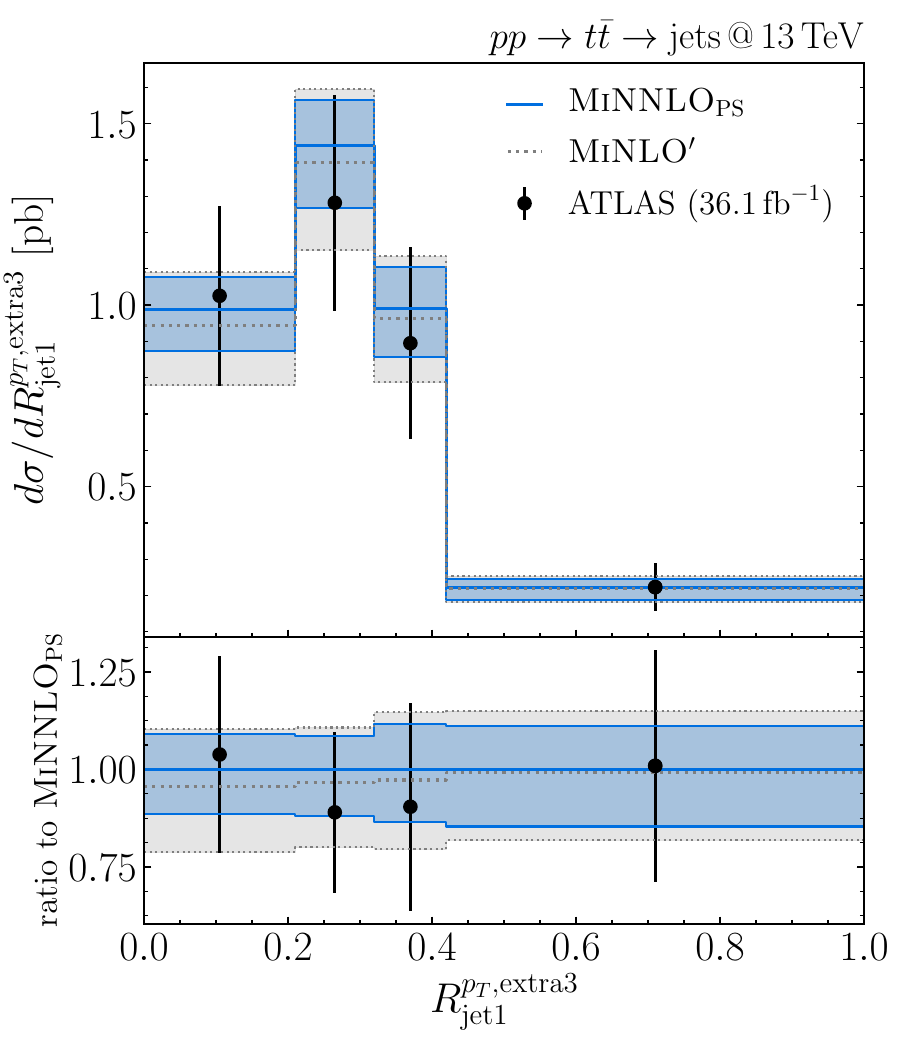}
\end{tabular}\caption{\label{fig:hadronic_appendix_2}  Comparison of \minnlo{} (blue, solid) and \minlo{} (black, dashed) predictions with ATLAS data \cite{ATLAS:2020ccu} (black points with errors) in \setuphadronic{}.}
\end{center}
\end{figure}

\begin{figure}[t!]
\begin{center}
\begin{tabular}{ccc}
\hspace{-.55cm}
\includegraphics[width=.34\textwidth,page=21]{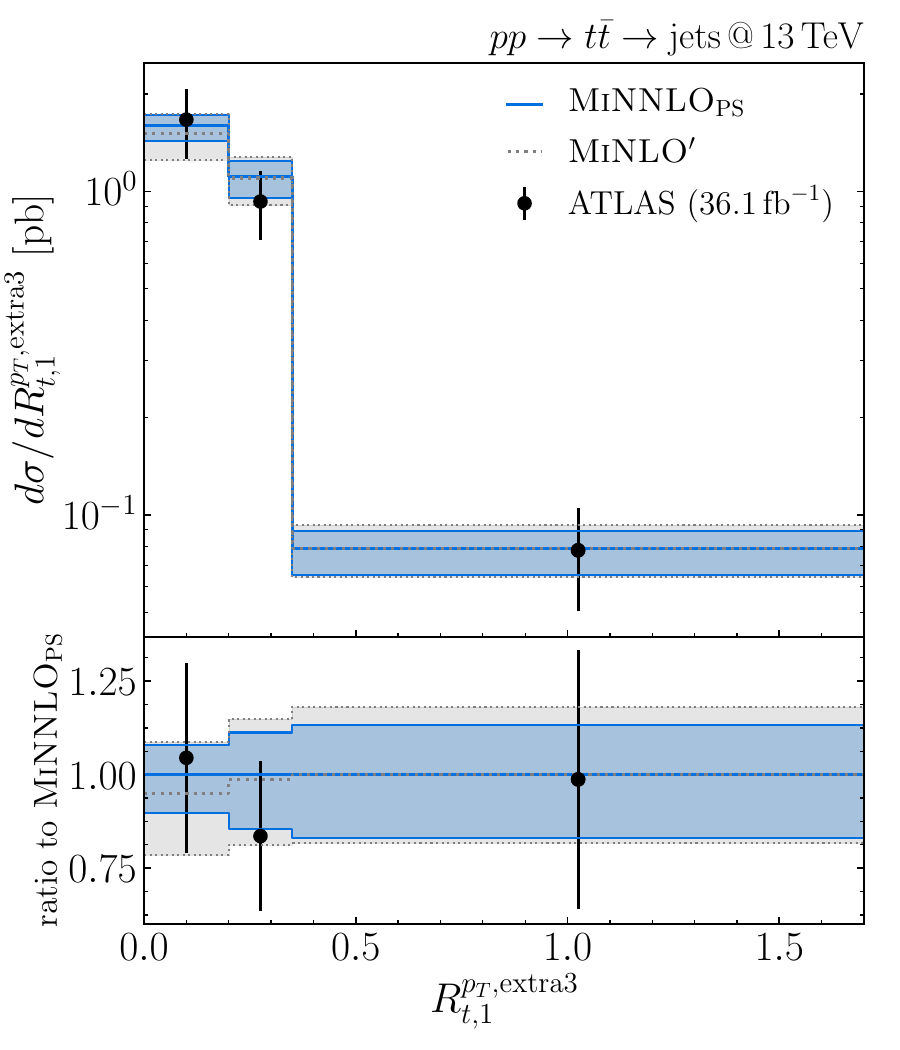}
&
\hspace{-0.6cm}
\includegraphics[width=.34\textwidth,page=22]{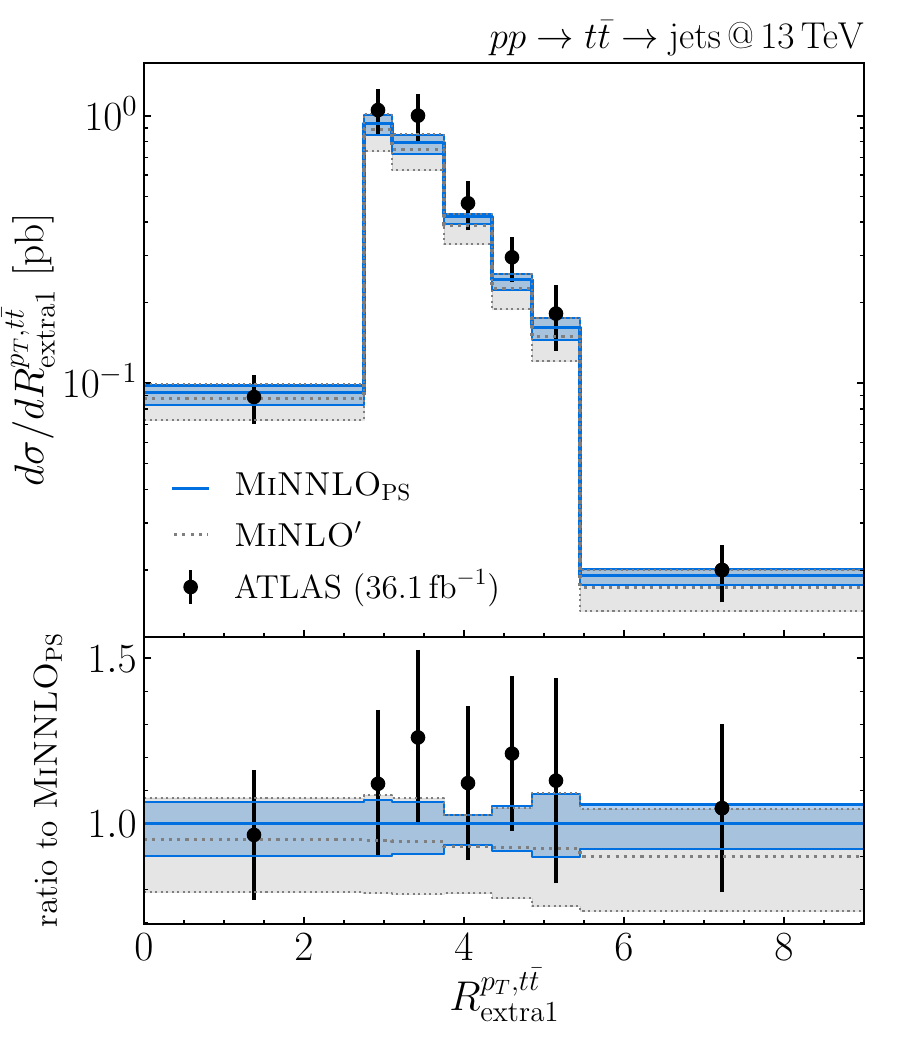} 
&
\hspace{-0.6cm}
\includegraphics[width=.34\textwidth,page=24]{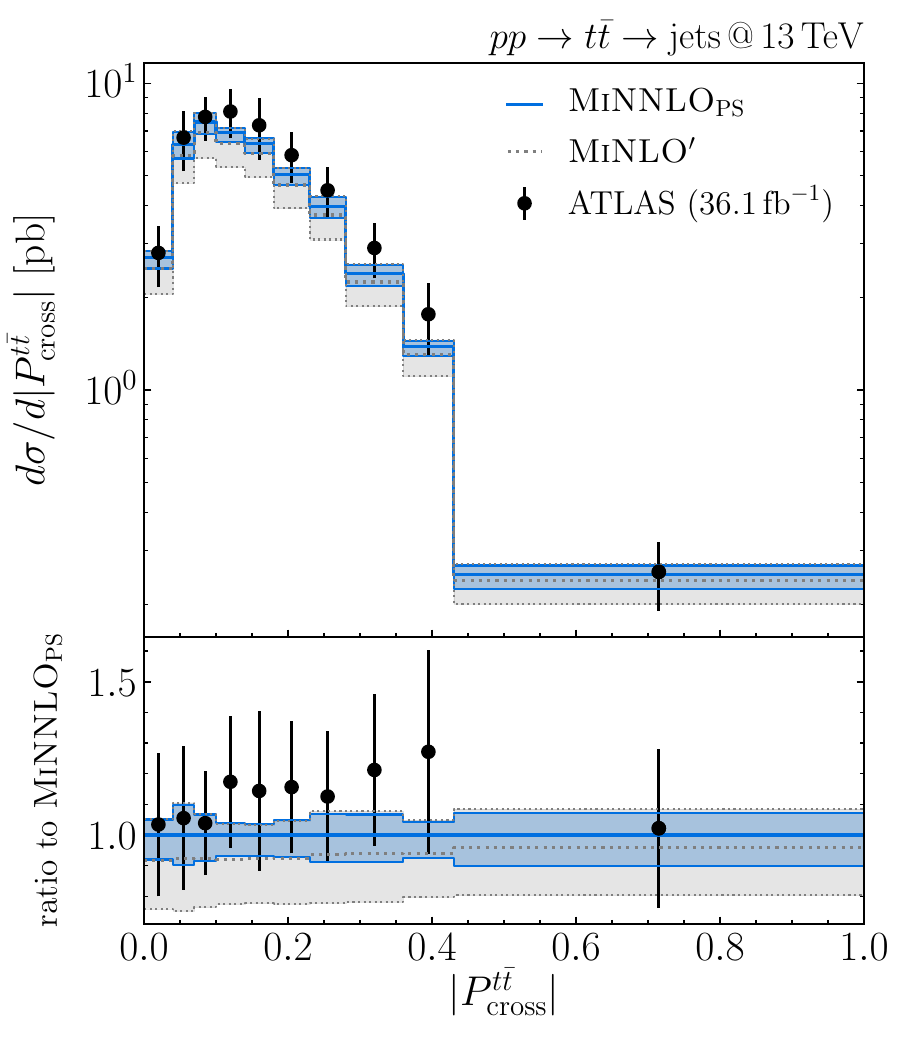}\\
\hspace{-.55cm}
\includegraphics[width=.34\textwidth,page=25]{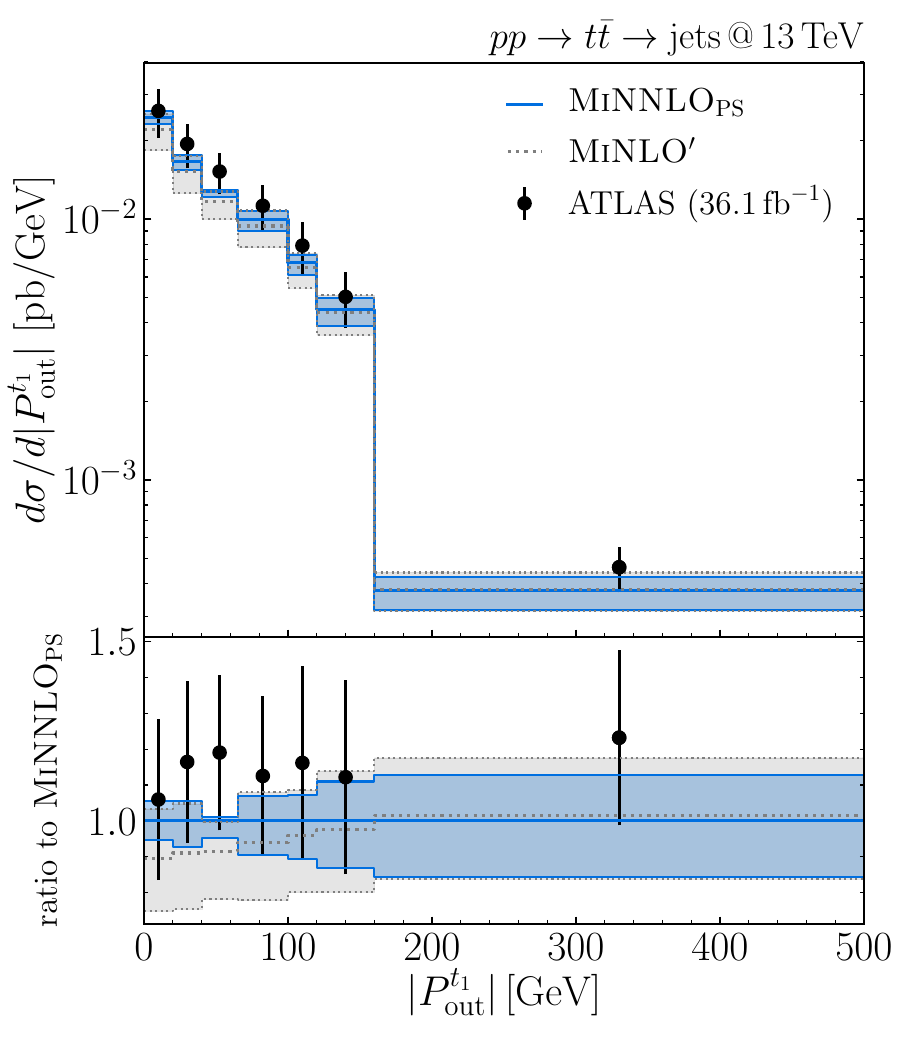}
&
\hspace{-0.6cm}
\includegraphics[width=.34\textwidth,page=26]{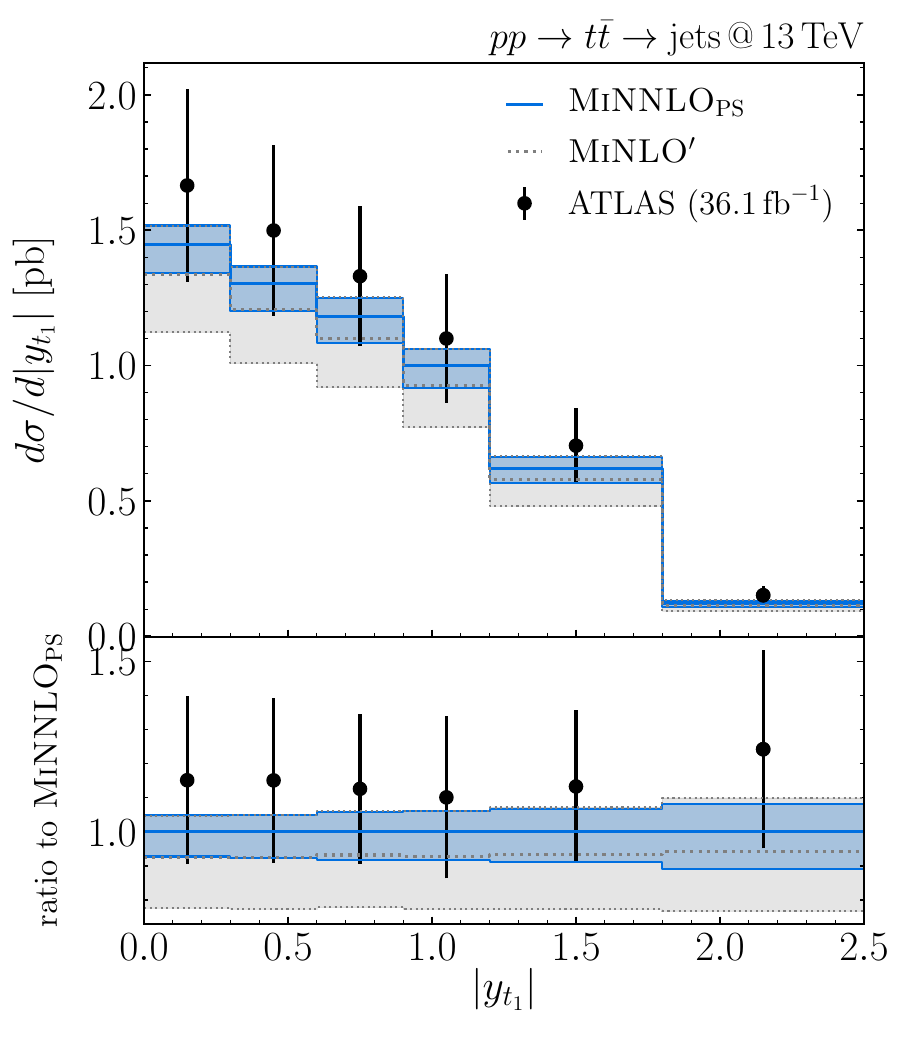} 
&
\hspace{-0.6cm}
\includegraphics[width=.34\textwidth,page=27]{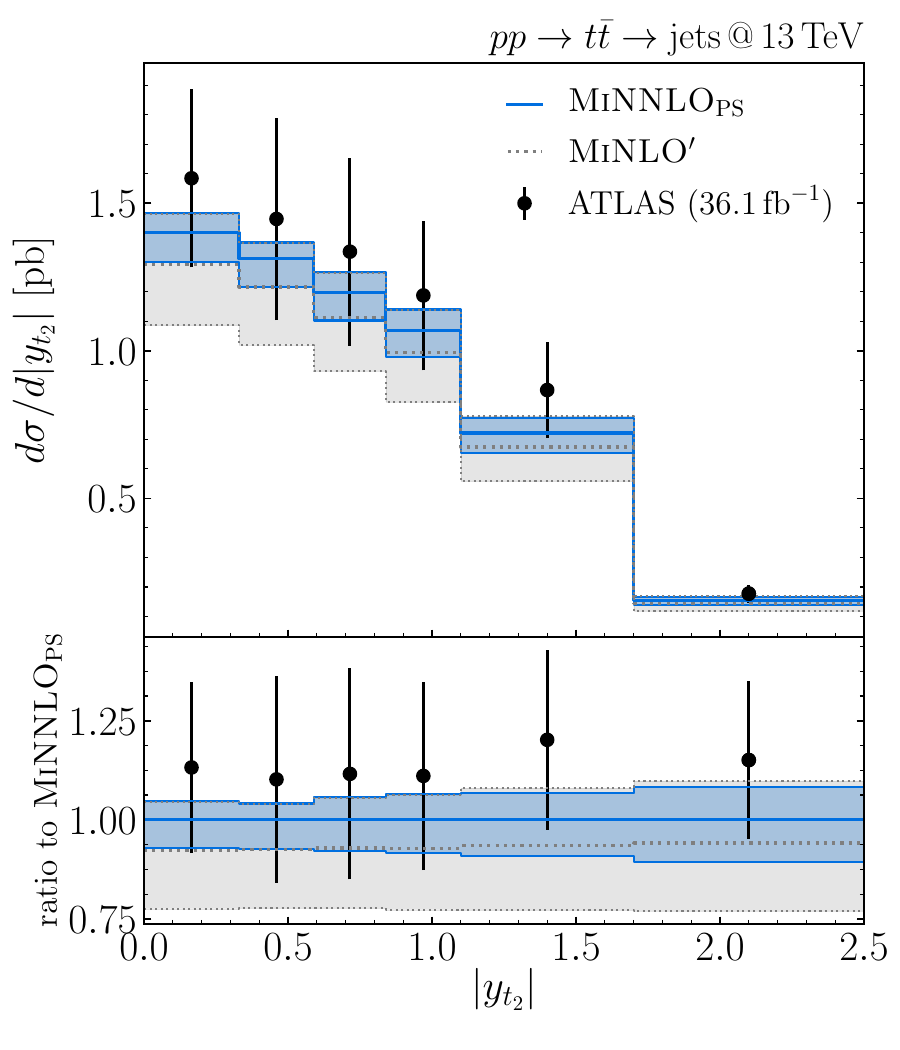}\\
\hspace{-.55cm}
\includegraphics[width=.34\textwidth,page=28]{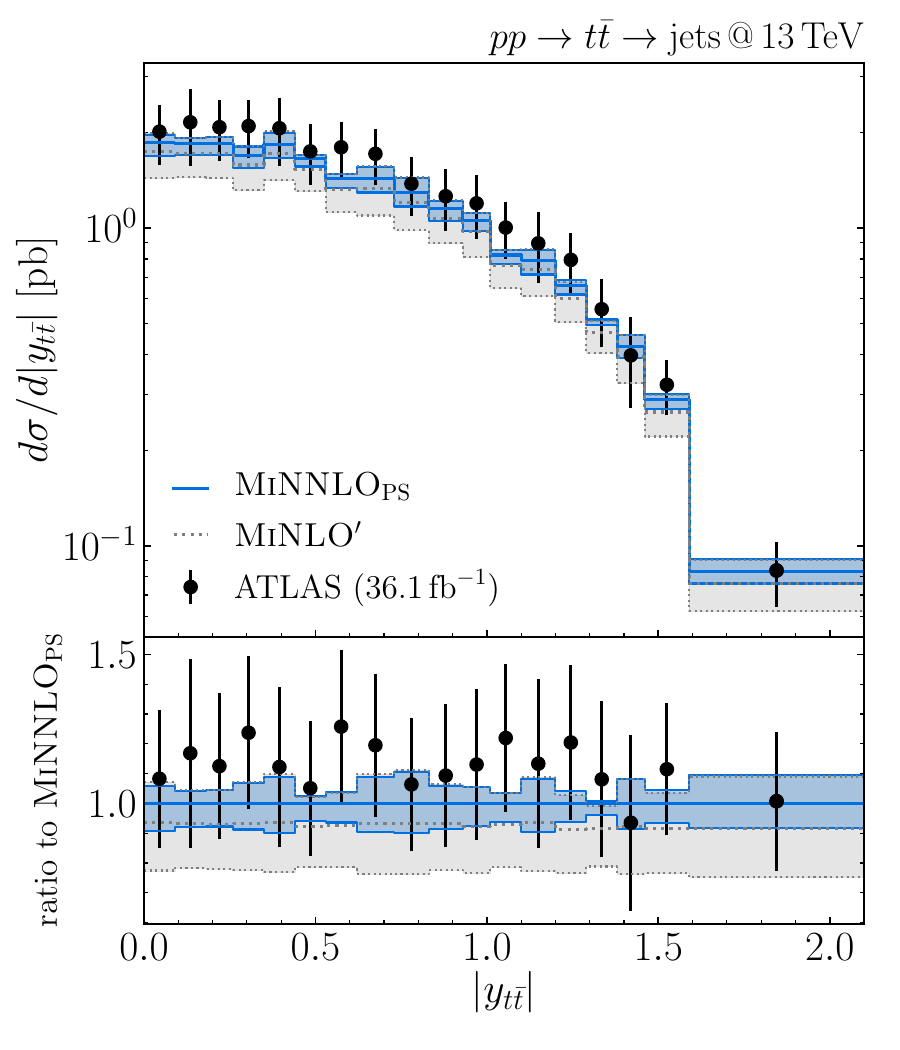}
&
\hspace{-0.6cm}
\includegraphics[width=.34\textwidth,page=33]{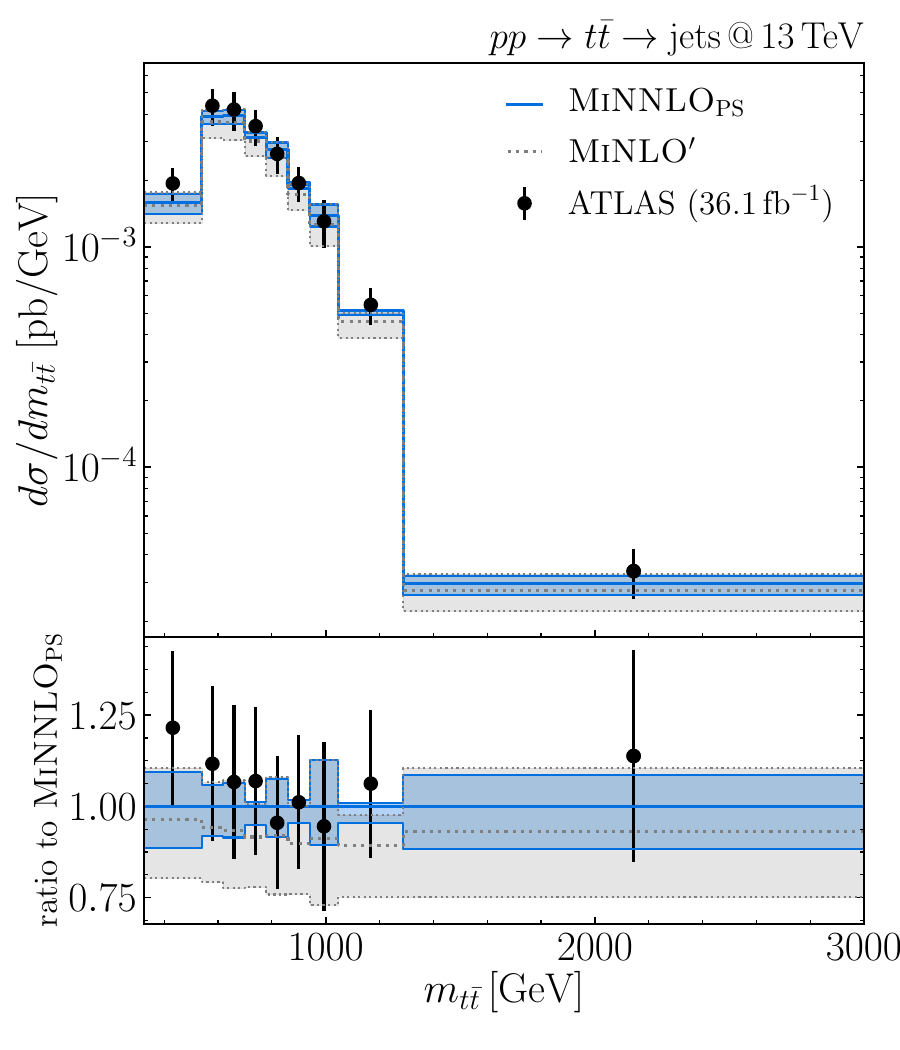} 
&
\hspace{-0.6cm}
\includegraphics[width=.34\textwidth,page=34]{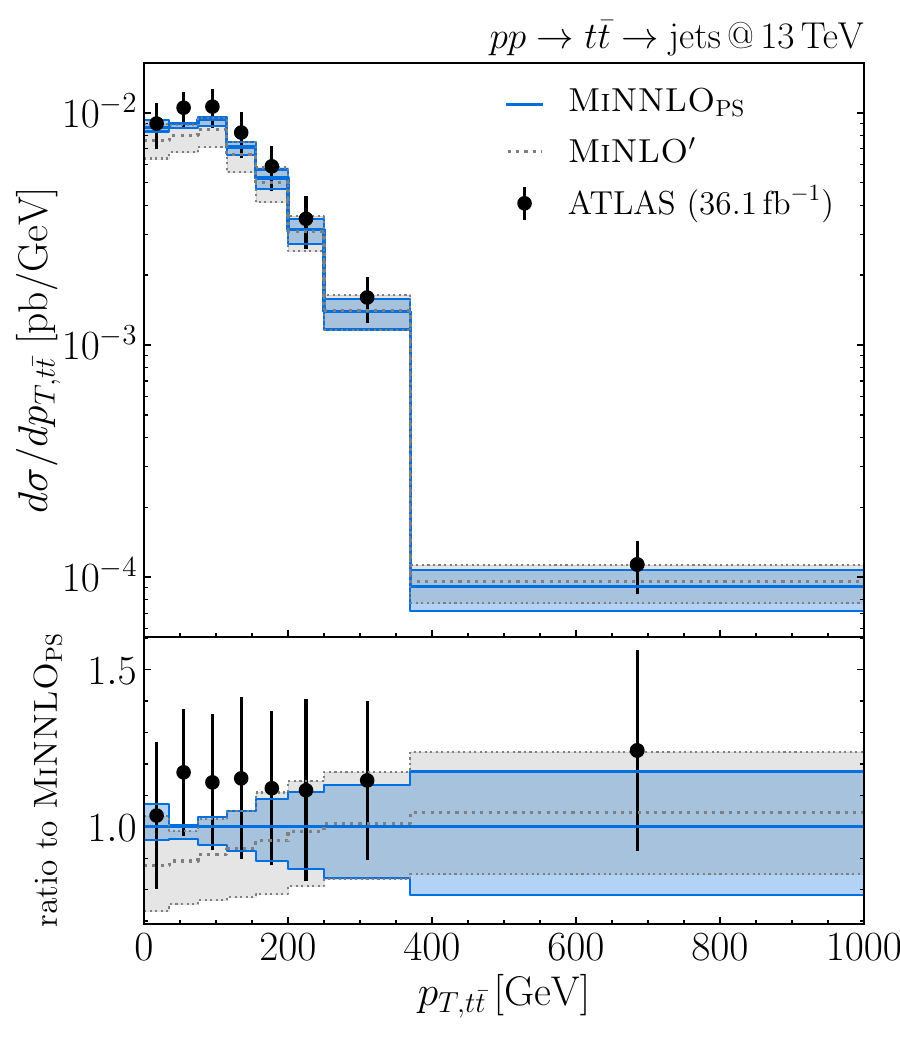}
\end{tabular}\caption{\label{fig:hadronic_appendix_3}  Comparison of \minnlo{} (blue, solid) and \minlo{} (black, dashed) predictions with ATLAS data \cite{ATLAS:2020ccu} (black points with errors) in \setuphadronic{}.}
\end{center}
\end{figure}

\clearpage

\newpage
\addcontentsline{toc}{section}{References}
\bibliography{MiNNLO}
\bibliographystyle{JHEP}

\end{document}